\documentclass[fleqn,usenatbib]{mnras}

\usepackage{mathptmx}

\usepackage[T1]{fontenc}

\DeclareRobustCommand{\VAN}[3]{#2}
\let\VANthebibliography\thebibliography
\def\thebibliography{\DeclareRobustCommand{\VAN}[3]{##3}\VANthebibliography}

\usepackage{graphicx}	
\usepackage{amsmath}	
\usepackage{txfonts}   
\usepackage{multirow}  
\usepackage{tablefootnote} 
\usepackage{threeparttable}  

\title[Binarity of B-type stars in young clusters]{Characterisation of B-type stars in four young Galactic open clusters. I. Stellar content and binary properties\thanks{Based on observations made with the Mercator Telescope operated on the island of La Palma by the Flemish Community, at the Spanish Observatorio del Roque de los Muchachos of the Instituto de Astrofisica de Canarias.}}

\author[F. Nardini et al.]{
F. Nardini$^{1,2}$,
J. Bodensteiner$^{3,4}$, 
H. Sana$^{2}$,
L. Mahy$^{5}$,
K. Deshmukh$^{2}$
and D. M. Bowman$^{1,2}$
\\
$^{1}$School of Mathematics, Statistics and Physics, Newcastle University, Newcastle upon Tyne, NE1 7RU, United Kingdom\\
$^{2}$Institute of Astronomy, KU Leuven, Celestijnenlaan 200D, 3001 Leuven, Belgium\\
$^{3}$Anton Pannekoek Institute for Astronomy, University of Amsterdam, 1090 GE Amsterdam, The Netherlands \\
$^{4}$European Southern Observatory, Karl-Schwarzschild-Strasse 2, Garching by Munich, Germany\\
$^{5}$Royal Observatory of Belgium, Avenue Circulaire/Ringlaan 3, 1180 Brussels, Belgium
}

\date{Accepted 2025 May 12. Received 2025 May 09; in original form 2025 January 28}

\pubyear{\the\year{}}

\begin{document}
\label{firstpage}
\pagerange{\pageref{firstpage}--\pageref{lastpage}}
\maketitle

\begin{abstract}
Multiplicity among massive stars represents a major uncertainty in stellar evolution theory, because the extra physical processes that it introduces significantly impacts each star's structure. While multiplicity of O-type stars is fairly well constrained, for B-type stars it is not. B-type stars are more common and have longer lifetimes, thus providing an opportunity to characterize multiplicity at different ages. 
Moreover, young open clusters are advantageous for studying coeval and chemically homogeneous environments. Using a multi-epoch spectroscopic campaign with the HERMES spectrograph we determine multiplicity properties and rotation rates of 74 B-type stars in four Galactic open clusters: $h$ and $\chi$ Persei, NGC~457, NGC~581, and NGC~1960. 
We measure radial velocities with a cross-correlation method and determine tentative orbital solutions for 26 of the 28 identified binaries. We detect several Be stars, five of them being members of binary systems. 
We correct the observed binary fractions for observational biases and determine an average intrinsic binary fraction of 79$_{-16}^{+19}\%$. The consistency in binary fraction among the four clusters, which are between 15 and 30~Myr, suggests a reasonably homogenous binary fraction across this age range.
We used \textsc{tlusty} atmosphere models to determine the projected rotational velocities, with an average value of 240~km\,s$^{-1}$ for both single and binary systems. Whereas, the Be stars show higher velocities between 200 and 360~km\,s$^{-1}$. 
\end{abstract}




\section{Introduction}

In recent years, the detection of gravitational waves from merging compact objects has increased significantly thanks to ground-based observatories including the LIGO/Virgo/KAGRA collaboration \citep[e.g. ][]{Abbott_2023}. Detections will significantly increase with the advent of the Laser Interferometer Space Antenna \citep[LISA; ][]{Amaro} and the Einstein telescope \citep{Maggiore_2020}. To fully understand compact-object mergers, we must also understand the physics of the progenitor stars. The majority of neutron star progenitors are early B-type stars (i.e. birth masses between 8~M$_{\odot}\lesssim$~M~$\lesssim$~16~M$_{\odot}$), which typically end their lives as core-collapse supernovae \citep{DeMink_2015}. 

The majority of massive stars are found in binary or multiple systems \citep{Kobulnicky_2012, sana2012, sana2013, sana_2014, Dunstall, Moe, banyard} and to constrain the multiplicity properties of OB-type stars, it is important to consider field stars as well as those in clusters and associations across different ages. Surveys of stars in clusters are volume-limited and have the advantage that the stars were born from the same molecular cloud, hence they can be assumed to be coeval and having the same metallicity \citep{Lada_2003}.
For O-type stars, binary fractions are generally well-established. Results for O-type stars in young Galactic clusters and in the 30 Doradus star-forming region in the Large and Small Magellanic Clouds (LMC and SMC) where they reported a binary fraction corrected for observational biases between 51 and 70$\%$ \citep{sana2012, sana2013, Kobulnicky}.
These results are consistent with what was found for Galactic field O-type stars from GOSSS \citep{Sota_2014}, IACOB \citep{simon-diaz}, and OWN \citep{Barba_2017}, and the Small Magellanic Cloud \citep[SMC; e.g. RIOTS survey; ][]{Lamb_2016} where they similarly reported observed spectroscopic binary fractions of around $43-65\%$ and $59\pm12\%$, respectively.

In contrast, the spectroscopic binary properties of B-type stars remain less constrained, especially across different ages. Some studies of (very) young Galactic clusters ($\lesssim 7$~Myr) and the 30 Doradus region in the LMC find binary fractions for B-type stars corrected for observational biases of 58$\pm11\%$ \citep{Dunstall} and $52\pm8\%$ \citep{banyard}, consistent with the surveys of O-type stars. 
Meanwhile, recent investigations of B-type stars in older clusters in the LMC and the SMC (i.e. ages between 40-100~Myr) report a significantly lower binary fraction of 25-35$\%$ \citep{Bodensteiner_I, Saracino_2023}, which might be an age or a mass effect since these clusters are populated by less massive stars.
The longer nuclear timescale of B-type stars with respect to O-type stars allows us to investigate their multiplicities across different ages.
Moreover, their prevalence provides critical insight into binary evolution and cluster dynamics across different environments and ages. Notably, studies have focused predominantly on very young clusters (i.e. $1-10$~Myr), leaving an important observational gap for studying multiplicity in B-type stars in clusters in the $15-30$~Myr age range.

Other than their coeval nature, a population of stars born in the same cluster provides an ideal environment for identifying binary systems that are likely to interact in the future, and also binaries that have interacted already, so-called binary interaction products.
Blue stragglers and Oe/Be-type stars are two examples of stars believed to be binary interaction products \citep{Marchant_bodensteiner}.
Blue stragglers are thought to be rejuvenated stars through either mergers or mass transfer \citep{Ferraro_1997, Schneider_2015}, whereas Oe/Be stars are thought to be spun up by mass and angular momentum transport \citep{Pols1991, Hastings2021}.
For Oe/Be stars, this hypothesis is supported by the low observed multiplicity fractions, suggesting that Oe/Be stars have different binary properties than OB stars \citep{Bodensteiner_beform, Bodensteiner_Bloem}. 
Additionally, these binary interaction products appear to occupy specific regions on the Hertzsprung-Russell (HR) diagram (and its observational equivalent being the colour-magnitude diagram; CMD).
Using binary population synthesis, \citet{wang, Wang2022, wang_2023} identified the typical locations of both pre- and post-interaction stars on these diagrams. Their findings revealed that after about 20 Myr, four different populations can be identified: (i) binary systems that never interact and pre-interacting binary systems that are on a single-star isochrone, defined as the red main sequence; (ii) mergers and semi-detached binary systems located above the main sequence turn-off; (iii) mass transfer products which are typically located to the right of the red main sequence towards cooler temperatures,  defined as extended red main sequence; and (iv) the blue main sequence located to the left of the red main sequence\footnote{We use the same terminology of blue and red main sequence to be consistent with \citet{wang, Wang2022, wang_2023}.}, comprising of the pre-main sequence mergers, which creates an apparent bifurcation in the main sequence.

The bifurcated main sequence for binary interaction products on the HR~diagram (or equivalently the CMD) is complemented by differences in rotational velocity. 
\citet{Gossage2019} and \cite{wang_2023} argue that stars on the blue main sequence are slowly rotating, whereas stars on the red main sequence are rotating at about 65\% of critical rotation, and stars on the extended red main sequence tend to have near-critical rotation, which is where we tend to find classical Be stars.
Therefore, binarity deeply affects the rotation of stars. Population synthesis simulations further reinforce this idea, indicating that fast rotators are likely the result of binary mass transfer \citep{Pols1991, demink2013, Hastings2021,wang_2023}. At the same time, some merger products and pre-interaction systems can show slower rotation rates  \citep{Schneider2019, Wang2022}.
To investigate the multiplicity of B-type stars in young clusters within the poorly studied intermediate-age range (i.e. $15-30$~Myr), where theoretical predictions suggest distinct binary and rotational velocity properties for pre- and post-interaction binaries, we performed a multi-epoch spectroscopic campaign of the B-star population of four different Galactic open clusters ($h$ and $\chi$ Persei, NGC~457, NGC~581 and NGC~1960).
This paper focuses on the spectroscopic characterisation of the B-type stars in the clusters to determine their multiplicity fractions, the distributions of the orbital parameters of the identified binaries, and the projected rotational velocities of the whole sample.
Section~\ref{sect:obs_data_red} summarises the target selection and the data reduction. Section~\ref{sect:spect_type} discusses how the spectral classification was performed and presents the distribution of spectral types of stars among the four clusters. The methodology for identifying binary systems is summarised in Sect.~\ref{sect:binary_detect} along with the methodology for determining the orbital parameters and the bias correction procedure. In Sect.~\ref{sect:discuss_binary} the results of the multiplicity analyses are discussed along with comparisons with previous works. In Sect.~\ref{sect:atm_parameters} both the methodology and the results of the projected rotational velocities are discussed. Lastly, we give a summary and conclude in Sect.~\ref{sect:conclusion}.

\section{Observations and data reduction}\label{sect:obs_data_red}

\subsection{Cluster and target selection}

The four Galactic open clusters ($h$ and $\chi$ Persei, NGC~457, NGC~581 and NGC~1960) were selected because their literature ages range from 15 to 30~Myr. These clusters are young enough to contain early-B-type stars yet also sample a significant age range. The characteristics of each cluster, reported in Table \ref{tab:Clusters_param}, were obtained from the GAIA DR2-based catalogue by \citet{Cantat_Gaudin_2020} and its subsequent update by \citet{Dias2021}, also DR2-based. Given the similarities in age, metallicity, distance and extinction, $h$ and $\chi$ Persei are often referred to as double cluster, hence they were combined and their constituent stars were analysed as members of a single group.

\begin{figure*}%
    \centering
    \includegraphics[width=0.45\textwidth]{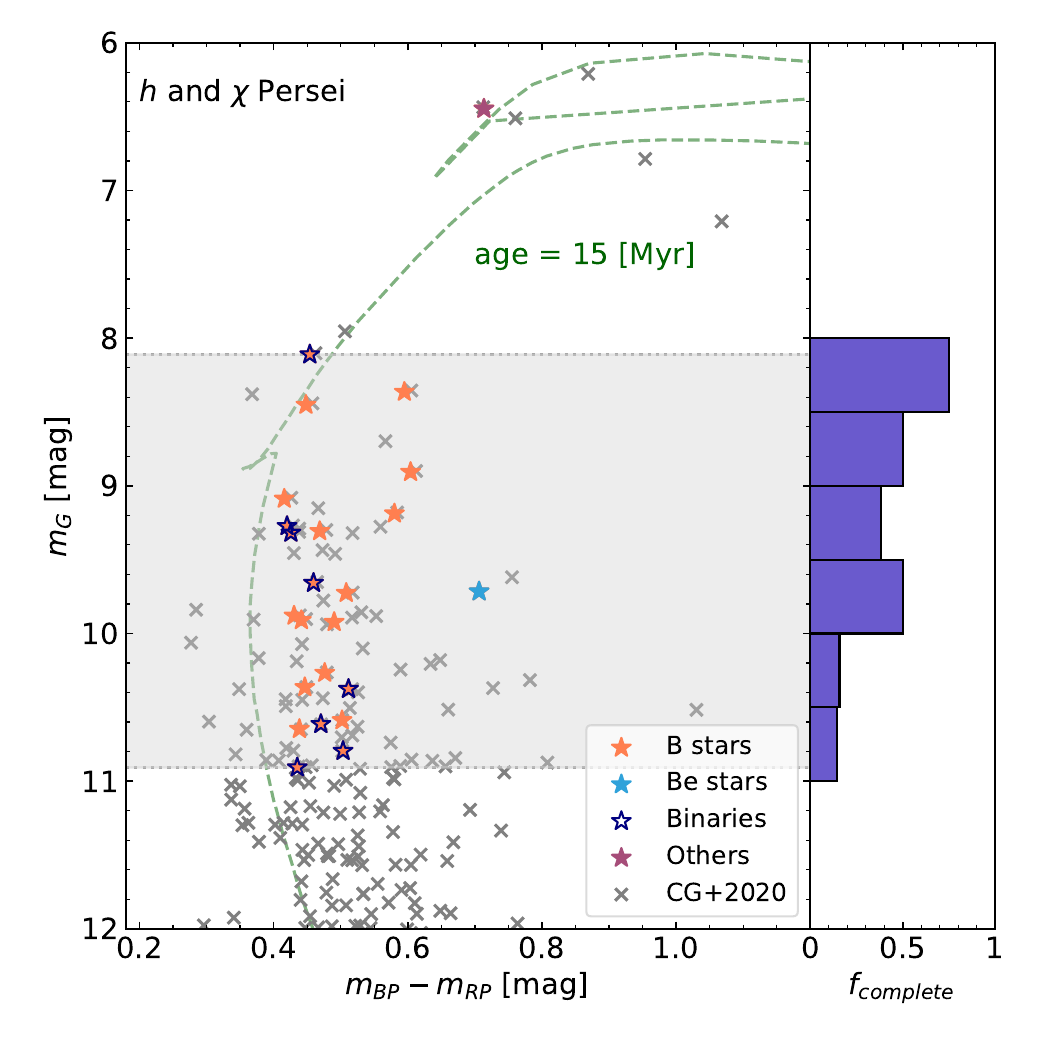}
    \includegraphics[width=8cm]{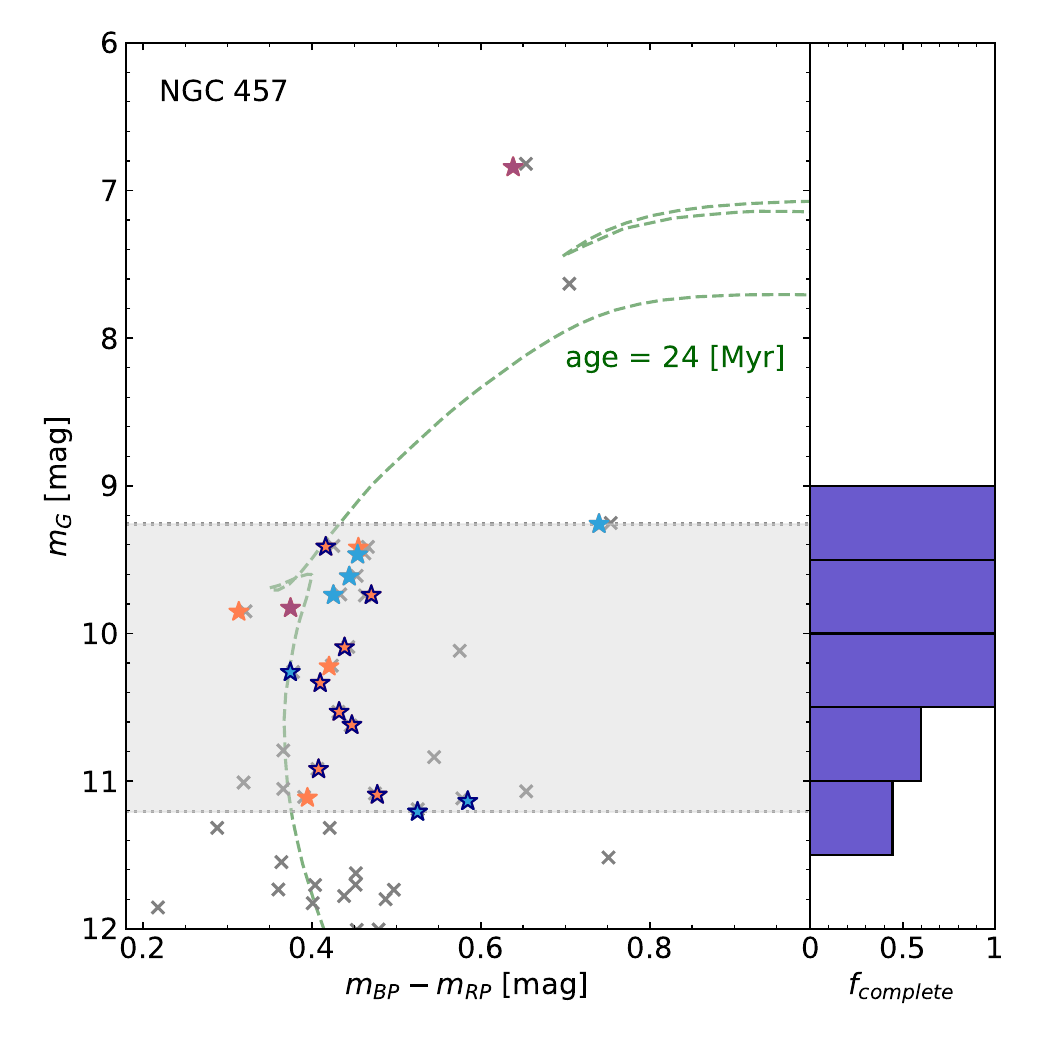}

    \includegraphics[width=8cm]{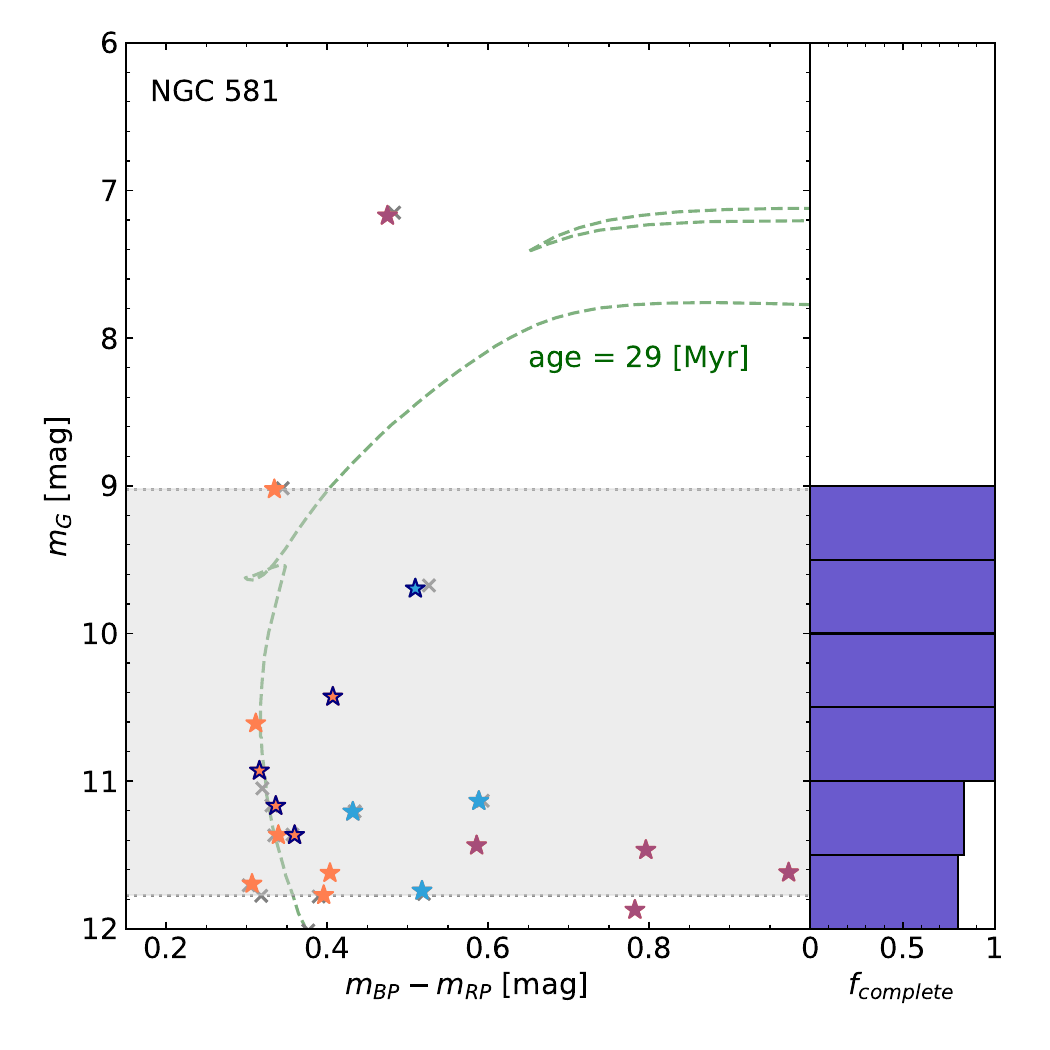}
    \includegraphics[width=8cm]{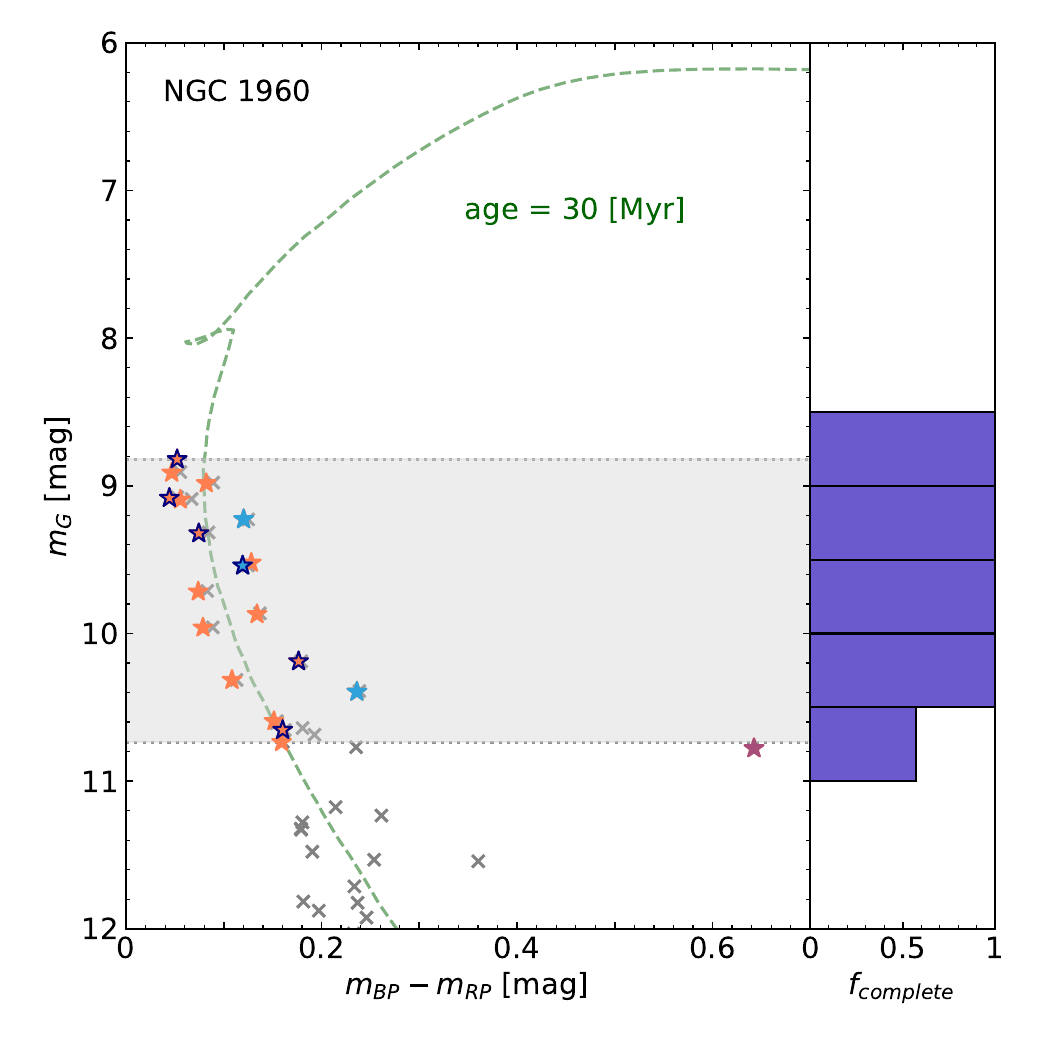}
    \vspace*{-1mm}
    \caption
    {CMD of the four clusters with different colours representing different types of stars, where "Others" represent the post-main sequence and the non-B-type stars, while the blue edge colour of a symbol represents the identified binaries. The vertical histograms show the completeness in bins of 0.5 magnitudes. The grey regions are the magnitude ranges where the completeness fractions for the B-type main-sequence stars in our sample were computed. The isochrones were computed using the Padova database of stellar evolution tracks and isochrones \citep{Bressan, Chen_parsec}. The legend in the first panel is the same for all four figures. We note that the offset between the catalogue and our sample is due to the different GAIA data releases used.}%
    \label{fig:cmd_completeness}%
\end{figure*}

Following the selection of the four clusters, the target star selection was performed using the astronomical database SIMBAD \citep{SIMBAD} and then expanded using the \citet{Cantat_Gaudin_2020} catalogue. We selected stars of spectral type B and V$<$12 mag to ensure feasible observations with the HERMES spectrograph \citep[see Section \ref{sec:data_red}; ][]{raskin_2011, Raskin}.

Our initial sample comprised 105 stars across all four clusters. The catalogue of \citet{Cantat_Gaudin_2020} was also used to determine the completeness of each cluster's sample. Table \ref{tab:Clusters_param} shows the completeness fractions for the clusters. These were computed by taking the ratio of the B-type main-sequence stars in our sample with the total number of B-type main sequence stars in the literature between the lowest and highest magnitude range since not all B-type main-sequence stars have available HERMES spectroscopy.  Moreover, stars identified as not having a spectral type B and those likely to be post-main sequence stars were removed from our sample.
Some of the members of NGC~581 and NGC~1960 are not listed in the catalogue of \citet{Cantat_Gaudin_2020}, because of their chosen proper motion threshold excluding stars with higher proper motion and hence not classifying them as cluster members. 
For NGC~581 three stars were reported as members by \citet{ngc581_1974}, while one star is not classified as a member in the literature but its properties are coherent with the properties of the cluster. For NGC~1960, the two stars affected by this were classified as members by \citet{ngc1960_1951}.

Figure~\ref{fig:cmd_completeness} shows the CMDs of the four clusters, as well as indicative isochrones computed using the Padova database of stellar evolution tracks \citep[PARSEC; ][]{Bressan, Chen_parsec} based on the ages determined by \citet{Dias2021}.
There is an offset between the magnitudes of the stars in our sample and the stars in the catalogue of \citet{Cantat_Gaudin_2020} which is due to the different Gaia data release version since we used the Gaia DR3 database.
In general, the literature errors associated with isochrone-derived ages are likely underestimated, therefore the ages of our clusters are subject to uncertainties. 
We emphasize that these isochrones are not fits to the CMD, but are shown in Fig.~\ref{fig:cmd_completeness} to guide the eye.

\begin{table*}
\footnotesize
\caption[Parameters of the four clusters analysed]%
{Right ascension (RA) and declination (DEC) of each cluster, as well as the distance, age and extinction from the \citet{Cantat_Gaudin_2020} and \citet{Dias2021} catalogues. The final columns show the number of stars in a cluster ($N_{\rm cluster}$), those that were filtered based on a brightness threshold for spectroscopic follow-up and their spectral type ($N_{\rm m_G}$), and those that comprise the sample of B-type stars in this study ($N_{\rm sample}$). The final column is the completeness for each cluster of our sample of main sequence B-type stars with respect to the main sequence B-type stars in the literature \cite[e.g.][]{Cantat_Gaudin_2020}.}
\begin{tabular}{ccccccccccc}
\hline \hline
Cluster    & RA     & DEC     & distance      & age          & $A_V$         & N$_{\rm cluster}$ & N$_{\rm m_G}$ & N$_{\rm sample}$    & Completeness            \\
           & [deg]  & [deg]   & [kpc]         & [Myr]        & [mag]         &                   &                     &                         \\ \hline
h Per      & 34.741 & +57.134 & 2.25$\pm$0.06 & 12.9$\pm$1.1 & 1.75$\pm$0.04 & 720      & \multirow{2}{*}{79}         & \multirow{2}{*}{23} & \multirow{2}{*}{29$\%$} \\
$\chi$ Per & 35.584 & +57.149 & 2.15$\pm$0.11 & 15.4$\pm$1.1 & 1.71$\pm$0.02 & 483               &                     &                         \\
NGC~457    & 19.887 & +58.278 & 2.54$\pm$0.13 & 23.6$\pm$1.2 & 1.61$\pm$0.02 & 594   &   23     & 19                  & 83$\%$                  \\
NGC~581    & 23.339 & +60.659 & 2.29$\pm$0.09 & 29.4$\pm$1.3 & 1.45$\pm$0.04 & 152    &  16    & 14                  & 88$\%$                 \\
NGC~1960   & 84.084 & +34.135 & 1.32$\pm$0.10 & 30.2$\pm$1.2 & 0.93$\pm$0.02 & 299    &  22      & 18                  & 82$\%$                 \\ \hline
\end{tabular}
\label{tab:Clusters_param}
\end{table*}

\subsection{Spectroscopic observations, reduction and normalisation}\label{sec:data_red}

Spectroscopic observations were performed using the state-of-the-art High Efficiency and Resolution Mercator Echelle Spectrograph (HERMES) spectrograph mounted on the 1.2-m Mercator telescope, located at the Roque de Los Muchachos Observatory on La Palma (Spain).
HERMES is a fibre-fed prism-cross-dispersed echelle spectrograph based on a white pupil layout \citep{raskin_2011, Raskin}. It has a detector size of 2048$\times$4608 13.5-\textmu m pixels, with a wavelength-dependent coating, a spectral resolving power of $R=85\,000$ (2.2 pixel sampling), and broad wavelength coverage of $3800-9000$~$\AA$ \citep{raskin_2011, Raskin}.

The raw spectra were automatically reduced by the HERMES reduction pipeline (HERMES-DRS, v.7.0). The reduction includes bias correction, flat-fielding, wavelength calibration, cosmic and dark removal and correction for barycentric motion. The normalisation was performed manually for all epochs of each star by fitting a spline to the continuum of the spectrum, with knot points chosen specifically to exclude spectral lines. Each normalised epoch was then inspected by eye to ensure robustness. 

Our {\sc HERMES} observations were done in a service-mode setup, and therefore the time coverage and time difference between epochs vary among the targets. On average we have ten epochs for each star, except for NGC~581 where in a few cases only four to six epochs are available. Our dedicated HERMES monitoring campaign spanned up to 3~yr for $h$ and $\chi$ Persei and NGC~1960, and up to 2~yr for NGC~457 and NGC~581. Additionally, the IACOB spectroscopic database \citep{IACOB} was checked for archival spectra such that snapshot observations since 2013 were utilised.
Only spectra with a minimum signal-to-noise ratio (S/N) of 40 per epoch were utilized, with a few exceptions.

\section{Spectral types}\label{sect:spect_type}

The spectral types of all stars were homogeneously determined. The broad wavelength coverage of the HERMES spectra includes the blue part of the optical spectral range (i.e. 3800--4700~$\AA$) which is usually used to perform spectral classification of early-type stars, thus allowing us to use the common classification schemes based on standard stars by \citet{gray_book}.

Table \ref{tab:NIST} provides the main spectral lines used for the spectral classification and also for radial velocity (RV) determination (see Sect.~\ref{sect:radial_vel}). Ionized helium (\ion{He}{II}) lines, which are present for stars earlier than spectral type B0.5 but are absent for later-type stars, along with the strength of \ion{He}{I} lines are important features for distinguishing between early-B and late-O stars. The ratio of \ion{He}{I} $\lambda$4471 to \ion{Mg}{II} $\lambda$4481 lines is used in the classification of both early- and late-B stars \citep{gray_book}.
The luminosity class was determined based on the wings of the Balmer lines, which are less pronounced for stars with lower surface gravity and indicate more evolved stars. However, this is degenerate with rotational and other forms of broadening mechanisms \citep{simon-diaz}. 

\begin{table}
\centering
\caption[caption]
{Spectral lines used for the spectral classification and the RVs determination. The values of the rest wavelength were obtained from the Atomic Spectra Database NIST\footnotemark.}
\begin{tabular}{cc}
\hline \hline
Line             & Rest wavelength [$\AA$] \\ \hline
\ion{He}{I}      & 4009.26                    \\
\ion{He}{I} + II & 4026.19                    \\
\ion{He}{I}      & 4143.76                    \\
\ion{C}{II}      & 4267.26                    \\
\ion{He}{I}      & 4387.93                    \\
\ion{He}{I}      & 4471.48                    \\
\ion{Mg}{II}     & 4481.13                    \\
\ion{Si}{III}    & 4552.62                    \\
\ion{He}{II}    & 4685.80                   \\
\ion{He}{I}      & 4921.93                   \\
\ion{He}{I}      & 5047.74                    \\
\ion{He}{I}      & 5875.62                    \\
\ion{He}{I}      & 6678.15                    \\ \hline
\end{tabular}
\label{tab:NIST}
\end{table}
\footnotetext{\url{https://physics.nist.gov/PhysRefData/ASD/lines_form.html?}}

Out of the 105 cluster member stars included in our {\sc HERMES} campaign, 74 were classified as B-type dwarf and giant stars. 
Five stars were classified as spectral type between B7 and B9 (i.e. birth masses approximately between 2 and 4~$\rm M_{\odot}$) in NGC 1960. We exclude these late-B stars in $N_{\rm sample}$ to ensure a homogeneous mass range across all four clusters. Three early B-type stars in $h$ and $\chi$ Persei were also excluded since they had only one spectrum available and it was not possible to determine their binary status.
The remaining 23 others were classified as spectral type A or post-main sequence stars (i.e. supergiants). The supergiants were not included in any further analysis since they are not the focus of this work and hence not included within $N_{\rm sample}$ in Table~\ref{tab:Clusters_param} either. We refer the reader to \citet{deburgos_1, deburgos_2} for an in-depth analysis of Galactic blue supergiants.

Be stars are classified if a spectrum displayed any hydrogen lines in emission. In our sample, 15 of the 74 stars were identified as Be stars, the majority of which are members of NGC~457, while only one Be star (BD+56~566) was found in $h$ and $\chi$ Persei. Most of the identified Be stars exhibited double-peaked emission lines in at least one line within the Balmer series. However, one exception (NGC 1960 101) presented a `bottle-shaped' \ion{H$\alpha$}{} emission line, indicating that the star is viewed (near) pole-on.

\begin{figure}
    \centering
    \includegraphics[width=0.45\textwidth]{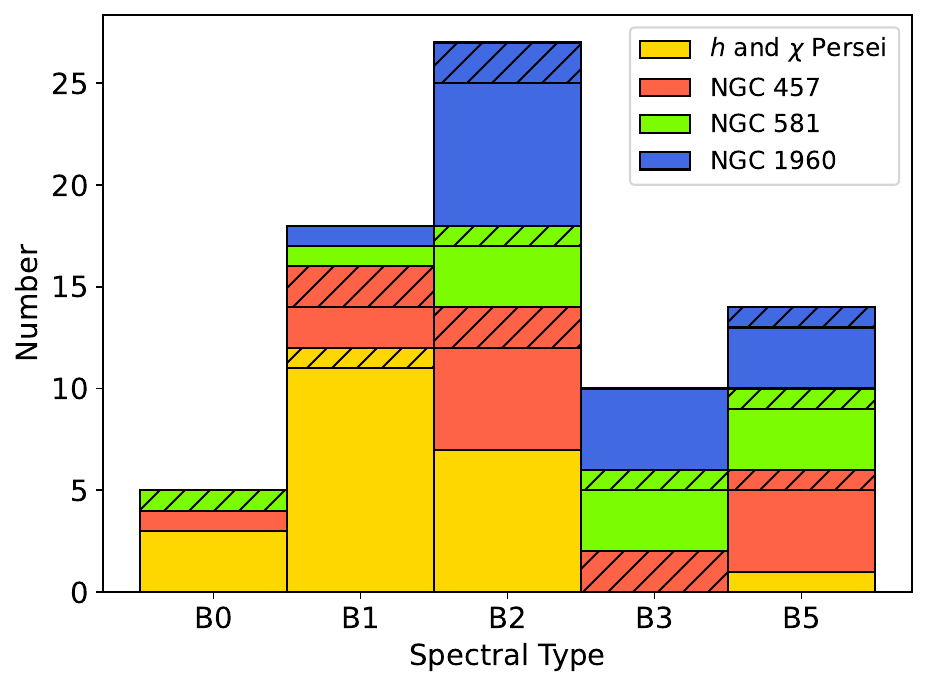}

    \includegraphics[width=0.45\textwidth]{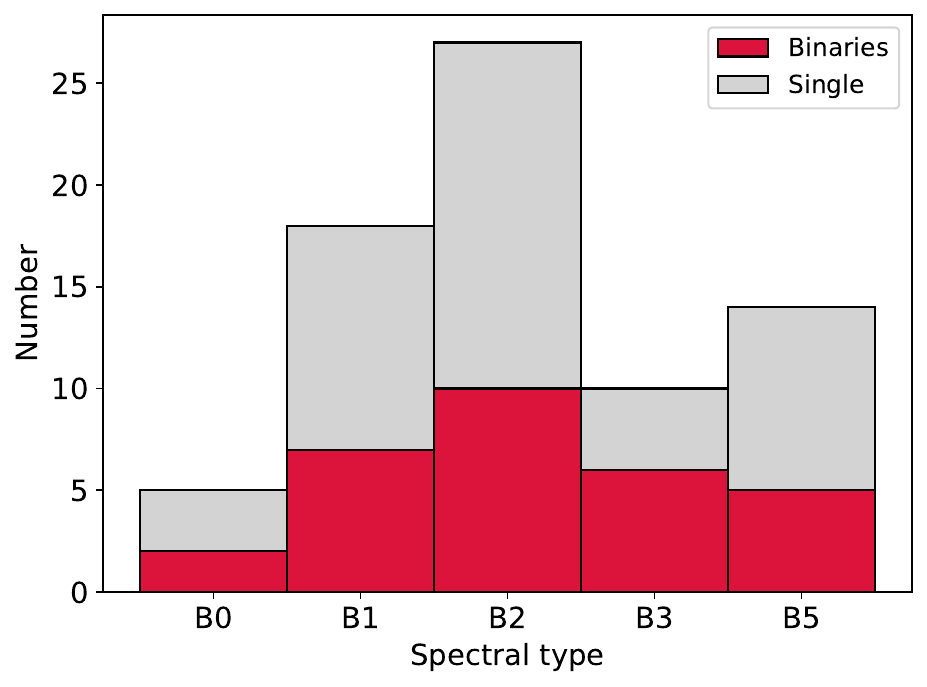}
    \caption[Distribution of the spectral types.]%
    {Histograms of the spectral types in the sample of 74 B-type stars in the four clusters. The top panel shows the distribution of spectral types per cluster (indicated by colour, see legend) and the hatches indicate the number of Be stars in each cluster. The bottom panel illustrates the occurrence of stars classified as binaries and singles in this work.}
    \label{fig:histsample}
\end{figure}

In summary, our sample comprises 74 B-type dwarf and giant B-type stars strictly earlier than B7 stars across the four clusters.  
The top panel of Fig.~\ref{fig:histsample} shows the distribution of spectral types of these stars.
Almost all of the stars in the double cluster $h$ and $\chi$ Persei are early-B stars, with the only exception being $\chi$~Per~2185 a B5 dwarf. NGC~457 and NGC~581 show a similar distribution of spectral types, while NGC~1960 is the only cluster to have no B0 stars.
The absence of earlier-type stars could be related either to the age of the cluster (it was reported to be the oldest cluster in our sample).

\section{Binary detection and characterisation }\label{sect:binary_detect}

\subsection{Identification of SB2 systems}

{\bf Of} the 74 B-type stars in our sample, three systems were visually identified as double-lined spectroscopic binaries (SB2s). These are $\chi$~Per~2392, NGC~457~85 and NGC~1960~008. We leave spectral disentangling and an in-depth study of these systems for future work, but visual inspection showed the double-lined feature predominantly in \ion{He}{I} lines for all three SB2s.
The RVs for these systems were derived using an analytical profile fit of the \ion{He}{I} 5876 line, with the sum of two Gaussian profiles as a fitting function. 
Out of the three systems, NGC 457 85 has been previously classified as an eclipsing binary by \citet{alfonso} and \citet{Zhang}. The two others, $\chi$ Per 2392 and NGC 1960 008, were previously unknown binaries. The periodograms and the phase-folded RV curves are shown in Appendix~\ref{app:binaries}. 
Two additional stars that showed a double-line feature in only one of the available spectra are flagged as possible SB2s in Table~\ref{tab: orbital solutions}.

\subsection{Radial velocities }\label{sect:radial_vel}

We used a cross-correlation technique to determine RVs for the remaining 71 single-lined B-type stars in our sample.
The cross-correlation method used in this work is based on \citet{Shenar_2019} and \citet{dsilva}, which is adapted from \citet{Zucker}.
We cross-correlated specific spectral lines to a reference spectrum, which initially was chosen to be the one with the highest S/N. This was done so that the impact of poor-quality data was minimised when measuring the first RV estimate. The different spectra were then co-added in the rest frame of the initial template to create a new, higher S/N template, which was used to determine the RVs again. The uncertainties were determined by fitting a parabola to the maximum region of the cross-correlation function.
For a more detailed derivation and discussion of the techniques, we refer to \citet{Zucker}.

Table~\ref{tab:NIST} shows the spectral lines that we used to determine the RVs.
In general, \ion{He}{I} absorption lines were mostly used. For stars with slower rotation rates and high S/N spectra, weaker metal lines such as \ion{C}{II} and \ion{Si}{III} were also used.
At least three absorption lines were used in the analysis to obtain more reliable and robust results since low S/N might affect the determination of the RVs.
Regarding the Be stars, these were all visually inspected to determine the level of contamination caused by the decretion disk. Balmer lines, \ion{He}{I} and metal lines that showed any emission component were discarded from the RV analysis since the disk can cause high spectral line variability.

\subsection{Multiplicity criteria}
To assess if a star is in a binary system based on its RV variability, we need to mitigate the impact of other types of line profile variability \citep[e.g. pulsations; ][]{Bowman_2020} and ensure any RV variation is statistically significant. For these reasons, two multiplicity criteria were adopted following previous applications in the literature \citep[see e.g.][]{sana2013, Bodensteiner_II, banyard, Mahy_multiplicity}.
To be classified as a binary system, at least one pair of RVs $(v_i,v_j)$ needed to satisfy both criteria simultaneously.

The first criterion indicates whether an object displays statistically significant RV variations and requires:
\begin{equation}
    \frac{|v_i - v_j|}{\sqrt{\sigma_i^2 + \sigma_j^2}} >4.0 ,
    \label{eqn:4sigma}
\end{equation}
with $v_i$ and $v_j$ being the RVs measurements obtained at two different epochs, and $\sigma_i$ and $\sigma_j$ being the 1-$\sigma$ measured uncertainties, respectively.
The threshold of four in Eqn.~(\ref{eqn:4sigma}) was chosen so that the expected number of false positives is smaller than one, given the number of RV pairs in our sample. 

The second criterion considers other processes that can cause variability in RVs (e.g. pulsations), which are also commonly present in binaries containing B-type dwarf stars \citep{shapley, Southworth_2020, Southworth_2021}.
Therefore two significantly different RV measurements needed to differ from one another by more than a fixed RV variation threshold:
\begin{equation}
   \Delta \rm RV = |v_i -v_j| > \Delta \rm RV_{min}   
\end{equation}
where the threshold chosen is large enough to separate variability due to orbital motion from other sources of variability. We adopt $\Delta \rm RV_{min} =20$ km\,s$^{-1}$ based on the fact that for typical RV variations caused by pulsations for B-type stars are commonly below 20 km\,s$^{-1}$ \citep{aerts, Bowman_2020}. This choice is in line with similar studies on main-sequence massive stars, both O-type \citep{sana2013} and B-type \citep{Bodensteiner_II, banyard}.

\subsection{Orbital parameters}
The orbital properties of the identified binaries with more than four available spectra were analysed to ascertain their distributions across the different cluster ages and stellar mass ranges.
A Lomb-Scargle periodogram \citep{Lomb, Scargle} was applied to the RV time series of all identified binaries, encompassing both SB1 and SB2 systems, utilising the Astropy package \texttt{timeseries} \citep{astropy1, astropy2, astropy3}. This technique is particularly suited for handling unevenly spaced astronomical data, which characterises our sample due to gaps of up to 1-2 yr in most observations. The periodogram peaks were deemed significant if their amplitude surpassed a $1\%$ false-alarm probability threshold.

To determine the orbital solutions of each binary system, we employed the SPectroscopic and INterferometric Orbital Solution \citep[\texttt{spinOS}\footnote{\url{https://github.com/matthiasfabry/spinOS}}; ][]{spinOS_fabry} \texttt{Python} software package.
This allows us to fit an orbital model and perform least-squares minimisation to determine the optimal orbital parameters of both SB1 and SB2 systems.
The fitting process requires an initial estimate of the parameters defining the binary orbit, including the period, eccentricity, argument of periastron of the secondary ($\omega$), time of periastron passage, semi-amplitude of the RV curve of the primary ($K_1$), and the systemic velocity of the system ($\gamma$). For SB2 systems, the semi-amplitude of the secondary ($K_2$) must also be provided. The \texttt{spinOS} software can employ a two-step fitting procedure. First, a Levenberg-Marquardt local non-linear least squares optimisation method is used. This initial step helps to obtain a good starting point for the subsequent Markov Chain Monte Carlo (MCMC) analysis. The MCMC algorithm samples the parameter space and explores different combinations of parameters that are consistent with the observed data and their uncertainties in order to estimate confidence intervals for all fitting parameters.

Different orbital period values were tested for binaries that exhibited multiple significant peaks in the periodogram of their RV time series. The highest peak was not always selected as the final orbital period. 
For systems with multiple significant orbital fits, the solution with the lowest $\chi^{2}$ value was ultimately chosen as the best orbital period.
For the SB2 system NGC 457 85 it was not possible to determine a significant peak with the Lomb-Scargle periodogram, so we imposed the photometric period determined by \citet{alfonso} as the initial solution for our spectroscopic analysis.

The periodograms and the phase-folded RVs for all binaries are shown in Appendix~\ref{app:binaries}. The first system shown in Fig.~\ref{app:chiper_2311} is an example of a binary with a clear period that well explains the orbit. Figure~\ref{app:ngc457_100} is an example where no peaks are above the 1\% false alarm probability and it was not possible to solve the orbit. Lastly, Fig.~\ref{app:ngc1960_109} is an example of an ambiguous binary where there is orbital motion detected through RV variability, but the indicative periods ($\sim$1.6~d) are plausibly pulsation periods.

\subsection{Correction for observational biases}\label{sect:obs_biases}

The identified SB1 and SB2 systems account only for the observed spectroscopic binary fraction, $f_{\rm bin,obs}$, which is a lower limit of the intrinsic spectroscopic binary fraction, $f_{\rm bin,int}$. The true spectroscopic binary fraction can be determined by taking into account detection biases. In this work, two bias analyses were performed.
The first one describes the probability of detecting significant RV shifts (as defined by our two binary criteria) given the orbital properties of a binary system, as well as the temporal sampling and the accuracy of RV measurements in our HERMES campaign. The second bias is related to line blending of the components and how the presence of a second set of spectral lines can change the composite lines, which is difficult to detect if the lines never deblend. This may reduce the apparent amplitude of RV variations and hence reduce the likelihood of detecting binaries \citep{Bodensteiner_II}.
This bias considers that SB2 systems are not always easily identifiable as such. In particular, if the wavelength shift is not large enough, the lines of the two different components are not distinguishable. The spectrum appears as a single-lined star, with perhaps a larger amount of line broadening \citep{Sana2011}, therefore the projected rotational velocity of both components plays an important role.
Both biases are related to the RV amplitude of the primary and secondary, the observation quality and the instrumental resolution. 

To determine the SB1 bias, the approach described by \citet{sana2013} was followed.
Using Monte Carlo simulations, we generate four populations of binary systems, each matching the size of the observed sample. The simulations account for the observational setup to estimate the sensitivity of detecting binaries in the data.
The first step is to randomly select the initial mass of the primary star $\rm M_1$ from the Salpeter initial mass function \citep{salpeter}. 
Two mass ranges were used based on the spectral type distribution of the stars within each cluster (see Fig.~\ref{fig:histsample}). For $h$ and $\chi$ Persei, NGC~457 and NGC~581 the mass of the primary was drawn between 5 and 15 M$_{\odot}$, while for NGC~1960 between 5 and 11 M$_{\odot}$.
Following \citet{sana2013}, \citet{Bodensteiner_II} and \citet{banyard}, we used power-laws distributions for orbital parameters: $f(e)\sim q^{\kappa}$, $f(\log_{10}(P/{\rm d}))\sim (\log_{10}(P))^{\pi}$ and $f(e)\sim e^{\eta}$.
For the period, a value of $\pi=-0.25\pm0.25$ is adopted, while for the two other indexes, the values adopted are from the Galactic O-star sample \citep{sana2012}, where $\kappa=-0.2\pm0.6$ for the mass ratio, and $\eta=-0.4\pm0.2$ for the eccentricity.
The inclination, $i$, the argument of periastron, $\omega$, and the reference in time for the ephemeris were randomly selected.

\citet{Bodensteiner_II} performed a series of simulations for low-resolution VLT/MUSE spectra using \textsc{tlusty}\footnote{\url{http://svo2.cab.inta-csic.es/theory/newov2/index.php?models=tlusty_bstarbin}} models \citep{tlusty_2007} to determine the threshold for RV shifts that allows us to distinguish SB2 systems. 
In this work, we repeated those simulations for high-resolution HERMES spectra.
\citet{Bodensteiner_II} degraded the simulated spectra to the MUSE resolution (between 1700 and 3700), but for this work, the degradation had a much lower impact given the high spectral resolution of HERMES (i.e. 80\,000; \citealt{raskin_2011}). We adopted a fixed S/N of 70 although the S/N distribution has a large variance in the sample (between 20 and 100).
Three sets of simulations are performed for a given B0 primary with rotational velocities of 100, 200 and 400 km\,s$^{-1}$, and a secondary with $v\,\sin\,i$ of 100 km\,s$^{-1}$ for each simulation. The spectrum of the secondary star is adapted given a mass ratio ranging from 0.1 to 1 in steps of 0.1. For each simulation, the RV of the primary is fixed from 0 to 300 km\,s$^{-1}$ in steps of 10 km\,s$^{-1}$, while the RV of the secondary is given by $v_2=-v_1/q$. The spectra of the two components are then coadded according to an estimated light ratio.

\begin{figure}
    \centering
    \includegraphics[scale=0.4]{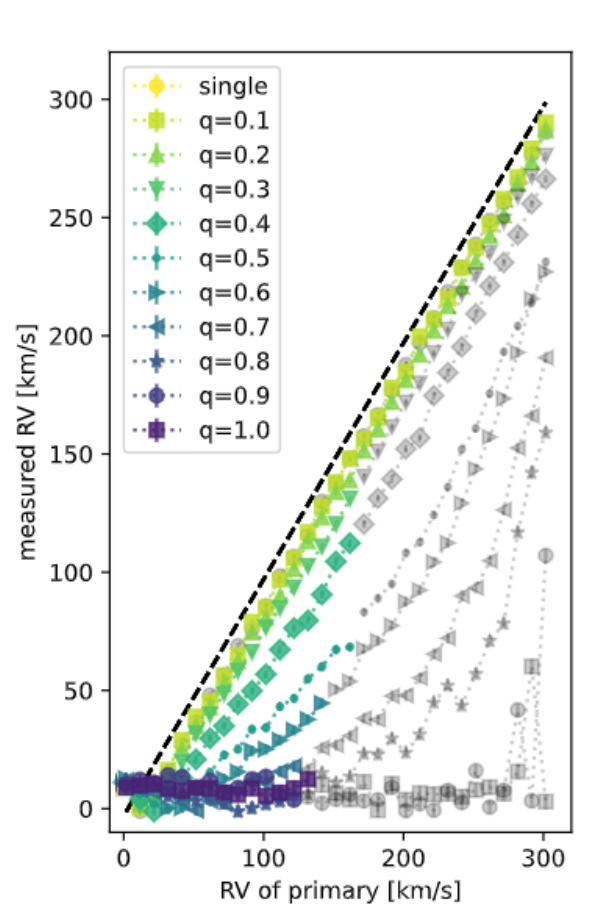}
    \caption[SB2 bias correction]%
    {Measured RVs as a function of RV input of the primary in the simulated systems. The $v\,\sin\,i$ of the primary, a B0-type star, is 200 km\,s$^{-1}$. The legend reports the ten different mass ratios that were simulated. Grey symbols indicate where the SB2 bias was lifted since a visual inspection allowed to correctly identify the SB2s as such.}
    \label{fig:sb2bias_mine}
\end{figure}

Similar to how we identified SB2 systems in our HERMES spectra, the SB2 detection is done by visual inspection and RV measurements are performed through an analytical fit, specifically a Gaussian fit of the \ion{He}{I} line at 4026$\AA$.
Figure \ref{fig:sb2bias_mine} shows one of three simulations for a primary B0 star of $v\sin i=200$ km\,s$^{-1}$. Our analysis yields similar results to those obtained by \cite{Bodensteiner_II}. Systems with $q<0.6$ are identifiable as binaries if the rotational velocity is not too large (i.e. below 100 km\,s$^{-1}$), but detected as SB1s rather than SB2s.
Systems with comparable component masses ($\rm q>0.6$) are challenging to identify as binaries because the light contribution and spectral line intensity of the components are similar, irrespective of their projected rotational velocity. 
For systems with large RV amplitudes, the line blending bias can be lifted by successfully identifying both components. However, for RV separations below 100~km\,s$^{-1}$, there is a strong bias that affects all systems, regardless of the nature of their components. These systems are often misclassified as SB1 systems or, in some cases, as single rotators.

\begin{figure*}
    \centering
    \includegraphics[width=0.7\linewidth]{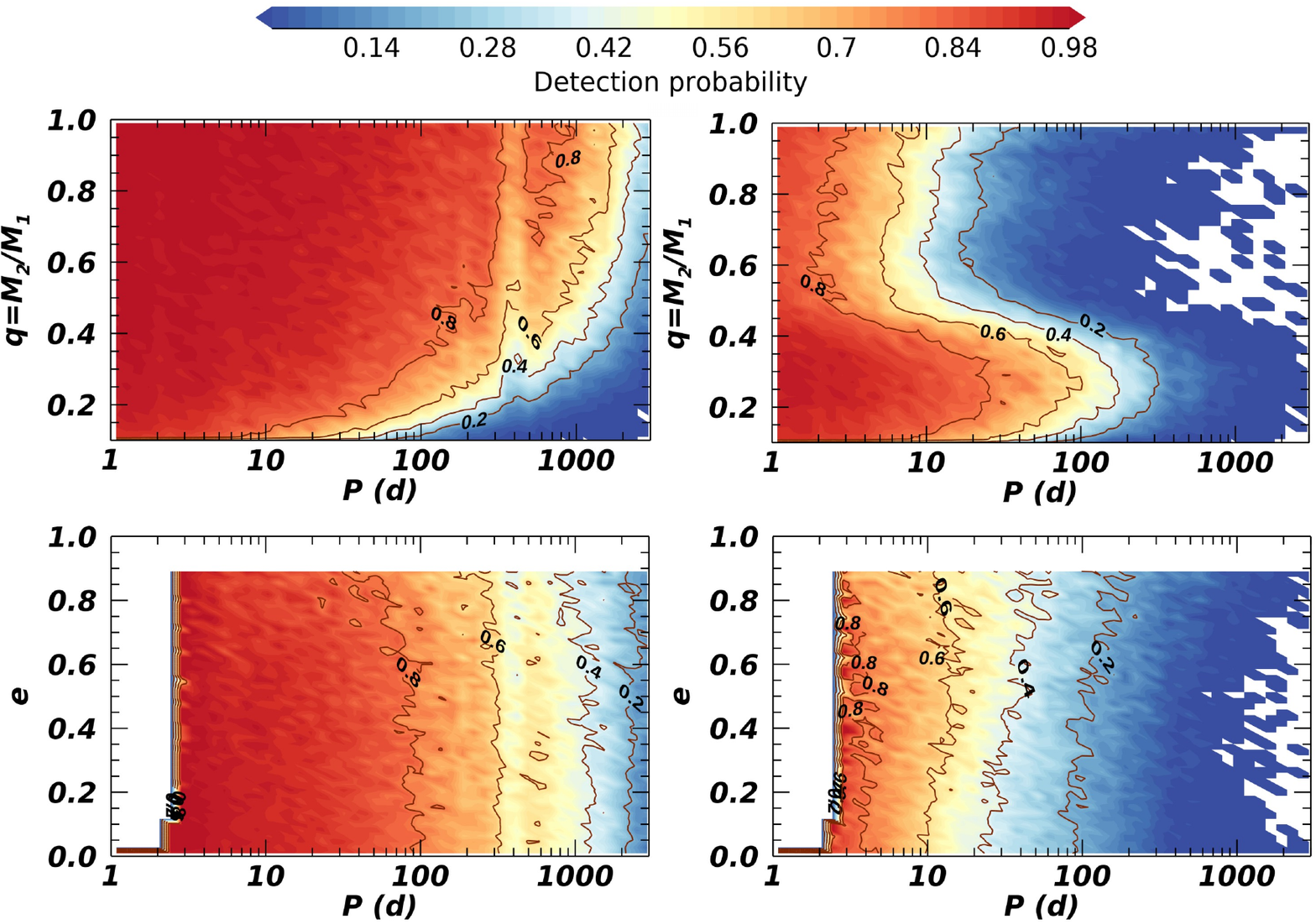}
    \caption{Binary detection probability maps for the double cluster $h$ and $\chi$ Persei. The left panels are computed for the SB1 bias and the right show the SB1+SB2 bias. The top row shows the mass ratio as a function of the period, while the bottom panels show the eccentricity as a function of the period.}
    \label{fig:bias_corr}
\end{figure*}

The Monte Carlo simulation is repeated to take into account the SB2 corrections.
The two multiplicity criteria are applied to the simulated time series of RVs, for which the associated representative errors are taken from the uncertainties of our observations. The probability ($P_{\rm detect}$) of detecting binaries is determined after repeating the simulation 10,000 times.
By dividing our previously obtained observed spectroscopic binary fraction with the detection probability, we thus obtain the intrinsic spectroscopic binary fraction of the cluster:
$
    f_{\rm bin,int} =  {f_{\rm bin,obs}}/{P_{\rm detect}}. 
$
The two-dimensional bias correction plots in Fig.~\ref{fig:bias_corr} show how taking into account either one or both biases influences the detection probability for the double cluster $h$ and $\chi$ Persei. When considering only the SB1 bias, the detection probability remains high for up to 100~d, independent of the mass ratio and the eccentricity of the system. Including the SB2 bias makes the overall probability drop significantly for systems with periods higher than 100~d at any eccentricity, and for systems with a mass ratio higher than 0.6.

\section{Discussion of binarity results}\label{sect:discuss_binary}
\subsection{Multiplicity fractions}
Through our RV analysis, we identify 30 of the 74 B-type stars as displaying large, significant RV variations, and hence we classify them as binaries.
The bottom panel of Fig.~\ref{fig:histsample} shows the distribution of single stars and binaries along the spectral types of the sample.
In Table \ref{tab:fractions} we report both the observed spectroscopic binary fraction and the intrinsic fraction after correcting for the SB1 and SB2 biases. 
Overall, we find an intrinsic binary fraction exceeding 60\% in NGC~581, and over 80\% in the other three clusters. Given their similarity, we combine the four clusters and find an overall intrinsic binary fraction of $f_{\rm bin,int} = 79^{+19}_{-16}\%$.
The intrinsic spectroscopic binary fractions of all four clusters agree within 1$\sigma$, indicating that there are no strong differences in the binary fractions of clusters with ages between 15 and 30~Myr.

\begin{table}
    \centering
    \caption{
    Observed spectroscopic binary fraction ($f_{\rm bin,obs}$) and the intrinsic (bias corrected) fraction ($f_{\rm bin,int}$) across the four clusters. Name of the cluster, age \citep[from ][]{Dias2021}, observed spectroscopic binary fraction and the intrinsic spectroscopic binary fraction corrected for both SB1 and SB2 bias. The detection probability in the last column ($P_{\rm detect}$), is considering both SB1 and SB2 biases. Error bars give the 68$\%$ confidence intervals and are capped if they exceed 100\%.
    }
    \begin{tabular}{lccccc}
        \hline 
    \hline
       Cluster  & age [Myr]   & $f_{\rm bin,obs}$ & $f_{\rm bin,int}$ & $P_{\rm detect}$\\
       \hline
       $h$ and $\chi$ Per & 15.4$\pm$1.1  & $35\pm10\%$ & $81^{+19}_{-30}\%$ & $43^{+18}_{-10}\%$ \\[0.05cm]
       NGC~457 & 23.6$\pm$1.2    & $58\pm11\%$ & $83^{+17}_{-22}\%$ & $70^{+19}_{-12}\%$\\[0.05cm]
       NGC~581 & 29.4$\pm$1.3 & $36\pm13\%$ & $69^{+31}_{-31}\%$ & $52^{+30}_{-14}\%$\\[0.05cm]
       NGC~1960 & 30.2$\pm$1.2 & $33\pm11\%$ & $81^{+19}_{-35}\%$ & $41^{+21}_{-11}\%$ \\[0.05cm]
       \hline    
       All & - & $41\pm6\%$ & $79^{+19}_{-16}\%$ & $51^{+10}_{-7}\%$\\[0.05cm]
       \hline
    \end{tabular}
    \label{tab:fractions}
\end{table}

We note that of the literature studies of binarity for B-type stars mentioned previously, only \citet{Bodensteiner_II} and \citet{banyard} take into account the SB2 bias correction. Therefore the intrinsic binary fractions of other B-type stars surveys might be underestimated. On the other hand, O-type stars are less affected by the SB2 bias, because of their generally higher RV amplitudes.
In the 30 Doradus region in the LMC \citep[age of $\sim$2 Myr; ][]{Crowther_2010}, \citet{Dunstall} analysed 408 B-type stars and found an observed and intrinsic binary fraction of $25\pm2$\% and 58$\pm11\%$, respectively.
\citet{banyard} analysed the B-type population in the Galactic cluster NGC~6231, which has an age between 2 and 7 Myr \citep{Kuhn_2017}, and found an observed binary fraction of $33\pm5$\%. The intrinsic binary fraction, corrected for SB1 and SB2 biases, is 52$\pm8\%$ \citep{banyard}.
Our intrinsic binary fractions, although higher, are consistent with the intrinsic binary fractions of the younger sample of \citet{Dunstall} within the errors. The somewhat lower intrinsic binary fraction detected by \citet{banyard} might be due to the presence of late B-type stars (down to spectral type B9) in their sample.
Since the SB2 bias is not applied in the study of \citet{Dunstall}, this presumably would yield a higher intrinsic spectroscopic binary fraction and therefore become compatible with our results.
Additionally, we find that our observed spectroscopic binary fractions are in agreement with what was found by Galactic field surveys on B-type stars \citep[45$\%$;][]{Sota_2014, simon-diaz, Barba_2017}. The different observational setups, however, make a direct comparison difficult.

Two older clusters populated by less massive stars were studied by \citet{Bodensteiner_II} and \citet{Saracino_2023}. For example, \citet{Bodensteiner_II} investigated the B-star population in the SMC open cluster NGC~330 \citep[age of $\sim$40 Myr; ][]{Patrick_2020} and found a lower intrinsic spectroscopic binary fraction of $34^{+8}_{-7}$\%, which may be due to the lower mass of the sample or its low metallicity. 
\citet{Saracino_2023} investigated the population of stars with birth masses between 2 and 5~M$_{\odot}$ in the $\sim100$~Myr-old cluster NGC~1850 in the LMC and found an intrinsic spectroscopic binary fraction of $24\pm5\%$.
This result implies that the decrease in the spectroscopic binary fraction is due to the mass range of the stars in a given sample, thus suggesting a difference in the binary fractions or period distributions of early-B and late-B stars \citep[see also][]{banyard}.

For more massive stars, \citet{Kobulnicky} analysed the Cygnus OB2 association, composed of 45 O- and 82 early B-type stars, and found an SB1 corrected fraction of 55$\%$ for orbital periods below 5000~d.
Similar values were reported by \citet{sana2012, sana2013}, with a spectroscopic binary fraction of 69$\pm9\%$ for the O-star population in six Galactic clusters, and 51$\pm4\%$ for the O-star population in the 30~Dor region of the LMC. Therefore, the spectroscopic binary fractions of the Galactic O-type star samples are consistent with the intrinsic binary fraction of the B-type stars in our four clusters.

\subsection{Orbital parameters}

We provide tentative orbits for 25 out of the 27 identified SB1 systems, and determined the orbital parameters, with the caveat that currently only sparse data sets are available.
Table \ref{tab: orbital solutions} reports the orbital parameters along with their binary mass function which sets a lower limit for the mass of the secondary component. The solutions for the three identified SB2 systems are reported in Table \ref{tab:orbital_solutions_sb2}.

\begin{figure}
    \centering
    \includegraphics[width=0.45\textwidth]{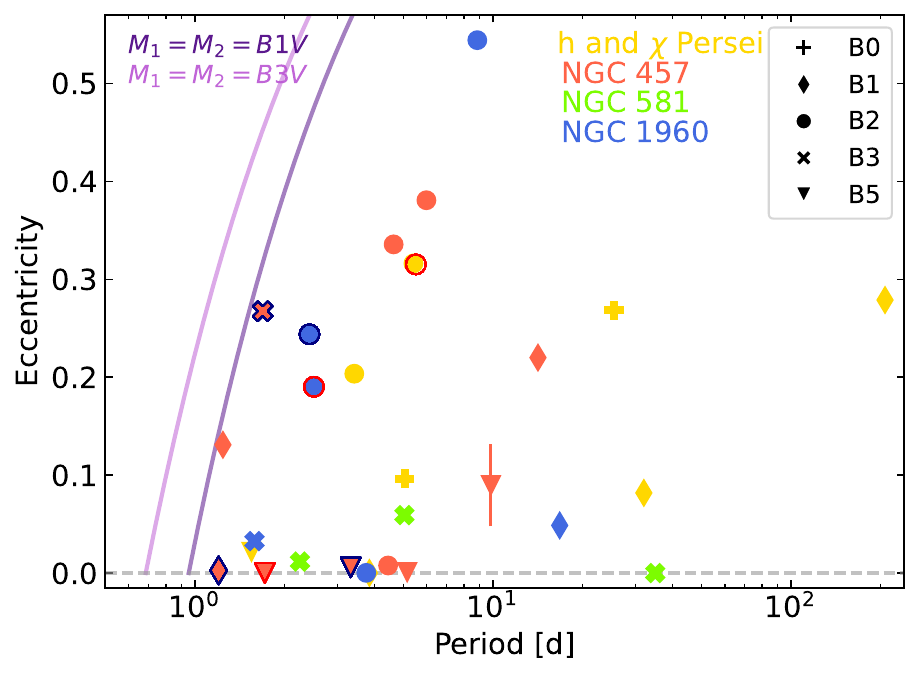}
    \caption{Eccentricity as a function of the period of all the 28 binaries identified in this work. Different members of the clusters are represented in different colours and the different shapes represent the spectral type of the primary. The blue edge colour of a symbol indicates Be stars found in binary systems while the red edge symbol colour indicates SB2 systems.
    The two solid lines correspond to the period and eccentricity a system should have for the semi-major axis to be equal to the sum of the radii of the two components. The darker purple colour is for a system composed of two B1 V stars and the lighter colour for two B3 V stars. The two systems close to the purple lines with short period and high eccentricity have tentative orbits.}

    \label{fig:P_vs_e}
\end{figure}

Figure \ref{fig:P_vs_e} shows the eccentricity as a function of period for all the identified binaries. 
Only two systems (NGC 457 91 and NGC 457 154) do not pass the Lucy-Sweeney test at the 5\% significance level (i.e. $e/\sigma_e \leq 2.49$). This means that the measured eccentricity is not significantly different from zero \citep{Lucy_sweeney}.
As indicated in Fig.~\ref{fig:bias_corr}, our {\sc HERMES} campaign is barely sensitive to detecting binaries with periods above 100~d, a limit that also depends on the mass ratio. This is also reflected in Fig.~\ref{fig:P_vs_e}, which shows that most detected binaries have periods below 100\,d, with a majority of systems being short-period binaries with periods below 5\,d. Especially for these short-period binaries, an independent verification with photometry (e.g. light curves; \citealt{Southworth_2020}) is subject of future work to determine if the identified period is the true orbital period or if it might be the pulsation period of the star.

To compare and analyse the orbital parameter distributions, especially period and eccentricity, of the binary systems in different clusters, we use the cumulative distribution function (CDF) and a Kuiper test \citep{Kuiper1960TestsCR}. 
In this study, a two-sample Kuiper test is performed using the \texttt{kuiper\_two} function using the package \texttt{astropy.stats} \citep{astropy1,astropy2,astropy3}. This allowed us to perform a statistical comparison of the CDFs of the period and eccentricity distributions among the different clusters.
The outputs are the test statistic value (D) which represents the sum of the largest positive ($\rm D^+$) and the negative ($\rm D^-$) difference between the two cumulative distributions, and the probability (fpp) that a value of D, as large as the one observed, is due to statistical fluctuations. Following \citet{banyard}, we consider any probability below a 10\% threshold to be a significant indication of a difference between the samples. We note that for small samples such as ours the value of fpp is quite approximate.

Our application of the Kuiper test shows that differences in the period and eccentricity distributions are not statistically significant and therefore the CDFs of the four clusters are all compatible with each other. 
Since no statistically significant difference was found among our four clusters, the orbital period and eccentricity samples were combined to obtain a larger sample for comparison to the literature. 

\begin{figure}
    \centering
    \includegraphics[width=0.47\textwidth]{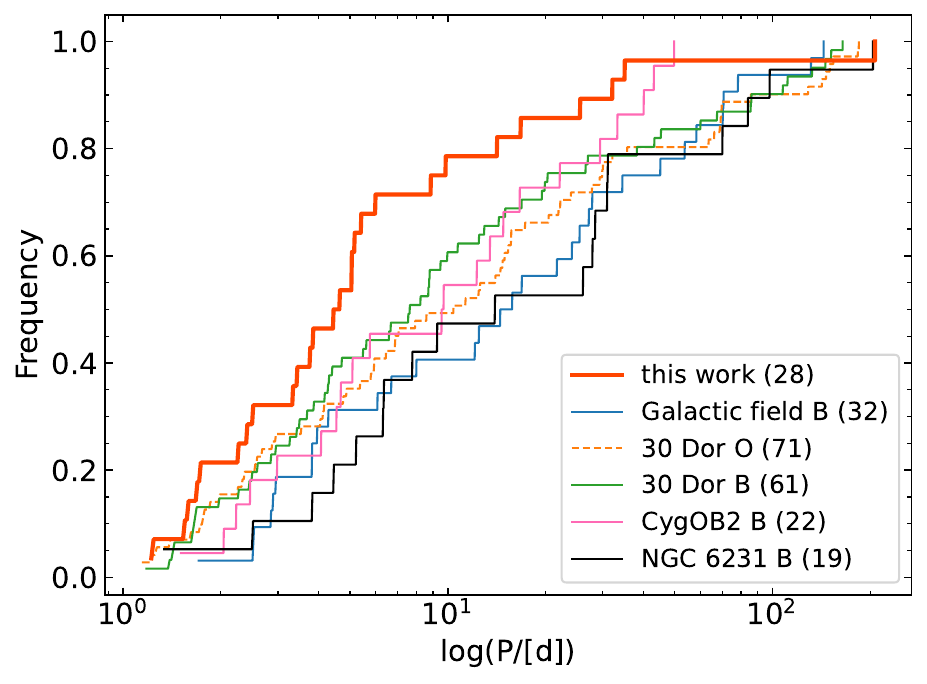}
    \includegraphics[width=0.47\textwidth]{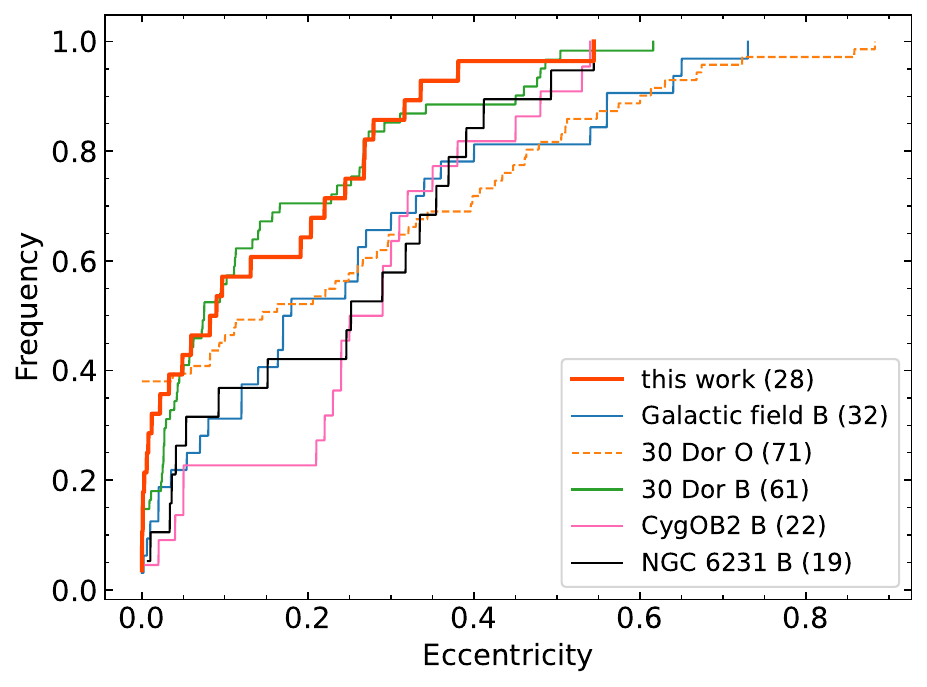}
    \caption[CDF of period and eccentricity of the binary systems.]%
    {CDFs of period and eccentricity are shown in the top and bottom panels, respectively. The bold orange line represents the analysis of this work, while the results obtained from previous works are plotted in different colours, with solid lines the B star samples and with dotted lines the O star. In order: Galactic B-type stars \citep{Abt_1990}, 30 Dor O-type stars \citep{Almeida}, 30 Dor B-type stars \citep{Villasenor}, Cyg OB2 \citep{Kobulnicky} and NGC~6231 B-type stars \citep{banyard}. The numbers in the parentheses represent the number of stars in each sample. }
    \label{fig:cdfs}
\end{figure}

The results comparing our works and the previous literature studies are shown in Table \ref{tab:kuiper_persei}.
The CDFs of the orbital period and eccentricity distributions are plotted in Fig.~\ref{fig:cdfs}, in which our results are being compared to previous surveys that include not only Galactic B-type stars \citep{banyard}, but extra-galactic massive stars, both O- and B-type, in the 30 Doradus region \citep{Almeida, Villasenor}. We note that we have performed a cut in the literature surveys' distributions to have a maximum orbital period compatible with our longest detected period ($\sim200$\,d) to allow a reasonable comparison.

We apply another Kuiper test to our combined four-cluster sample.
Neither the period nor eccentricity distributions are consistent with the \citet{Almeida} sample of O-type stars in the 30 Doradus region, which might be due to the different mass ranges.
There are period distribution differences to the field B-type star sample of \citet{Abt_1990}, and with the B-star population in NGC~6231 \citep{banyard}. This might be due to the differences in mass ranges considered in our and their studies, since they go down to B9 stars.
Our eccentricity distribution is inconsistent with the B-star population in the Cygnus OB2 association \citep{Kobulnicky}, which consisted of 26 binary systems from spectral type B0 to B2, including dwarfs, giants and supergiants.


\begin{table*}
\caption[]%
{Results of the Kuiper test for the period and eccentricity distribution performed on the orbital period and eccentricity of the 28 identified binaries for which it was possible to solve the orbit (Sample 1) and the comparison literature sample (Sample 2) with their spectral types specified in the fourth column. }
\begin{tabular}{llclcc|cc}
\hline \hline
Sample 1  & Sample 2       & SpT & Ref              & \multicolumn{2}{c|}{Kuiper test Period} & \multicolumn{2}{c}{Kuiper test Eccentricity} \\ \cline{5-8} 
          &                &     &                  & D                  & fpp                & D                     & fpp                  \\ \hline
This work & 30 Dor         & O   & \citet{Almeida}    & 0.35               & 9\%                & 0.65                  & 0\%                  \\
          & Galactic field & B   & \citet{Abt_1990}   & 0.47               & 4\%                & 0.29                  & 54\%                 \\
          & 30 Dor         & B   & \citet{Villasenor} & 0.31               & 24\%               & 0.32                  & 17\%                 \\
          & CygOB2         & B   & \citet{Kobulnicky} & 0.33               & 45\%               & 0.50                  & 2\%                  \\
          & NGC 6231       & B   & \citet{banyard}    & 0.47               & 7\%                & 0.31                  & 68\%                 \\ \hline
\end{tabular}\label{tab:kuiper_persei}
\end{table*}

\section{Projected rotational velocity}\label{sect:atm_parameters}

\subsection{Base grid models}
To determine the projected rotational velocities of our stars, we performed an atmospheric parameter fitting. The methodology is based on \citet{Bodensteiner_II} but using Galactic metallicity models.
Using a grid-based search of BSTAR2006 \textsc{tlusty} models for B-type stars \citep{tlusty_1995, tlusty_2007} and a $\chi^2$ minimisation, the best fitting $v\,\sin\,i$ was determined for each star of the sample.
The temperature range of the \textsc{tlusty} B-type star grid spans from 15\,000 to 30\,000~K, with additional models computed available at temperatures as low as 9000~K, in steps of 1000~K. 
For these additional models, the $\log g$ values range from 3.0 to 5.0 in steps of 0.25. In the original \textsc{tlusty} grid, the highest $\log g$ value is also 5.0, while the starting $\log g$ value varies with temperature: it begins at 1.75 for models at 15000~K, increases to 2.0 for 16000 and 18000~K, to 2.25 for 19000 and 20000~K, to 2.5 for 21000 to 24000~K, and to 2.75 for 25000 to 28000~K. For models at 29000 and 30000~K, the starting $\log\,g$ returns to 3.0.

To consider the line broadening caused by the rotational velocity, each atmospheric \textsc{tlusty} model was broadened using the \texttt{PyAstronomy} package \texttt{rotBroad}\footnote{\url{https://pyastronomy.readthedocs.io/en/latest/pyaslDoc/aslDoc/rotBroad.html}} with values ranging from 0 to 500 km\,s$^{-1}$ in steps of 20 km\,s$^{-1}$.
The microturbulence velocity is fixed to 2 km~s$^{-1}$ since there is little evidence that B-type dwarfs and giants have large values of microturbulence \citep{Dufton1981, Landstreet}. The macroturbulence velocity was not considered, even though it can be the order of tens of km~s$^{-1}$ for B-type dwarfs \citep{simon-diaz, Gebruers}. The result of this approximation is that the measured projected rotational velocities are an upper limit. 

The \textsc{tlusty} spectral range in the grid is available from 900 to 10000~$\AA$, but only specific diagnostic lines were considered in the fitting process.
Mainly \ion{He}{I}, \ion{He}{II} for B0 stars and Balmer lines were used as diagnostic lines and when the S/N was high enough, metallic lines like \ion{Mg}{II}, \ion{Si}{}, \ion{O}{}, \ion{N}{} and \ion{C}{} were also used. For Be stars, problems arose since in some instances, the disk caused not only the Balmer lines to be in emission but also several helium and metallic lines. In those cases only the lines that did not show any emission component were used.
The $\chi^2$ minimisation was performed between the normalised \textsc{tlusty} models and the co-added observed HERMES spectra for the single stars and the SB1 binaries. We note that dilution due to binarity is not taken into account for these systems.
The $v\,\sin\,i$ of the SB2 systems were not determined since the spectra are composite.

\subsection{Discussion of the projected rotational velocities}

\begin{figure}
    \centering
    \includegraphics[width=0.45\textwidth]{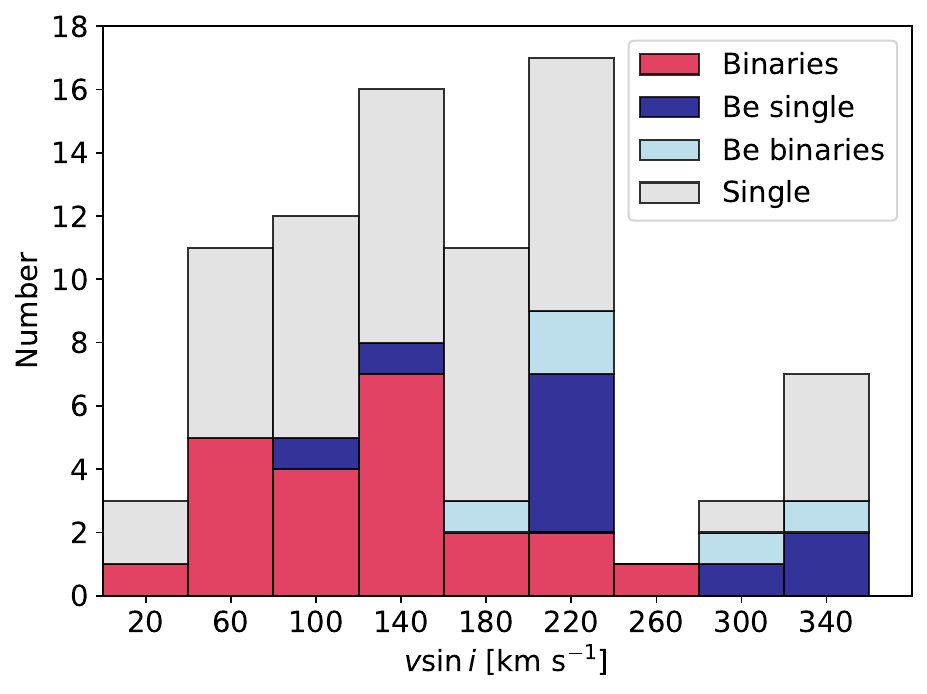}
    \caption[]%
    {Distribution of the projected rotational velocities ($v\,\sin\,i$) of all the stars in the sample. The bin size was set to 40~km\,s$^{-1}$ so that each sample bar has two or more stars.   }
    \label{fig:histvsini}
\end{figure}

The distribution of $v\,\sin\,i$ for the combined four-cluster sample is shown in Fig. \ref{fig:histvsini}. We divided our sample based on their binary status and whether they were classified as Be stars or not.
Single stars show a flat distribution with most of them having $v\,\sin\,i<240$~km\,s$^{-1}$.
Stars in binaries show a similar distribution to the single stars, with most of them being not particularly fast rotators and having projected rotational velocities between 40 and 160 km\,s$^{-1}$. 
The distributions show somewhat of an excess at 200 km\,s$^{-1}$, and are dominated by slow-to-moderate rotators. 
On the contrary, Be stars, both single and in binaries, are rapid rotators with projected rotational velocities above 200 km\,s$^{-1}$ in general. 

\begin{figure}
    \centering
    
    \includegraphics[width=0.45\textwidth]{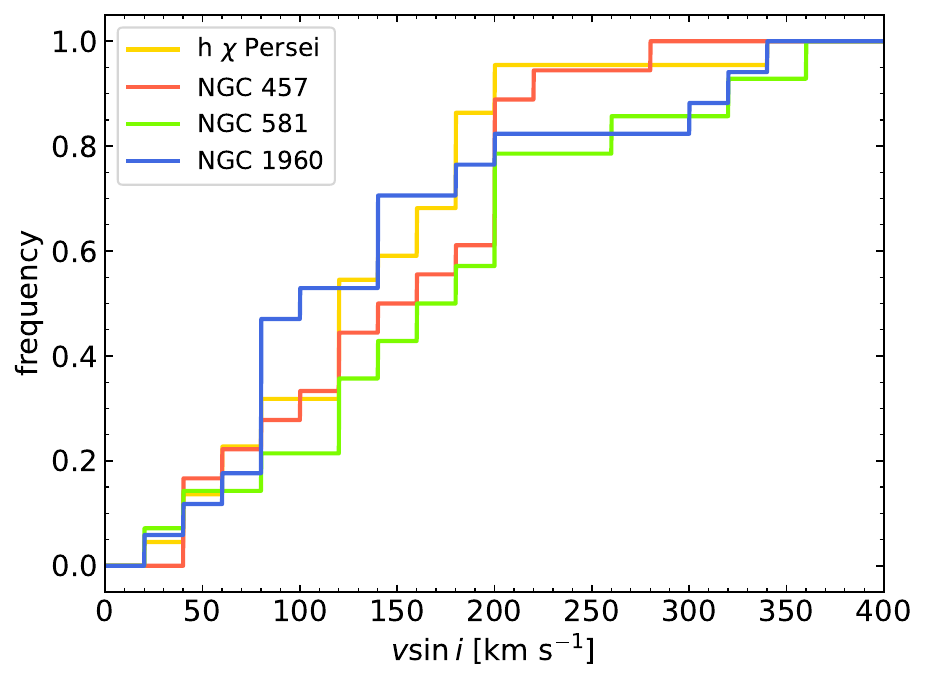}
    \caption[]%
    {CDF of the projected rotational velocity ($v\,\sin\,i$) for the four clusters in different colours.}
 \label{fig:cdf_vsini}
\end{figure}

Figure \ref{fig:cdf_vsini} shows the CDF of the projected rotational velocities for the four clusters.
A Kuiper test was performed to infer if there is a statistically significant difference in the $v\,\sin\,i$ distributions, similar to what was done for the orbital period and eccentricity. 
From this Kuiper test, we find no significant difference in the $v\,\sin\,i$ distributions of our four clusters.

To determine if there was a significant difference in the $v\,\sin\,i$ distribution among the four cluster populations of apparently single stars, the sample was divided into two smaller sets based on their spectral type (i.e. based on their mass). 
The statistical test showed that no statistically significant difference was found. Therefore we conclude that the distribution of $v\,\sin\,i$ is homogeneous among all B-type stars in all of our four clusters.

\citet{Dufton} estimated the projected rotational velocities for a sample of 334 single early B-type stars in the 30 Dor region of the LMC and found that they range up to 450 km\,s$^{-1}$, and also that the distribution shows a bi-modal structure.  
About one-quarter of their sample showed to have deconvolved rotational velocities $v<100$ km\,s$^{-1}$, while the other components had values in the range of $200<v<350$ km\,s$^{-1}$.
Meanwhile, \citet{bodensteiner_vsini} determined the $v\,\sin\,i$ in a sample of more than 200 B-type stars, both singles and binaries, in the $\sim40$~Myr-old cluster NGC~330 located in the SMC. They demonstrated that their sample of B-type stars have $v\,\sin\,i$ values of around 100-250 km\,s$^{-1}$, while the projected rotational velocities of the Be stars are around 200-400 km\,s$^{-1}$. We do not see any bi-modality in the distribution of our projected rotational velocities. On average, the $v\,\sin\,i$ values in our work are lower than what has been measured in the LMC and SMC, and we do not detect ultra-fast rotators ($v\,\sin\,i>360$~km\,s$^{-1}$) that have been instead detected in the two extragalactic surveys.

\section{Conclusions}\label{sect:conclusion}
In this work, we performed a multi-epoch high-resolution spectroscopic study on 74 B-type stars spread over four 15-30~Myr-old Galactic open clusters: $h$ and $\chi$ Persei, NGC~457, NGC~581 and NGC~1960. The observations were obtained using the high-resolution spectrograph HERMES mounted on the Mercator telescope in La Palma.
The RVs were measured by cross-correlating the spectra. We classified the stars as presumably single or binary based on two RV variability criteria. SB2s were detected by visually inspecting the spectra.

We find that the observed spectroscopic binary fractions of the main sequence B-star population in the four clusters are: $31\pm9\%$ for $h$ and $\chi$ Persei, $58\pm11\%$ for NGC~457, $33\pm12\%$ for NGC~581 and $33\pm11\%$ for NGC~1960. The overall observed spectroscopic binary fraction of our sample is $41\pm6\%$.
These fractions were then corrected to take into account the SB1 and SB2 biases. 
By combining the two biases we determined the detection probability which allowed us to identify the intrinsic binary fractions accounting for orbital periods up to about 8~years ($10^{3.5}$ d). For $h$ and $\chi$ Persei, the intrinsic spectroscopic binary fraction is $81^{+19}_{-30}\%$, for NGC~457 it is $83^{+17}_{-22}\%$, for NGC~581 it is $69^{+31}_{-31}\%$ and for NGC~1960 it is $81^{+19}_{-35}\%$. The overall average intrinsic spectroscopic binary fraction is $79^{+19}_{-16}\%$.

We compared our results with those determined for stars in different mass ranges (O- and B-type) and metallicities (Galactic, LMC and SMC). We find that our intrinsic spectroscopic fractions are somewhat higher than other surveys of B-type stars but consistent within the errors. Older clusters populated by less massive stars have significantly lower intrinsic spectroscopic binary fractions (25-35$\%$), suggesting that the spectroscopic binary fraction of a given sample depends on the mass range of the stars (or on its age).  
Overall, our work shows that there are no statistically significant differences between our binary fractions and those of other O- and B-type stars in different metallicity environments. Therefore, our work demonstrates that the majority of B-type stars are found in multiple systems in Galactic metallicity young clusters.

We tentatively solved the orbit for 28 out of 30 identified binaries, which comprise 25 SB1s and three SB2 systems. 
Most of the identified binary systems show orbital periods shorter than 10~d.
We do not observe systems with $P>210$~d and $e>0.6$, which is in line with the binary detection probability we estimate for our campaign.
We compared the period and eccentricity distributions of our sample with those in the literature. Regarding the period distribution, our work is not compatible with the B star populations in the field, and the Galactic cluster NGC~6231. Our eccentricity distribution is not compatible with the B stars in the Cyg~OB2 association. Neither period nor eccentricity are compatible with the O~stars in the 30 Doradus region. These inconsistencies might be related to the difference in mass ranges considered in the different surveys.

The projected rotational velocities of the SB1 binaries and single stars were determined using a grid of BSTAR2006 \textsc{tlusty} stellar models computed at solar metallicity.
Both single and binary stars show a mostly flat $v\,\sin\,i$ distribution, with the first peaking at around 220~km\,s$^{-1}$, while stars in binaries peak at 140~km\,s$^{-1}$. Meanwhile, most of the Be stars show rapid rotation with values typically above 200~km\,s$^{-1}$.
The Kuiper test results show that the $v\,\sin\,i$ distributions are statistically consistent across the four clusters and between different spectral type (i.e. mass) sub-samples, indicating a homogeneous distribution among all B-type stars in our study.
On average, our measured projected rotational velocities are lower compared to what has been determined in the LMC and SMC.

In a future study, we will analyse light curves of stars in our sample assembled by the NASA Transiting Exoplanet Survey Satellite \citep[TESS; ][]{Ricker2015}. We will independently assess binarity within our sample by searching for photometric signatures of binarity, such as ellipsoidal modulation and eclipses. This analysis is particularly important for distinguishing between orbital motions and pulsations in short-period binaries (i.e. periods of order days) that we have detected, which can have similar pulsation periods \citep[e.g.; ][]{Southworth2022}. Moreover, asteroseismic modelling of detected pulsations in binary systems will determine the physical properties and evolutionary status of stars within our sample that pulsate.

\section*{acknowledgements}

The authors gratefully acknowledge UK Research and Innovation (UKRI) in the form of a Frontier Research grant under the UK government’s ERC Horizon Europe funding guarantee (SYMPHONY; PI Bowman; grant number: EP/Y031059/1), and a Royal Society University Research Fellowship (PI Bowman; grant number: URF{\textbackslash}R1{\textbackslash}231631). The research leading to these results has received funding from the European Research Council (ERC) under the European Union’s Horizon 2020 research and innovation programme (MULTIPLES; PI Sana; grant agreement number 772225). This research has used data obtained at the Mercator Observatory which receives funding from the Research Foundation – Flanders (FWO) (grant agreements I000325N and I000521N). This work presents results from the European Space Agency space mission Gaia. Gaia data are being processed by the Gaia Data Processing and Analysis Consortium (DPAC). Funding for the DPAC is provided by national institutions, in particular the institutions participating in the Gaia multilateral agreement.

\section*{Data Availability}

For the purpose of open access, the authors have applied a CC BY licence to the author accepted manuscript version: \url{https://arxiv.org/abs/TBD}. Data products that support the results in this paper are publicly available via the Zenodo repository: \url{https://doi.org/TBD}. The GAIA data used in this work are publicly available via the Gaia website: \url{https://gea.esac.esa.int/archive/}. The Padova database of stellar evolutionary tracks and isochrones is publicly available via the website: \url{https://stev.oapd.inaf.it/cgi-bin/cmd}. This research has made use of the SIMBAD database, operated at CDS, Strasbourg, France; the SAO/NASA Astrophysics Data System; and the VizieR catalogue access tool, CDS, Strasbourg, France. This research has made use of the following software packages: {\tt matplotlib} \citep{Hunter_2007}, {\tt numpy} \citep{harris2020array}, {\tt pandas} \citep{reback2020pandas, mckinney-proc-scipy-2010}, {\tt astropy} \citep{astropy1, astropy2, astropy3}, {\tt scipy} \citep{2020SciPy-NMeth}, {\tt pyAstronomy} \citep{pya}, {\tt TLUSTY} \citep{tlusty_1995, tlusty_2007}, {\tt spinOS} \citep{spinOS_fabry}.



\bibliographystyle{mnras}
\bibliography{bibliography.bib} 

\begin{thebibliography}{}
\makeatletter
\relax
\def\mn@urlcharsother{\let\do\@makeother \do\$\do\&\do\#\do\^\do\_\do\%\do\~}
\def\mn@doi{\begingroup\mn@urlcharsother \@ifnextchar [ {\mn@doi@}
  {\mn@doi@[]}}
\def\mn@doi@[#1]#2{\def\@tempa{#1}\ifx\@tempa\@empty \href
  {http://dx.doi.org/#2} {doi:#2}\else \href {http://dx.doi.org/#2} {#1}\fi
  \endgroup}
\def\mn@eprint#1#2{\mn@eprint@#1:#2::\@nil}
\def\mn@eprint@arXiv#1{\href {http://arxiv.org/abs/#1} {{\tt arXiv:#1}}}
\def\mn@eprint@dblp#1{\href {http://dblp.uni-trier.de/rec/bibtex/#1.xml}
  {dblp:#1}}
\def\mn@eprint@#1:#2:#3:#4\@nil{\def\@tempa {#1}\def\@tempb {#2}\def\@tempc
  {#3}\ifx \@tempc \@empty \let \@tempc \@tempb \let \@tempb \@tempa \fi \ifx
  \@tempb \@empty \def\@tempb {arXiv}\fi \@ifundefined
  {mn@eprint@\@tempb}{\@tempb:\@tempc}{\expandafter \expandafter \csname
  mn@eprint@\@tempb\endcsname \expandafter{\@tempc}}}

\bibitem[\protect\citeauthoryear{{Abbott} et~al.,}{{Abbott}
  et~al.}{2023}]{Abbott_2023}
{Abbott} R.,  et~al., 2023, \mn@doi [Physical Review X]
  {10.1103/PhysRevX.13.041039}, \href
  {https://ui.adsabs.harvard.edu/abs/2023PhRvX..13d1039A} {13, 041039}

\bibitem[\protect\citeauthoryear{{Abt}, {Gomez}  \& {Levy}}{{Abt}
  et~al.}{1990}]{Abt_1990}
{Abt} H.~A.,  {Gomez} A.~E.,   {Levy} S.~G.,  1990, \mn@doi [\apjs]
  {10.1086/191508}, \href
  {https://ui.adsabs.harvard.edu/abs/1990ApJS...74..551A} {74, 551}

\bibitem[\protect\citeauthoryear{{Aerts}, {Puls}, {Godart}  \&
  {Dupret}}{{Aerts} et~al.}{2009}]{aerts}
{Aerts} C.,  {Puls} J.,  {Godart} M.,   {Dupret} M.~A.,  2009, \mn@doi [\aap]
  {10.1051/0004-6361/200810471}, \href
  {https://ui.adsabs.harvard.edu/abs/2009A&A...508..409A} {508, 409}

\bibitem[\protect\citeauthoryear{{Alfonso-Garz{\'o}n}, {Domingo}, {Mas-Hesse}
  \& {Gim{\'e}nez}}{{Alfonso-Garz{\'o}n} et~al.}{2012}]{alfonso}
{Alfonso-Garz{\'o}n} J.,  {Domingo} A.,  {Mas-Hesse} J.~M.,   {Gim{\'e}nez} A.,
   2012, \mn@doi [\aap] {10.1051/0004-6361/201220095}, \href
  {https://ui.adsabs.harvard.edu/abs/2012A&A...548A..79A} {548, A79}

\bibitem[\protect\citeauthoryear{{Almeida} et~al.,}{{Almeida}
  et~al.}{2017}]{Almeida}
{Almeida} L.~A.,  et~al., 2017, \mn@doi [\aap] {10.1051/0004-6361/201629844},
  \href {https://ui.adsabs.harvard.edu/abs/2017A&A...598A..84A} {598, A84}

\bibitem[\protect\citeauthoryear{{Amaro-Seoane} et~al.,}{{Amaro-Seoane}
  et~al.}{2017}]{Amaro}
{Amaro-Seoane} P.,  et~al., 2017, \mn@doi [arXiv e-prints]
  {10.48550/arXiv.1702.00786}, \href
  {https://ui.adsabs.harvard.edu/abs/2017arXiv170200786A} {p. arXiv:1702.00786}

\bibitem[\protect\citeauthoryear{{Astropy Collaboration} et~al.,}{{Astropy
  Collaboration} et~al.}{2013}]{astropy1}
{Astropy Collaboration} et~al., 2013, \mn@doi [\aap]
  {10.1051/0004-6361/201322068}, \href
  {https://ui.adsabs.harvard.edu/abs/2013A&A...558A..33A} {558, A33}

\bibitem[\protect\citeauthoryear{{Astropy Collaboration} et~al.,}{{Astropy
  Collaboration} et~al.}{2018}]{astropy2}
{Astropy Collaboration} et~al., 2018, \mn@doi [\aj] {10.3847/1538-3881/aabc4f},
  \href {https://ui.adsabs.harvard.edu/abs/2018AJ....156..123A} {156, 123}

\bibitem[\protect\citeauthoryear{{Astropy Collaboration} et~al.,}{{Astropy
  Collaboration} et~al.}{2022}]{astropy3}
{Astropy Collaboration} et~al., 2022, \mn@doi [\apj]
  {10.3847/1538-4357/ac7c74}, \href
  {https://ui.adsabs.harvard.edu/abs/2022ApJ...935..167A} {935, 167}

\bibitem[\protect\citeauthoryear{{Banyard}, {Sana}, {Mahy}, {Bodensteiner},
  {Villase{\~n}or}  \& {Evans}}{{Banyard} et~al.}{2022}]{banyard}
{Banyard} G.,  {Sana} H.,  {Mahy} L.,  {Bodensteiner} J.,  {Villase{\~n}or}
  J.~I.,   {Evans} C.~J.,  2022, \mn@doi [\aap] {10.1051/0004-6361/202141037},
  \href {https://ui.adsabs.harvard.edu/abs/2022A&A...658A..69B} {658, A69}

\bibitem[\protect\citeauthoryear{{Barb{\'a}}, {Gamen}, {Arias}  \&
  {Morrell}}{{Barb{\'a}} et~al.}{2017}]{Barba_2017}
{Barb{\'a}} R.~H.,  {Gamen} R.,  {Arias} J.~I.,   {Morrell} N.~I.,  2017, in
  {Eldridge} J.~J.,  {Bray} J.~C.,  {McClelland} L.~A.~S.,   {Xiao} L.,  eds,
  IAU Symposium Vol. 329, The Lives and Death-Throes of Massive Stars. pp
  89--96, \mn@doi{10.1017/S1743921317003258}

\bibitem[\protect\citeauthoryear{{Boden}}{{Boden}}{1951}]{ngc1960_1951}
{Boden} E.,  1951, Uppsala Astronomical Observatory Annals, \href
  {https://ui.adsabs.harvard.edu/abs/1951UppAn...3d...1B} {3, 1}

\bibitem[\protect\citeauthoryear{{Bodensteiner} et~al.,}{{Bodensteiner}
  et~al.}{2020a}]{Bodensteiner_I}
{Bodensteiner} J.,  et~al., 2020a, \mn@doi [\aap]
  {10.1051/0004-6361/201936743}, \href
  {https://ui.adsabs.harvard.edu/abs/2020A&A...634A..51B} {634, A51}

\bibitem[\protect\citeauthoryear{{Bodensteiner}, {Shenar}  \&
  {Sana}}{{Bodensteiner} et~al.}{2020b}]{Bodensteiner_beform}
{Bodensteiner} J.,  {Shenar} T.,   {Sana} H.,  2020b, \mn@doi [\aap]
  {10.1051/0004-6361/202037640}, \href
  {https://ui.adsabs.harvard.edu/abs/2020A&A...641A..42B} {641, A42}

\bibitem[\protect\citeauthoryear{{Bodensteiner} et~al.,}{{Bodensteiner}
  et~al.}{2021}]{Bodensteiner_II}
{Bodensteiner} J.,  et~al., 2021, \mn@doi [\aap] {10.1051/0004-6361/202140507},
  \href {https://ui.adsabs.harvard.edu/abs/2021A&A...652A..70B} {652, A70}

\bibitem[\protect\citeauthoryear{{Bodensteiner} et~al.,}{{Bodensteiner}
  et~al.}{2023}]{bodensteiner_vsini}
{Bodensteiner} J.,  et~al., 2023, \mn@doi [\aap] {10.1051/0004-6361/202345950},
  \href {https://ui.adsabs.harvard.edu/abs/2023A&A...680A..32B} {680, A32}

\bibitem[\protect\citeauthoryear{{Bodensteiner} et~al.,}{{Bodensteiner}
  et~al.}{2025}]{Bodensteiner_Bloem}
{Bodensteiner} J.,  et~al., 2025, \mn@doi [arXiv e-prints]
  {10.48550/arXiv.2502.02641}, \href
  {https://ui.adsabs.harvard.edu/abs/2025arXiv250202641B} {p. arXiv:2502.02641}

\bibitem[\protect\citeauthoryear{{Bowman}}{{Bowman}}{2020}]{Bowman_2020}
{Bowman} D.~M.,  2020, \mn@doi [Frontiers in Astronomy and Space Sciences]
  {10.3389/fspas.2020.578584}, \href
  {https://ui.adsabs.harvard.edu/abs/2020FrASS...7...70B} {7, 70}

\bibitem[\protect\citeauthoryear{{Bressan}, {Marigo}, {Girardi}, {Salasnich},
  {Dal Cero}, {Rubele}  \& {Nanni}}{{Bressan} et~al.}{2012}]{Bressan}
{Bressan} A.,  {Marigo} P.,  {Girardi} L.,  {Salasnich} B.,  {Dal Cero} C.,
  {Rubele} S.,   {Nanni} A.,  2012, \mn@doi [\mnras]
  {10.1111/j.1365-2966.2012.21948.x}, \href
  {https://ui.adsabs.harvard.edu/abs/2012MNRAS.427..127B} {427, 127}

\bibitem[\protect\citeauthoryear{Cantat-Gaudin \& Anders}{Cantat-Gaudin \&
  Anders}{2020}]{Cantat_Gaudin_2020}
Cantat-Gaudin T.,  Anders F.,  2020, \mn@doi [Astronomy \& Astrophysics]
  {10.1051/0004-6361/201936691}, 633, A99

\bibitem[\protect\citeauthoryear{{Chen}, {Bressan}, {Girardi}, {Marigo}, {Kong}
   \& {Lanza}}{{Chen} et~al.}{2015}]{Chen_parsec}
{Chen} Y.,  {Bressan} A.,  {Girardi} L.,  {Marigo} P.,  {Kong} X.,   {Lanza}
  A.,  2015, \mn@doi [\mnras] {10.1093/mnras/stv1281}, \href
  {https://ui.adsabs.harvard.edu/abs/2015MNRAS.452.1068C} {452, 1068}

\bibitem[\protect\citeauthoryear{Crowther, Schnurr, Hirschi, Yusof, Parker,
  Goodwin  \& Kassim}{Crowther et~al.}{2010}]{Crowther_2010}
Crowther P.~A.,  Schnurr O.,  Hirschi R.,  Yusof N.,  Parker R.~J.,  Goodwin
  S.~P.,   Kassim H.~A.,  2010, \mn@doi [Monthly Notices of the Royal
  Astronomical Society] {10.1111/j.1365-2966.2010.17167.x}, 408, 731

\bibitem[\protect\citeauthoryear{{Czesla}, {Schr{\"o}ter}, {Schneider},
  {Huber}, {Pfeifer}, {Andreasen}  \& {Zechmeister}}{{Czesla}
  et~al.}{2019}]{pya}
{Czesla} S.,  {Schr{\"o}ter} S.,  {Schneider} C.~P.,  {Huber} K.~F.,  {Pfeifer}
  F.,  {Andreasen} D.~T.,   {Zechmeister} M.,  2019, {PyA: Python
  astronomy-related packages} (\mn@eprint {ascl} {1906.010})

\bibitem[\protect\citeauthoryear{{Dias}, {Monteiro}, {Moitinho}, {L{\'e}pine},
  {Carraro}, {Paunzen}, {Alessi}  \& {Villela}}{{Dias} et~al.}{2021}]{Dias2021}
{Dias} W.~S.,  {Monteiro} H.,  {Moitinho} A.,  {L{\'e}pine} J.~R.~D.,
  {Carraro} G.,  {Paunzen} E.,  {Alessi} B.,   {Villela} L.,  2021, \mn@doi
  [\mnras] {10.1093/mnras/stab770}, \href
  {https://ui.adsabs.harvard.edu/abs/2021MNRAS.504..356D} {504, 356}

\bibitem[\protect\citeauthoryear{{Dsilva}, {Shenar}, {Sana}  \&
  {Marchant}}{{Dsilva} et~al.}{2020}]{dsilva}
{Dsilva} K.,  {Shenar} T.,  {Sana} H.,   {Marchant} P.,  2020, \mn@doi [\aap]
  {10.1051/0004-6361/202038446}, \href
  {https://ui.adsabs.harvard.edu/abs/2020A&A...641A..26D} {641, A26}

\bibitem[\protect\citeauthoryear{{Dufton}, {Durrant}  \& {Durrant}}{{Dufton}
  et~al.}{1981}]{Dufton1981}
{Dufton} P.~L.,  {Durrant} C.~J.,   {Durrant} A.~C.,  1981, \aap, \href
  {https://ui.adsabs.harvard.edu/abs/1981A&A....97...10D} {97, 10}

\bibitem[\protect\citeauthoryear{{Dufton} et~al.,}{{Dufton}
  et~al.}{2013}]{Dufton}
{Dufton} P.~L.,  et~al., 2013, \mn@doi [\aap] {10.1051/0004-6361/201220273},
  \href {https://ui.adsabs.harvard.edu/abs/2013A&A...550A.109D} {550, A109}

\bibitem[\protect\citeauthoryear{{Dunstall} et~al.,}{{Dunstall}
  et~al.}{2015}]{Dunstall}
{Dunstall} P.~R.,  et~al., 2015, \mn@doi [\aap] {10.1051/0004-6361/201526192},
  \href {https://ui.adsabs.harvard.edu/abs/2015A&A...580A..93D} {580, A93}

\bibitem[\protect\citeauthoryear{{Fabry}, {Hawcroft}, {Mahy}, {Marchant}, {Le
  Bouquin}  \& {Sana}}{{Fabry} et~al.}{2021}]{spinOS_fabry}
{Fabry} M.,  {Hawcroft} C.,  {Mahy} L.,  {Marchant} P.,  {Le Bouquin} J.-B.,
  {Sana} H.,  2021, {spinOS: SPectroscopic and INterferometric Orbital Solution
  finder}, Astrophysics Source Code Library, record ascl:2102.001 (\mn@eprint
  {ascl} {2102.001})

\bibitem[\protect\citeauthoryear{{Ferraro} et~al.,}{{Ferraro}
  et~al.}{1997}]{Ferraro_1997}
{Ferraro} F.~R.,  et~al., 1997, \mn@doi [\aap]
  {10.48550/arXiv.astro-ph/9703026}, \href
  {https://ui.adsabs.harvard.edu/abs/1997A&A...324..915F} {324, 915}

\bibitem[\protect\citeauthoryear{{Gebruers} et~al.,}{{Gebruers}
  et~al.}{2022}]{Gebruers}
{Gebruers} S.,  et~al., 2022, \mn@doi [\aap] {10.1051/0004-6361/202243839},
  \href {https://ui.adsabs.harvard.edu/abs/2022A&A...665A..36G} {665, A36}

\bibitem[\protect\citeauthoryear{{Gossage} et~al.,}{{Gossage}
  et~al.}{2019}]{Gossage2019}
{Gossage} S.,  et~al., 2019, \mn@doi [\apj] {10.3847/1538-4357/ab5717}, \href
  {https://ui.adsabs.harvard.edu/abs/2019ApJ...887..199G} {887, 199}

\bibitem[\protect\citeauthoryear{{Gray} \& {Corbally}}{{Gray} \&
  {Corbally}}{2009}]{gray_book}
{Gray} R.~O.,  {Corbally} Christopher J.,  2009, {Stellar Spectral
  Classification}

\bibitem[\protect\citeauthoryear{Harris et~al.,}{Harris
  et~al.}{2020}]{harris2020array}
Harris C.~R.,  et~al., 2020, \mn@doi [Nature] {10.1038/s41586-020-2649-2}, 585,
  357

\bibitem[\protect\citeauthoryear{{Hastings}, {Langer}, {Wang}, {Schootemeijer}
  \& {Milone}}{{Hastings} et~al.}{2021}]{Hastings2021}
{Hastings} B.,  {Langer} N.,  {Wang} C.,  {Schootemeijer} A.,   {Milone} A.~P.,
   2021, \mn@doi [\aap] {10.1051/0004-6361/202141269}, \href
  {https://ui.adsabs.harvard.edu/abs/2021A&A...653A.144H} {653, A144}

\bibitem[\protect\citeauthoryear{{Hubeny} \& {Lanz}}{{Hubeny} \&
  {Lanz}}{1995}]{tlusty_1995}
{Hubeny} I.,  {Lanz} T.,  1995, \mn@doi [\apj] {10.1086/175226}, \href
  {https://ui.adsabs.harvard.edu/abs/1995ApJ...439..875H} {439, 875}

\bibitem[\protect\citeauthoryear{Hunter}{Hunter}{2007}]{Hunter_2007}
Hunter J.~D.,  2007, \mn@doi [Computing in Science \& Engineering]
  {10.1109/MCSE.2007.55}, 9, 90

\bibitem[\protect\citeauthoryear{{Kobulnicky} et~al.,}{{Kobulnicky}
  et~al.}{2012}]{Kobulnicky_2012}
{Kobulnicky} H.~A.,  et~al., 2012, \mn@doi [\apj] {10.1088/0004-637X/756/1/50},
  \href {https://ui.adsabs.harvard.edu/abs/2012ApJ...756...50K} {756, 50}

\bibitem[\protect\citeauthoryear{{Kobulnicky} et~al.,}{{Kobulnicky}
  et~al.}{2014}]{Kobulnicky}
{Kobulnicky} H.~A.,  et~al., 2014, \mn@doi [\apjs]
  {10.1088/0067-0049/213/2/34}, \href
  {https://ui.adsabs.harvard.edu/abs/2014ApJS..213...34K} {213, 34}

\bibitem[\protect\citeauthoryear{{Kuhn}, {Getman}, {Feigelson}, {Sills},
  {Gromadzki}, {Medina}, {Borissova}  \& {Kurtev}}{{Kuhn}
  et~al.}{2017}]{Kuhn_2017}
{Kuhn} M.~A.,  {Getman} K.~V.,  {Feigelson} E.~D.,  {Sills} A.,  {Gromadzki}
  M.,  {Medina} N.,  {Borissova} J.,   {Kurtev} R.,  2017, \mn@doi [\aj]
  {10.3847/1538-3881/aa9177}, \href
  {https://ui.adsabs.harvard.edu/abs/2017AJ....154..214K} {154, 214}

\bibitem[\protect\citeauthoryear{Kuiper}{Kuiper}{1960}]{Kuiper1960TestsCR}
Kuiper N.~H.,  1960.

\bibitem[\protect\citeauthoryear{{Lada} \& {Lada}}{{Lada} \&
  {Lada}}{2003}]{Lada_2003}
{Lada} C.~J.,  {Lada} E.~A.,  2003, \mn@doi [\araa]
  {10.1146/annurev.astro.41.011802.094844}, \href
  {https://ui.adsabs.harvard.edu/abs/2003ARA&A..41...57L} {41, 57}

\bibitem[\protect\citeauthoryear{{Lamb}, {Oey}, {Segura-Cox}, {Graus},
  {Kiminki}, {Golden-Marx}  \& {Parker}}{{Lamb} et~al.}{2016}]{Lamb_2016}
{Lamb} J.~B.,  {Oey} M.~S.,  {Segura-Cox} D.~M.,  {Graus} A.~S.,  {Kiminki}
  D.~C.,  {Golden-Marx} J.~B.,   {Parker} J.~W.,  2016, \mn@doi [\apj]
  {10.3847/0004-637X/817/2/113}, \href
  {https://ui.adsabs.harvard.edu/abs/2016ApJ...817..113L} {817, 113}

\bibitem[\protect\citeauthoryear{{Landstreet}, {Kupka}, {Ford}, {Officer},
  {Sigut}, {Silaj}, {Strasser}  \& {Townshend}}{{Landstreet}
  et~al.}{2009}]{Landstreet}
{Landstreet} J.~D.,  {Kupka} F.,  {Ford} H.~A.,  {Officer} T.,  {Sigut}
  T.~A.~A.,  {Silaj} J.,  {Strasser} S.,   {Townshend} A.,  2009, \mn@doi
  [\aap] {10.1051/0004-6361/200912083}, \href
  {https://ui.adsabs.harvard.edu/abs/2009A&A...503..973L} {503, 973}

\bibitem[\protect\citeauthoryear{{Lanz} \& {Hubeny}}{{Lanz} \&
  {Hubeny}}{2007}]{tlusty_2007}
{Lanz} T.,  {Hubeny} I.,  2007, \mn@doi [\apjs] {10.1086/511270}, \href
  {https://ui.adsabs.harvard.edu/abs/2007ApJS..169...83L} {169, 83}

\bibitem[\protect\citeauthoryear{{Lomb}}{{Lomb}}{1976}]{Lomb}
{Lomb} N.~R.,  1976, \mn@doi [\apss] {10.1007/BF00648343}, \href
  {https://ui.adsabs.harvard.edu/abs/1976Ap&SS..39..447L} {39, 447}

\bibitem[\protect\citeauthoryear{{Lucy} \& {Sweeney}}{{Lucy} \&
  {Sweeney}}{1971}]{Lucy_sweeney}
{Lucy} L.~B.,  {Sweeney} M.~A.,  1971, \mn@doi [\aj] {10.1086/111159}, \href
  {https://ui.adsabs.harvard.edu/abs/1971AJ.....76..544L} {76, 544}

\bibitem[\protect\citeauthoryear{Maggiore et~al.,}{Maggiore
  et~al.}{2020}]{Maggiore_2020}
Maggiore M.,  et~al., 2020, \mn@doi [Journal of Cosmology and Astroparticle
  Physics] {10.1088/1475-7516/2020/03/050}, 2020, 050–050

\bibitem[\protect\citeauthoryear{{Mahy} et~al.,}{{Mahy}
  et~al.}{2022}]{Mahy_multiplicity}
{Mahy} L.,  et~al., 2022, \mn@doi [\aap] {10.1051/0004-6361/202040062}, \href
  {https://ui.adsabs.harvard.edu/abs/2022A&A...657A...4M} {657, A4}

\bibitem[\protect\citeauthoryear{{Marchant} \& {Bodensteiner}}{{Marchant} \&
  {Bodensteiner}}{2024}]{Marchant_bodensteiner}
{Marchant} P.,  {Bodensteiner} J.,  2024, \mn@doi [\araa]
  {10.1146/annurev-astro-052722-105936}, \href
  {https://ui.adsabs.harvard.edu/abs/2024ARA&A..62...21M} {62, 21}

\bibitem[\protect\citeauthoryear{{Moe} \& {Di Stefano}}{{Moe} \& {Di
  Stefano}}{2017}]{Moe}
{Moe} M.,  {Di Stefano} R.,  2017, \mn@doi [\apjs] {10.3847/1538-4365/aa6fb6},
  \href {https://ui.adsabs.harvard.edu/abs/2017ApJS..230...15M} {230, 15}

\bibitem[\protect\citeauthoryear{{Patrick} et~al.,}{{Patrick}
  et~al.}{2020}]{Patrick_2020}
{Patrick} L.~R.,  et~al., 2020, in XIV.0 Scientific Meeting (virtual) of the
  Spanish Astronomical Society. p.~175

\bibitem[\protect\citeauthoryear{{Pols}, {Cote}, {Waters}  \& {Heise}}{{Pols}
  et~al.}{1991}]{Pols1991}
{Pols} O.~R.,  {Cote} J.,  {Waters} L.~B.~F.~M.,   {Heise} J.,  1991, \aap,
  \href {https://ui.adsabs.harvard.edu/abs/1991A&A...241..419P} {241, 419}

\bibitem[\protect\citeauthoryear{{Raskin } \& {Van Winckel}}{{Raskin } \& {Van
  Winckel}}{2014}]{Raskin}
{Raskin } G.,  {Van Winckel} H.,  2014, \mn@doi [Astronomische Nachrichten]
  {10.1002/asna.201312009}, \href
  {https://ui.adsabs.harvard.edu/abs/2014AN....335...32R} {335, 32}

\bibitem[\protect\citeauthoryear{{Raskin} et~al.,}{{Raskin}
  et~al.}{2011}]{raskin_2011}
{Raskin} G.,  et~al., 2011, \mn@doi [\aap] {10.1051/0004-6361/201015435}, \href
  {https://ui.adsabs.harvard.edu/abs/2011A&A...526A..69R} {526, A69}

\bibitem[\protect\citeauthoryear{{Ricker} et~al.,}{{Ricker}
  et~al.}{2015}]{Ricker2015}
{Ricker} G.~R.,  et~al., 2015, \mn@doi [Journal of Astronomical Telescopes,
  Instruments, and Systems] {10.1117/1.JATIS.1.1.014003}, \href
  {https://ui.adsabs.harvard.edu/abs/2015JATIS...1a4003R} {1, 014003}

\bibitem[\protect\citeauthoryear{{Salpeter}}{{Salpeter}}{1955}]{salpeter}
{Salpeter} E.~E.,  1955, \mn@doi [\apj] {10.1086/145971}, \href
  {https://ui.adsabs.harvard.edu/abs/1955ApJ...121..161S} {121, 161}

\bibitem[\protect\citeauthoryear{{Sana}, {James}  \& {Gosset}}{{Sana}
  et~al.}{2011}]{Sana2011}
{Sana} H.,  {James} G.,   {Gosset} E.,  2011, \mn@doi [\mnras]
  {10.1111/j.1365-2966.2011.18698.x}, \href
  {https://ui.adsabs.harvard.edu/abs/2011MNRAS.416..817S} {416, 817}

\bibitem[\protect\citeauthoryear{{Sana} et~al.,}{{Sana}
  et~al.}{2012}]{sana2012}
{Sana} H.,  et~al., 2012, \mn@doi [Science] {10.1126/science.1223344}, \href
  {https://ui.adsabs.harvard.edu/abs/2012Sci...337..444S} {337, 444}

\bibitem[\protect\citeauthoryear{{Sana} et~al.,}{{Sana}
  et~al.}{2013}]{sana2013}
{Sana} H.,  et~al., 2013, \mn@doi [\aap] {10.1051/0004-6361/201219621}, \href
  {https://ui.adsabs.harvard.edu/abs/2013A&A...550A.107S} {550, A107}

\bibitem[\protect\citeauthoryear{{Sana} et~al.,}{{Sana}
  et~al.}{2014}]{sana_2014}
{Sana} H.,  et~al., 2014, \mn@doi [\apjs] {10.1088/0067-0049/215/1/15}, \href
  {https://ui.adsabs.harvard.edu/abs/2014ApJS..215...15S} {215, 15}

\bibitem[\protect\citeauthoryear{{Saracino} et~al.,}{{Saracino}
  et~al.}{2023}]{Saracino_2023}
{Saracino} S.,  et~al., 2023, \mn@doi [\mnras] {10.1093/mnras/stad2706}, \href
  {https://ui.adsabs.harvard.edu/abs/2023MNRAS.526..299S} {526, 299}

\bibitem[\protect\citeauthoryear{{Scargle}}{{Scargle}}{1982}]{Scargle}
{Scargle} J.~D.,  1982, \mn@doi [\apj] {10.1086/160554}, \href
  {https://ui.adsabs.harvard.edu/abs/1982ApJ...263..835S} {263, 835}

\bibitem[\protect\citeauthoryear{{Schneider}, {Izzard}, {Langer}  \& {de
  Mink}}{{Schneider} et~al.}{2015}]{Schneider_2015}
{Schneider} F.~R.~N.,  {Izzard} R.~G.,  {Langer} N.,   {de Mink} S.~E.,  2015,
  \mn@doi [\apj] {10.1088/0004-637X/805/1/20}, \href
  {https://ui.adsabs.harvard.edu/abs/2015ApJ...805...20S} {805, 20}

\bibitem[\protect\citeauthoryear{{Schneider}, {Ohlmann}, {Podsiadlowski},
  {R{\"o}pke}, {Balbus}, {Pakmor}  \& {Springel}}{{Schneider}
  et~al.}{2019}]{Schneider2019}
{Schneider} F. R.~N.,  {Ohlmann} S.~T.,  {Podsiadlowski} P.,  {R{\"o}pke}
  F.~K.,  {Balbus} S.~A.,  {Pakmor} R.,   {Springel} V.,  2019, \mn@doi [\nat]
  {10.1038/s41586-019-1621-5}, \href
  {https://ui.adsabs.harvard.edu/abs/2019Natur.574..211S} {574, 211}

\bibitem[\protect\citeauthoryear{{Shapley} \& {Nicholson}}{{Shapley} \&
  {Nicholson}}{1919}]{shapley}
{Shapley} H.,  {Nicholson} S.~B.,  1919, \mn@doi [Proceedings of the National
  Academy of Science] {10.1073/pnas.5.10.417}, \href
  {https://ui.adsabs.harvard.edu/abs/1919PNAS....5..417S} {5, 417}

\bibitem[\protect\citeauthoryear{{Shenar} et~al.,}{{Shenar}
  et~al.}{2019}]{Shenar_2019}
{Shenar} T.,  et~al., 2019, \mn@doi [\aap] {10.1051/0004-6361/201935684}, \href
  {https://ui.adsabs.harvard.edu/abs/2019A&A...627A.151S} {627, A151}

\bibitem[\protect\citeauthoryear{{Sim{\'o}n-D{\'\i}az}, {Castro}, {Garcia},
  {Herrero}  \& {Markova}}{{Sim{\'o}n-D{\'\i}az} et~al.}{2011}]{IACOB}
{Sim{\'o}n-D{\'\i}az} S.,  {Castro} N.,  {Garcia} M.,  {Herrero} A.,
  {Markova} N.,  2011, \mn@doi [Bulletin de la Societe Royale des Sciences de
  Liege] {10.48550/arXiv.1009.5824}, \href
  {https://ui.adsabs.harvard.edu/abs/2011BSRSL..80..514S} {80, 514}

\bibitem[\protect\citeauthoryear{{Sim{\'o}n-D{\'\i}az}, {Godart}, {Castro},
  {Herrero}, {Aerts}, {Puls}, {Telting}  \&
  {Grassitelli}}{{Sim{\'o}n-D{\'\i}az} et~al.}{2017}]{simon-diaz}
{Sim{\'o}n-D{\'\i}az} S.,  {Godart} M.,  {Castro} N.,  {Herrero} A.,  {Aerts}
  C.,  {Puls} J.,  {Telting} J.,   {Grassitelli} L.,  2017, \mn@doi [\aap]
  {10.1051/0004-6361/201628541}, \href
  {https://ui.adsabs.harvard.edu/abs/2017A&A...597A..22S} {597, A22}

\bibitem[\protect\citeauthoryear{{Sota}, {Ma{\'\i}z Apell{\'a}niz}, {Morrell},
  {Barb{\'a}}, {Walborn}, {Gamen}, {Arias}  \& {Alfaro}}{{Sota}
  et~al.}{2014}]{Sota_2014}
{Sota} A.,  {Ma{\'\i}z Apell{\'a}niz} J.,  {Morrell} N.~I.,  {Barb{\'a}} R.~H.,
   {Walborn} N.~R.,  {Gamen} R.~C.,  {Arias} J.~I.,   {Alfaro} E.~J.,  2014,
  \mn@doi [\apjs] {10.1088/0067-0049/211/1/10}, \href
  {https://ui.adsabs.harvard.edu/abs/2014ApJS..211...10S} {211, 10}

\bibitem[\protect\citeauthoryear{{Southworth} \& {Bowman}}{{Southworth} \&
  {Bowman}}{2022}]{Southworth2022}
{Southworth} J.,  {Bowman} D.~M.,  2022, \mn@doi [\mnras]
  {10.1093/mnras/stac875}, \href
  {https://ui.adsabs.harvard.edu/abs/2022MNRAS.513.3191S} {513, 3191}

\bibitem[\protect\citeauthoryear{{Southworth}, {Bowman}, {Tkachenko}  \&
  {Pavlovski}}{{Southworth} et~al.}{2020}]{Southworth_2020}
{Southworth} J.,  {Bowman} D.~M.,  {Tkachenko} A.,   {Pavlovski} K.,  2020,
  \mn@doi [\mnras] {10.1093/mnrasl/slaa091}, \href
  {https://ui.adsabs.harvard.edu/abs/2020MNRAS.497L..19S} {497, L19}

\bibitem[\protect\citeauthoryear{{Southworth}, {Bowman}  \&
  {Pavlovski}}{{Southworth} et~al.}{2021}]{Southworth_2021}
{Southworth} J.,  {Bowman} D.~M.,   {Pavlovski} K.,  2021, \mn@doi [\mnras]
  {10.1093/mnrasl/slaa197}, \href
  {https://ui.adsabs.harvard.edu/abs/2021MNRAS.501L..65S} {501, L65}

\bibitem[\protect\citeauthoryear{{Steppe}}{{Steppe}}{1974}]{ngc581_1974}
{Steppe} H.,  1974, \aaps, \href
  {https://ui.adsabs.harvard.edu/abs/1974A&AS...15...91S} {15, 91}

\bibitem[\protect\citeauthoryear{{Villase{\~n}or} et~al.,}{{Villase{\~n}or}
  et~al.}{2021}]{Villasenor}
{Villase{\~n}or} J.~I.,  et~al., 2021, \mn@doi [\mnras]
  {10.1093/mnras/stab2197}, \href
  {https://ui.adsabs.harvard.edu/abs/2021MNRAS.507.5348V} {507, 5348}

\bibitem[\protect\citeauthoryear{Virtanen et~al.,}{Virtanen
  et~al.}{2020}]{2020SciPy-NMeth}
Virtanen P.,  et~al., 2020, \mn@doi [Nature Methods]
  {10.1038/s41592-019-0686-2}, \href {https://rdcu.be/b08Wh} {17, 261}

\bibitem[\protect\citeauthoryear{{Wang}, {Langer}, {Schootemeijer}, {Castro},
  {Adscheid}, {Marchant}  \& {Hastings}}{{Wang} et~al.}{2020}]{wang}
{Wang} C.,  {Langer} N.,  {Schootemeijer} A.,  {Castro} N.,  {Adscheid} S.,
  {Marchant} P.,   {Hastings} B.,  2020, \mn@doi [\apjl]
  {10.3847/2041-8213/ab6171}, \href
  {https://ui.adsabs.harvard.edu/abs/2020ApJ...888L..12W} {888, L12}

\bibitem[\protect\citeauthoryear{{Wang} et~al.,}{{Wang}
  et~al.}{2022}]{Wang2022}
{Wang} C.,  et~al., 2022, \mn@doi [Nature Astronomy]
  {10.1038/s41550-021-01597-5}, \href
  {https://ui.adsabs.harvard.edu/abs/2022NatAs...6..480W} {6, 480}

\bibitem[\protect\citeauthoryear{{Wang} et~al.,}{{Wang}
  et~al.}{2023}]{wang_2023}
{Wang} C.,  et~al., 2023, \mn@doi [\aap] {10.1051/0004-6361/202245413}, \href
  {https://ui.adsabs.harvard.edu/abs/2023A&A...670A..43W} {670, A43}

\bibitem[\protect\citeauthoryear{{Wenger} et~al.,}{{Wenger}
  et~al.}{2000}]{SIMBAD}
{Wenger} M.,  et~al., 2000, \mn@doi [\aaps] {10.1051/aas:2000332}, \href
  {https://ui.adsabs.harvard.edu/abs/2000A&AS..143....9W} {143, 9}

\bibitem[\protect\citeauthoryear{{W}es {M}c{K}inney}{{W}es
  {M}c{K}inney}{2010}]{mckinney-proc-scipy-2010}
{W}es {M}c{K}inney 2010, in {S}t\'efan van~der {W}alt {J}arrod {M}illman eds,
  {P}roceedings of the 9th {P}ython in {S}cience {C}onference. pp 56 -- 61,
  \mn@doi{10.25080/Majora-92bf1922-00a}

\bibitem[\protect\citeauthoryear{{Zhang}, {Luo}  \& {Fu}}{{Zhang}
  et~al.}{2012}]{Zhang}
{Zhang} X.~B.,  {Luo} C.~Q.,   {Fu} J.~N.,  2012, \mn@doi [\aj]
  {10.1088/0004-6256/144/3/86}, \href
  {https://ui.adsabs.harvard.edu/abs/2012AJ....144...86Z} {144, 86}

\bibitem[\protect\citeauthoryear{{Zucker}}{{Zucker}}{2003}]{Zucker}
{Zucker} S.,  2003, \mn@doi [\apj] {10.1086/506900}, \href
  {https://arxiv.org/abs/astro-ph/0303426v1} {650, 916}

\bibitem[\protect\citeauthoryear{{de Burgos}, {Sim{\'o}n-D{\'\i}az}, {Urbaneja}
   \& {Negueruela}}{{de Burgos} et~al.}{2023}]{deburgos_1}
{de Burgos} A.,  {Sim{\'o}n-D{\'\i}az} S.,  {Urbaneja} M.~A.,   {Negueruela}
  I.,  2023, \mn@doi [\aap] {10.1051/0004-6361/202346179}, \href
  {https://ui.adsabs.harvard.edu/abs/2023A&A...674A.212D} {674, A212}

\bibitem[\protect\citeauthoryear{{de Burgos}, {Sim{\'o}n-D{\'\i}az}, {Urbaneja}
   \& {Puls}}{{de Burgos} et~al.}{2024}]{deburgos_2}
{de Burgos} A.,  {Sim{\'o}n-D{\'\i}az} S.,  {Urbaneja} M.~A.,   {Puls} J.,
  2024, \mn@doi [\aap] {10.1051/0004-6361/202348808}, \href
  {https://ui.adsabs.harvard.edu/abs/2024A&A...687A.228D} {687, A228}

\bibitem[\protect\citeauthoryear{{de Mink} \& {Belczynski}}{{de Mink} \&
  {Belczynski}}{2015}]{DeMink_2015}
{de Mink} S.~E.,  {Belczynski} K.,  2015, \mn@doi [\apj]
  {10.1088/0004-637X/814/1/58}, \href
  {https://ui.adsabs.harvard.edu/abs/2015ApJ...814...58D} {814, 58}

\bibitem[\protect\citeauthoryear{{de Mink}, {Langer}, {Izzard}, {Sana}  \& {de
  Koter}}{{de Mink} et~al.}{2013}]{demink2013}
{de Mink} S.~E.,  {Langer} N.,  {Izzard} R.~G.,  {Sana} H.,   {de Koter} A.,
  2013, \mn@doi [\apj] {10.1088/0004-637X/764/2/166}, \href
  {https://ui.adsabs.harvard.edu/abs/2013ApJ...764..166D} {764, 166}

\bibitem[\protect\citeauthoryear{pandas~development team}{pandas~development
  team}{2020}]{reback2020pandas}
pandas~development team T.,  2020, pandas-dev/pandas: Pandas,
  \mn@doi{10.5281/zenodo.3509134}, \url
  {https://doi.org/10.5281/zenodo.3509134}

\makeatother
\end{thebibliography}




\appendix
\section{Tables}
\begin{table*}
\small
\centering
\caption{In the columns are reported the ID  of the 25 identified SB1s and their orbital parameters, which are: period, eccentricity, the semi-amplitude of the RV curve, the time of periastron passage and the argument of periastron, the binary mass function of the systems and reduced $\chi^2$ from the minimisation procedure. In the flag column, * indicates that the orbital solution is ambiguous, as the chosen period does not correspond to the tallest peak in the periodogram. ** indicates a poorer solution where multiple peaks have been tested, and the one with the lowest $\chi^2_{\rm red}$ value has been selected. Eccentricities lower than 0.001 were fixed to zero.}
\begin{tabular}{lcccccccc}
\hline \hline
SIMBAD ID       & P                   & e                   & $K_1$           & $T_0$       & $\omega$         & $f(M)$            & $\chi_{\rm red}^2$ & Flag \\
                & [d]                 &                     & [km\,s$^{-1}$]  & [MJD]       & [°]              & [M$_\odot$]       &                    &      \\ \hline
$\chi$ Per 2185 & 1.5483$\pm$0.0002   & 0.02185$\pm$0.00005 & 22.36$\pm$0.06  & 7655$\pm$4  & 80.77$\pm$0.01   & 0.0018$\pm$0.0001 & 24.08              &      \\
$\chi$ Per 2311 & 25.5237$\pm$0.0054  & 0.26821$\pm$0.00006 & 63.02$\pm$0.02  & 8538$\pm$3  & 55.06$\pm$2.83   & 0.6579$\pm$0.0006 & 74.11              &      \\
$h$ Per 0843    & 206.7662$\pm$0.3694 & 0.27885$\pm$0.00678 & 16.72$\pm$0.22  & 7480$\pm$6  & 93.54$\pm$1.90   & 0.0995$\pm$0.0016 & 7.17               & *    \\
$h$ Per 0980    & 5.0494$\pm$0.0004   & 0.09655$\pm$0.00002 & 50.65$\pm$0.01  & 7756$\pm$2  & 89.67$\pm$0.02   & 0.0676$\pm$0.0002 & 4378.73            & **   \\
$h$ Per 1004    & 3.4248$\pm$0.0007   & 0.20360$\pm$0.00075 & 23.02$\pm$0.15  & 7882$\pm$1  & 41.86$\pm$0.23   & 0.0043$\pm$0.0002 & 16.62              & *    \\
$h$ Per 1085    & 3.8491$\pm$0.0004   & $<10^{-5}$$^{(\ddagger)}$ & 20.00$\pm$0.01  & 7553$\pm$8  & 294.93$\pm$0.14  & 0.0032$\pm$0.0001 & 17.11              &      \\
HD 14052        & 32.0775$\pm$0.0024  & 0.08184$\pm$0.00002 & 55.62$\pm$0.01  & 6576$\pm$11 & 261.81$\pm$0.06  & 0.5686$\pm$0.0005 & 136.20             &      \\ \hline
NGC~457 6       & 1.6880$\pm$0.0001   & 0.26742$\pm$0.00013 & 44.54$\pm$0.03  & 7864$\pm$11 & 105.28$\pm$0.05  & 0.0154$\pm$0.0001 & 11.20              & **   \\
NGC~457 7       & 5.1563$\pm$0.0009   & $<10^{-5}$$^{(\ddagger)}$ & 25.01$\pm$0.11  & 7574$\pm$38 & 70.02$\pm$0.36   & 0.0083$\pm$0.0003 & 2.93               &      \\
NGC~457 8       & 1.2413$\pm$0.0001   & 0.13100$\pm$0.00012 & 31.69$\pm$0.03  & 9537$\pm$5  & 264.10$\pm$0.22  & 0.0041$\pm$0.0001 & 103.90             &      \\
NGC~457 14      & 1.2008$\pm$0.0001   & $<10^{-5}$$^{(\ddagger)}$ & 10.97$\pm$0.05  & 8401$\pm$17 & 16.56$\pm$0.12   & 0.0002$\pm$0.0001 & 25.32              & SB2? \\
NGC~457 19      & 14.1574$\pm$0.0018  & 0.22000$\pm$0.00003 & 41.41$\pm$0.01  & 8016$\pm$2  & 45.00$\pm$0.01   & 0.1035$\pm$0.0002 & 39.60              &      \\
NGC~457 37      & 4.4475$\pm$0.0003   & $<10^{-5}$$^{(\ddagger)}$ & 113.69$\pm$0.02 & 4333$\pm$8  & 191.15$\pm$0.06  & 0.6732$\pm$0.0005 & 275.24             &      \\
NGC~457 54      & 4.6414$\pm$0.0004   & 0.33571$\pm$0.00020 & 22.18$\pm$0.01  & 8234$\pm$14 & 31.52$\pm$0.01   & 0.0052$\pm$0.0001 & 40.97              & SB2? \\
NGC~457 91      & 3.3328$\pm$0.0014   & 0.0$^{(\dagger)}$   & 19.19$\pm$3.37  & 3888$\pm$19 & 149.74$\pm$81.53 & 0.0024$\pm$0.0009 & 1.08               & **   \\
NGC~457 100     & 5.9789$\pm$0.0033   & 0.38080$\pm$0.00159 & 10.32$\pm$0.05  & 9068$\pm$2  & 103.98$\pm$0.67  & 0.0007$\pm$0.0001 & 4.20               & *    \\
NGC~457 154     & 9.8417$\pm$0.0177   & 0.0$^{(\dagger)}$   & 11.29$\pm$2.20  & 8487$\pm$93 & 250.22$\pm$76.86 & 0.0015$\pm$0.0008 & 0.99               & *    \\ \hline
EM* GGA 52      & 2.2507$\pm$0.0002   & 0.01174$\pm$0.00001 & 56.78$\pm$0.02  & 8727$\pm$6  & 127.31$\pm$0.06  & 0.0425$\pm$0.0002 & 79.26              &      \\
NGC 581 35      & 5.0450$\pm$0.0048   & 0.05920$\pm$0.00290 & 16.57$\pm$0.73  & 7844$\pm$15 & 261.73$\pm$19.41 & 0.0024$\pm$0.0005 & 238.14             &      \\
NGC 581 59      & 35.0365$\pm$0.4761  & $<10^{-5}$$^{(\ddagger)}$ & 18.46$\pm$3.77  & 4315$\pm$85 & 313.83$\pm$32.62 & 0.0227$\pm$0.0031 & 52.64              &      \\ \hline
NGC 1960 009    & 16.7445$\pm$0.0029  & 0.04862$\pm$0.00001 & 34.07$\pm$0.01  & 7541$\pm$3  & 131.30$\pm$0.03  & 0.0682$\pm$0.0002 & 14.14              &      \\
NGC 1960 016    & 3.7573$\pm$0.0001   & $<10^{-5}$$^{(\ddagger)}$ & 24.87$\pm$0.01  & 972$\pm$1   & 263.15$\pm$0.09  & 0.0060$\pm$0.0001 & 82.34              &      \\
NGC 1960 027    & 2.4125$\pm$0.0029   & 0.24487$\pm$0.00472 & 12.67$\pm$0.22  & 7842$\pm$79 & 267.17$\pm$3.02  & 0.0005$\pm$0.0001 & 45.33              & **   \\
NGC~1960 109    & 1.5842$\pm$0.0001   & 0.03259$\pm$0.00002 & 19.49$\pm$0.01  & 9452$\pm$2  & 309.82$\pm$0.13  & 0.0012$\pm$0.0001 & 29.14              &      \\
NGC~1960 134    & 8.8600$\pm$0.0021   & 0.54431$\pm$0.00015 & 39.49$\pm$0.01  & 7811$\pm$1  & 113.60$\pm$0.04  & 0.0562$\pm$0.0002 & 9.07               &      \\ \hline
\end{tabular}
    \begin{tablenotes}
    \item \textbf{Notes.} The errors correspond to 1$\sigma$.
        $^{(\dagger)}$ Eccentricities that do not pass the Lucy and Sweeney test ($e/\sigma_e\leq2.49$).
        $^{(\ddagger)}$ Systems have significant but very small non-zero eccentricity.
    \end{tablenotes}
\label{tab: orbital solutions}
\end{table*}

\begin{table*}
\small
\centering
\caption{In the columns are reported the ID  of the three identified SB2s and their orbital parameters, which are: period, eccentricity, the semi-amplitude of the RV curve of the primary, the semi-amplitude of the RV curve of the secondary, the time of periastron passage and the argument of periastron, the binary mass function of the systems and reduced $\chi^2$ from the minimisation procedure.}
\begin{tabular}{lcccccccc}
\hline \hline
SIMBAD ID       & P                   & e                     & $K_1$             & $K_2$             & $T_0$       & $\omega $         & $f(M)$              & $\chi^2_{\rm red}$ \\
                & [d]                 &                       & [km\,s$^{-1}$]    & [km\,s$^{-1}$]    & [MJD] & [°]               & [M$_\odot$]         &                    \\ \hline
$\chi$ Per 2392 & 5.4090$\pm$0.0005 & 0.31625$\pm$0.00006 & 35.03$\pm$0.01  & 53.55$\pm$0.01  & 7654$\pm$1  & 6.59$\pm$0.01   & 0.0239$\pm$0.0001 & 134.24    \\
NGC~457 85      & 1.7160$\pm$0.0002 & $<10^{-5}$$^{(\ddagger)}$ & 150.01$\pm$0.03 & 249.99$\pm$0.06 & 8699$\pm$17 & 180.01$\pm$0.06 & 0.5968$\pm$0.0006 & 13.69        \\
NGC~1960 008    & 2.4973$\pm$0.0002 & 0.19119$\pm$0.00004 & 115.59$\pm$0.03 & 175.03$\pm$0.04 & 7648$\pm$4  & 18.57$\pm$0.01  & 0.3974$\pm$0.0004 & 5.87 \\ \hline
\end{tabular}
 \begin{tablenotes}
    \item \textbf{Notes.} The errors correspond to 1$\sigma$.
     $^{(\ddagger)}$ Systems have significant but very small non-zero eccentricity.

    \end{tablenotes}
\label{tab:orbital_solutions_sb2}
\end{table*}


\clearpage
\begin{table*}
\caption{Parameters for the 74 stars in the sample. The first and second columns list the SIMBAD and Gaia IDs, respectively. The third column provides the V-band magnitude from SIMBAD, while the fourth gives the spectral type as determined in this study. The fifth column shows the number of HERMES observation epochs. The ‘Flag’ column indicates the binary status of each star. The last column presents the projected rotational velocities with associated 2$\sigma$ uncertainties.} 
\begin{tabular}{llccccc}
\hline \hline
SIMBAD ID          & Gaia DR3 ID         & V     & SpT       & Epochs & Flag   & $v \sin i$        \\
                   &                     & [mag] &           &        &        & [km s$^{-1}$]     \\ \hline
\hline
BD+56 566 & 458454880886508032  & 10.16 & B1e V     & 2      & Single & 120$^{+40}_{-26}$ \\
$\chi$ Per 2185    & 458406124415716224  & 10.92 & B5 V      & 9      & SB1    & 180$^{+24}_{-30}$ \\
$\chi$ Per 2235    & 458454606008607232  & 8.40  & B1 V      & 18     & Single & 160$^{+8}_{-24}$  \\
$\chi$ Per 2246    & 458454606008606336  & 9.98  & B1V       & 19     & Single & 120$^{+6}_{-20}$  \\
$\chi$ Per 2255    & 458454709087817344  & 10.71 & B2 V      & 7      & Single & 340$^{+28}_{-44}$ \\
$\chi$ Per 2296    & 458407601884442496  & 8.53  & B1 II-III & 17     & Single & 180$^{+6}_{-26}$  \\
$\chi$ Per 2299    & 458407670603899648  & 8.32  & B1 II-III & 17     & Single & 120$^{+6}_{-18}$  \\
$\chi$ Per 2311    & 458407601884432384  & 9.40  & B0 V      & 17     & SB1    & 20$^{+14}_{-2}$   \\
$\chi$ Per 2392    & 458408323438859520  & 10.71 & B2 V      & 11     & SB2    & -                 \\
HD 14052           & 458379014584350592  & 7.79  & B1 II-III & 7      & SB1    & 60$^{}_{-22}$     \\
$h$ Per 0843       & 458377674554774016  & 8.38  & B1 V      & 18     & SB1    & 120$^{+6}_{-26}$  \\
$h$ Per 0864       & 458371695960355072  & 9.16  & B2 V      & 19     & Single & 160$^{+12}_{-50}$ \\
$h$ Per 0929       & 458377811993695232  & 10.32 & B2 V      & 9      & Single & 180$^{+18}_{-26}$ \\
$h$ Per 0936       & 458377708914487168  & 10.44 & B0 V      & 18     & Single & 40$^{+4}_{-8}$    \\
$h$ Per 0978       & 458377743274216192  & 10.68 & B2 II III & 7      & Single & 60$^{+2}_{-22}$   \\
$h$ Per 0980       & 458377743274231936  & 9.75  & B0 V      & 16     & SB1    & 40$^{+18}_{-2}$   \\
$h$ Per 0992       & 458377743274230784  & 10.00 & B1 II III & 13     & Single & 180$^{+30}_{-44}$ \\
$h$ Per 1004       & 458377743274223872  & 10.91 & B2 V      & 9      & SB1    & 140$^{+18}_{-22}$ \\
$h$ Per 1078       & 458374719617277184  & 9.82  & B1 V      & 19     & Single & 200$^{+18}_{-28}$ \\
$h$ Per 1085       & 458374822697357696  & 10.47 & B1 V      & 10     & SB1    & 80$^{+16}_{-10}$  \\
$h$ Per 1116       & 458374719617260672  & 9.29  & B1 V      & 20     & Single & 120$^{+14}_{-10}$ \\
$h$ Per 1132       & 458374719617266048  & 8.48  & B2 II-III & 11     & Single & 80$^{+24}_{-4}$   \\
$h$ Per 1133       & 458374719617270272  & 9.04  & B1 II-III & 14     & Single & 200$^{+16}_{-20}$ \\ \hline
NGC~457 6          & 413876247091175040  & 10.60 & B3e V     & 12     & SB1    & 200$^{+58}_{-54}$ \\
NGC~457 7          & 413876247091172480  & 10.95 & B5 V      & 11     & SB1    & 140$^{+46}_{-28}$ \\
NGC~457 8          & 413874769624097280  & 10.02 & B1 V      & 11     & SB1    & 60$^{+10}_{-12}$  \\
NGC~457 14         & 413874735262711040  & 10.20 & B1e V     & 10     & SB1    & 180$^{+42}_{-26}$ \\
NGC~457 19         & 413874524796674048  & 9.51  & B1 II-III & 12     & SB1    & 40$^{+16}_{-2}$   \\
NGC~457 33         & 413876414582276608  & 10.33 & B2 V      & 9      & Single & 100$^{+24}_{-6}$  \\
NGC~457 34         & 413876247091165824  & 10.74 & B5 V      & 7      & Single & 80$^{+32}_{-20}$  \\
NGC~457 37         & 413876311504679168  & 9.83  & B2 V      & 11     & SB1    & 120$^{+12}_{-10}$ \\
NGC~457 54         & 413874494744555136  & 10.18 & B2 V      & 12     & SB1    & 120$^{+26}_{-12}$ \\
NGC 457 85         & 413875525536666368  & 10.78 & B5 V      & 8      & SB2    & -                 \\
NGC~457 91         & 413874254226395008  & 11.30 & B5e V     & 12     & SB1    & 280$^{+60}_{-60}$ \\
NGC~457 100        & 413827387540175360  & 10.61 & B2 V      & 12     & SB1    & 160$^{+26}_{-20}$ \\
NGC~457 120        & 413877380962478336  & 9.93  & B0 V      & 15     & Single & 40$^{+4}_{-6}$    \\
NGC~457 128        & 413875010140633344  & 9.72  & B2e V     & 11     & Single & 200$^{+30}_{-18}$ \\
NGC~457 153        & 413851336279382272  & 9.48  & B1e V     & 20     & Single & 200$^{+52}_{-26}$ \\
NGC~457 154        & 413851370639124736  & 11.18 & B5 V      & 12     & SB1    & 220$^{+34}_{-32}$ \\
NGC~457 198        & 413879197721080960  & 9.46  & B2e V     & 18     & Single & 200$^{+16}_{-28}$ \\
NGC~457 275        & 413848381341936768  & 9.85  & B3e V     & 12     & Single & 200$^{+26}_{-32}$ \\
HD 236695          & 413980906852870912  & 9.50  & B2 II III & 3      & Single & 40$^{+10}_{-4}$   \\ \hline
NGC~581 49         & 509863062443025152  & 11.76 & B2e V     & 3      & Single & 320$^{}_{-136}$   \\
NGC~581 35$^{*}$         & 509862340890419328  & 10.45 & B3 V      & 7      & SB1    & 200$^{+30}_{-26}$ \\
NGC~581 59         & 509862890644331520  & 11.45 & B3 V      & 6      & SB1    & 260$^{+50}_{-30}$ \\
NGC~581 70         & 509863028083270656  & 11.74 & B5 V      & 3      & Single & 160$^{+38}_{-42}$ \\
NGC~581 73$^{*}$         & 509862989414917376  & 10.59 & B5 V      & 7      & Single & 180$^{+26}_{-20}$ \\
NGC~581 162        & 509861413175947008  & 11.22 & B2 V      & 2      & Single & 40$^{+16}_{-8}$   \\
NGC~581 111        & 509863509113178624  & 11.84 & B5 V      & 3      & Single & 140$^{+58}_{-42}$ \\
NGC~581 ZUG 8$^{*}$      & 509862169090117760  & 11.00 & B2 V      & 4      & SB1    & 80$^{+10}_{-18}$  \\
TYC 4031 2100 1$^{\dagger}$    & 509848523965083520  & 11.65 & B2 V      & 6      & Single & 20$^{+8}_{-20}$   \\
BD+59 273          & 509862031651190016  & 9.03  & B1 V      & 11     & Single & 120$^{+16}_{-8}$  \\
EM* GGA 52         & 509862478327484288  & 11.22 & B3 V      & 7      & SB1    & 120$^{+10}_{-26}$ \\
EM* GGA 54         & 509862924996841984  & 11.27 & B0e V     & 4      & Single & 200$^{+76}_{-56}$ \\
EM* GGA 56         & 509863852710461568  & 11.35 & B5e V     & 8      & Single & 360$^{100}_{-80}$ \\
V* V1122 Cas       & 509863195576354816  & 9.71  & B3e V     & 2      & SB1    & 200$^{+18}_{-36}$ \\ 
\hline
\end{tabular}
\label{tab:atmospheric_param}
\end{table*}

\begin{table*}
\contcaption{A table continued from the previous one.}
    \begin{tabular}{llccccc}
    \hline \hline
SIMBAD ID          & Gaia DR3 ID         & V     & SpT       & Epochs & Flag   & $v \sin i$        \\
                   &                     & [mag] &           &        &        & [km s$^{-1}$]     \\ \hline
NGC 1960 008       & 3449518312525804032 & 9.36  & B2 V      & 19     & SB2    & -                 \\
NGC~1960 009       & 3449518346885542272 & 9.13  & B1 V      & 16     & SB1    & 40$^{+12}_{-4}$   \\
NGC1960 016$^{*}$        & 3449518587403714432 & 8.79  & B2 V      & 9      & SB1    & 80$^{+16}_{-8}$   \\
NGC~1960 018       & 3449518243806341504 & 10.78 & B5 III    & 8      & Single & 140$^{+24}_{-40}$ \\
NGC~1960 021       & 3449524492981239296 & 9.60  & B2 V      & 7      & Single & 80$^{+14}_{-6}$   \\
NGC~1960 023       & 3449524497278695936 & 8.96  & B3 V      & 7      & Single & 20$^{+2}_{-18}$   \\
NGC~1960 027       & 3449521267463295488 & 9.58  & B2e V     & 12     & SB1    & 340$^{+50}_{-30}$ \\
NGC~1960 038       & 3449515185789622400 & 9.92  & B3 V      & 7      & Single & 80$^{+24}_{-16}$  \\
NGC~1960 047       & 3449518518684250752 & 10.44 & B5e V     & 9      & Single & 300$^{+70}_{-54}$ \\
NGC~1960 048$^{*}$       & 3449518518684245632 & 9.36  & B3 V      & 7      & Single & 60$^{+14}_{-6}$   \\
NGC~1960 061       & 3449524359839733632 & 9.09  & B2 V      & 7      & Single & 200$^{+32}_{-14}$ \\
NGC~1960 081       & 3449518106367398784 & 9.99  & B2 V      & 7      & Single & 140$^{+10}_{-22}$ \\
NGC~1960 087       & 3449519549476397056 & 10.65 & B5 V      & 7      & Single & 180$^{+36}_{-34}$ \\
NGC~1960 091       & 3449519515117517824 & 10.37 & B5 V      & 6      & Single & 320$^{+42}_{-50}$ \\
NGC~1960 101       & 3449521847281383168 & 9.14  & B2e V     & 8      & Single & 80$^{+6}_{-24}$   \\
NGC~1960 109       & 3449514258076685696 & 10.70 & B3 V      & 8      & SB1    & 140$^{+34}_{-26}$ \\
NGC~1960 134       & 3449520717707461888 & 10.39 & B2 V      & 16     & SB1    & 80$^{+18}_{-10}$  \\
NGC~1960 138       & 3449514223716942464 & 8.94  & B2 V      & 15     & Single & 100$^{+12}_{-10}$ \\ 
\hline
    \end{tabular}
     \begin{tablenotes}
    \item \textbf{Notes. } $^{*}$ stars not in CG+2020 catalogue.    
    $^{\dagger}$ star not (yet) officially classified as a cluster member.
    \end{tablenotes}
\end{table*}

\section{Additional figures}\label{app:binaries} 

\begin{figure}%
    \centering
    \includegraphics[width=7cm]{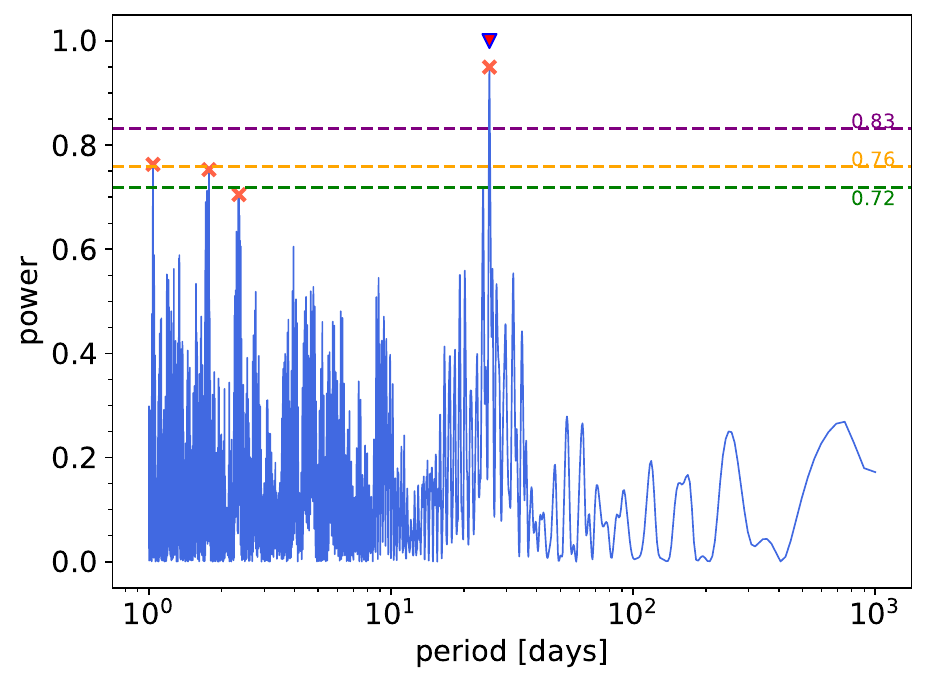} 
    \includegraphics[width=8.7cm]{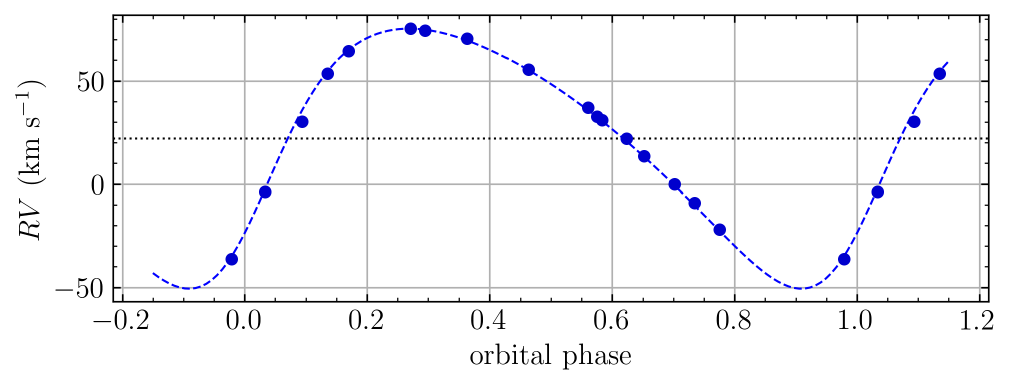} %
    \caption[Example of a Periodogram and orbital solution]%
    {Orbital parameter determination for the star $\chi$ Per 2311. The top panel shows the periodogram with the red crosses indicating the most significant peaks. The triangle indicates the period that was selected for the orbital solution. The horizontal dashed lines are the false alarm probabilities at three different levels: 0.1 (green), 0.05 (orange) and 0.01 (purple). The bottom panel shows the orbital solution obtained using \textsc{spinOS}. Each dot represents one RV measurement, the dashed blue curve is the best-fit solution and the horizontal black dashed line is the systemic velocity. }%
    \label{app:chiper_2311}%
\end{figure}

\begin{figure}%
    \centering
    \includegraphics[width=7cm]{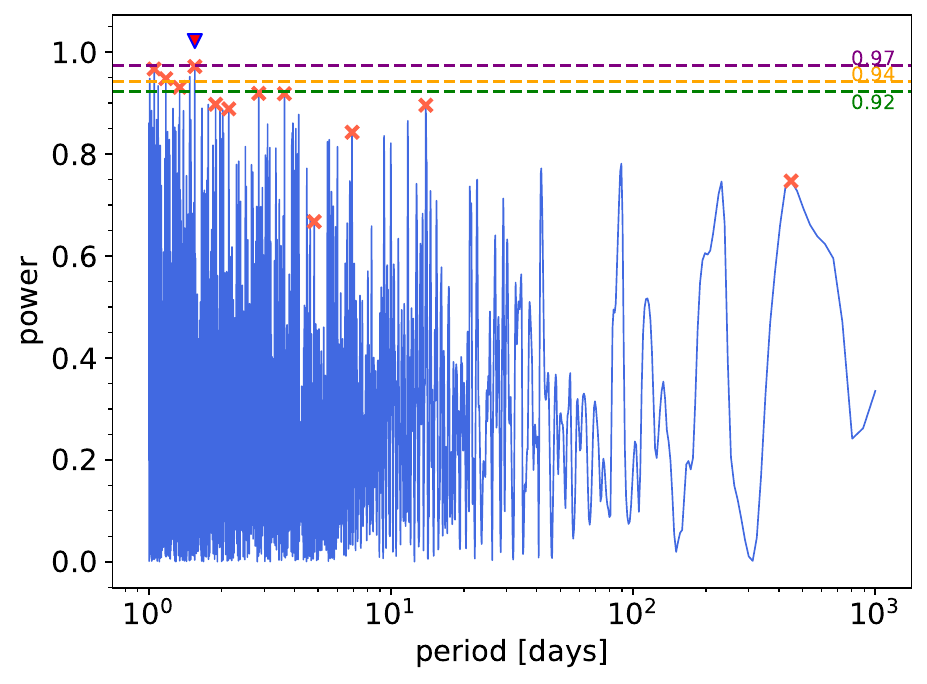} 
    \includegraphics[width=8.7cm]{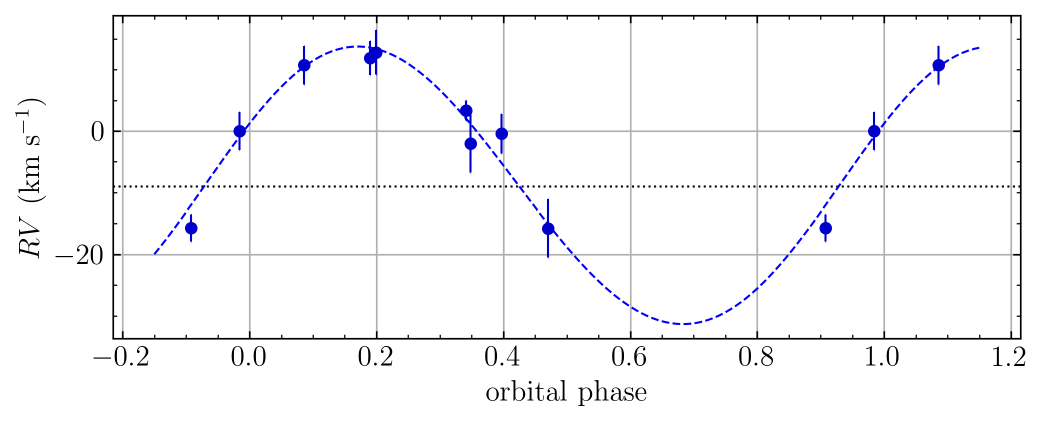} %
    \caption[]%
    { Same figure as Fig.~\ref{app:chiper_2311} but for $\chi$ per 2185.}%
    \label{}%
\end{figure}

\begin{figure}%
    \centering
    \includegraphics[width=7cm]{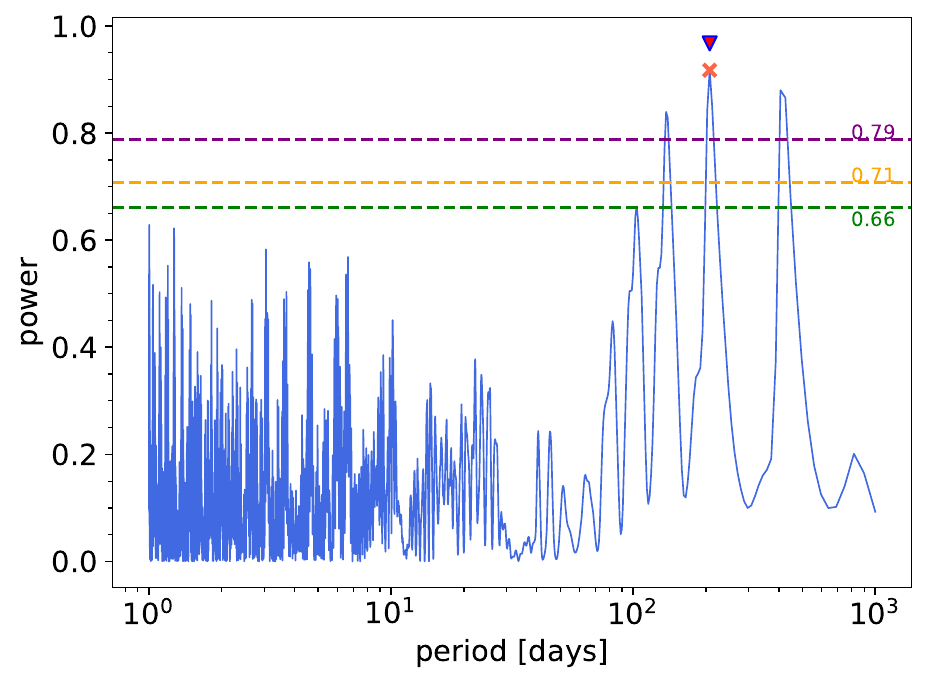} 
    \includegraphics[width=8.7cm]{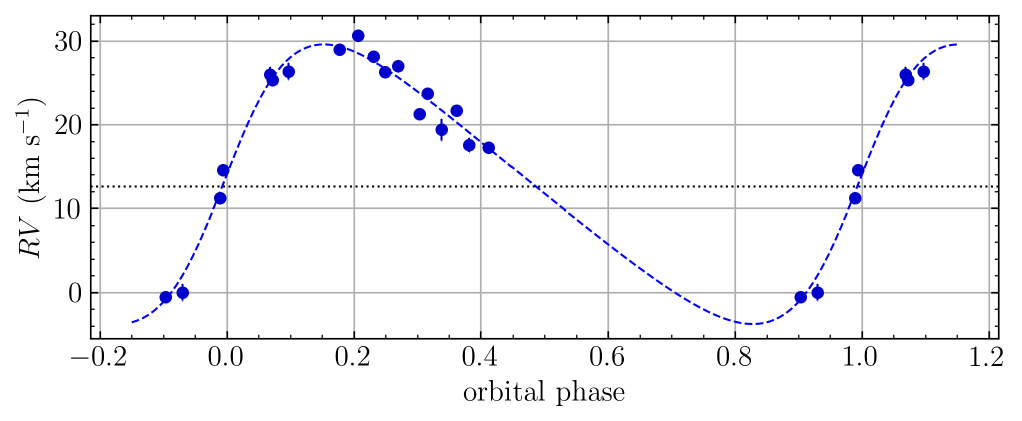} %
    \caption[]%
    {Same figure as Fig.~\ref{app:chiper_2311} but for $h$ Per 0843. }%
    \label{}%
\end{figure}

\begin{figure}%
    \centering
    \includegraphics[width=7cm]{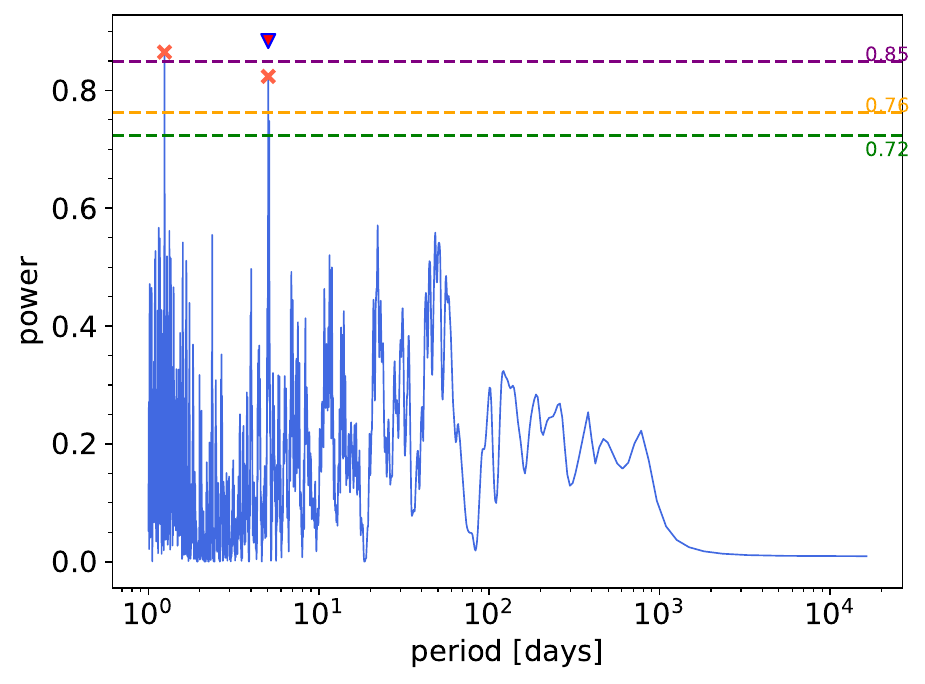} 
    \includegraphics[width=8.7cm]{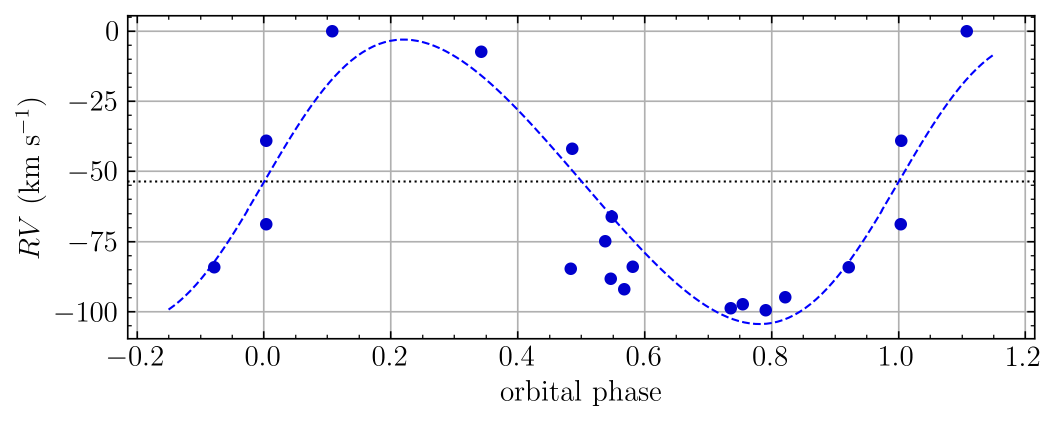} %
    \caption[]%
    {Same figure as Fig.~\ref{app:chiper_2311} but for $h$ Per 0980. }%
    \label{}%
\end{figure}

\begin{figure}%
    \centering
    \includegraphics[width=7cm]{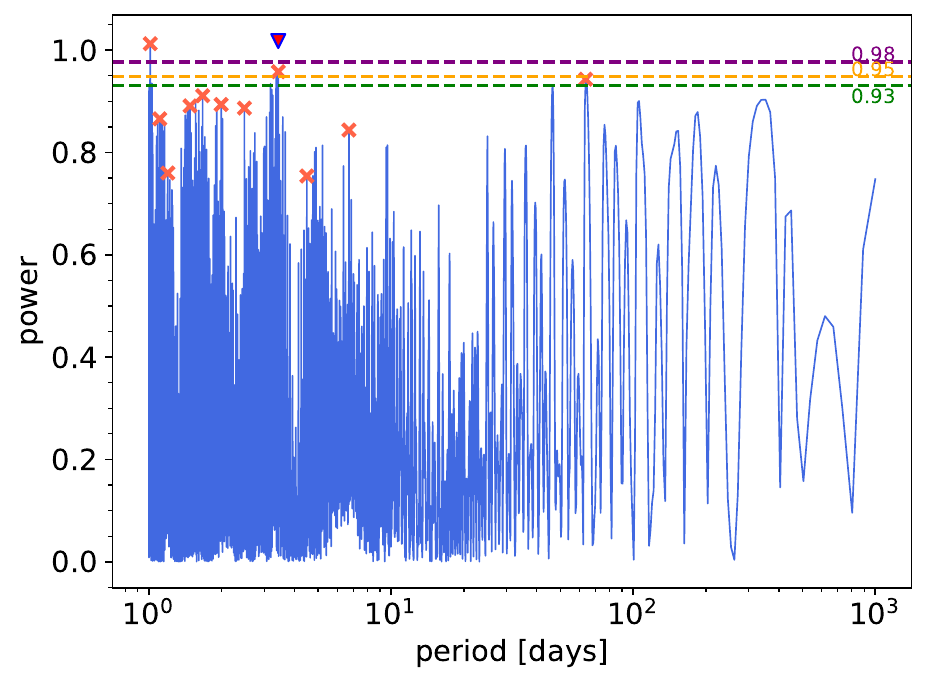} 
    \includegraphics[width=8.7cm]{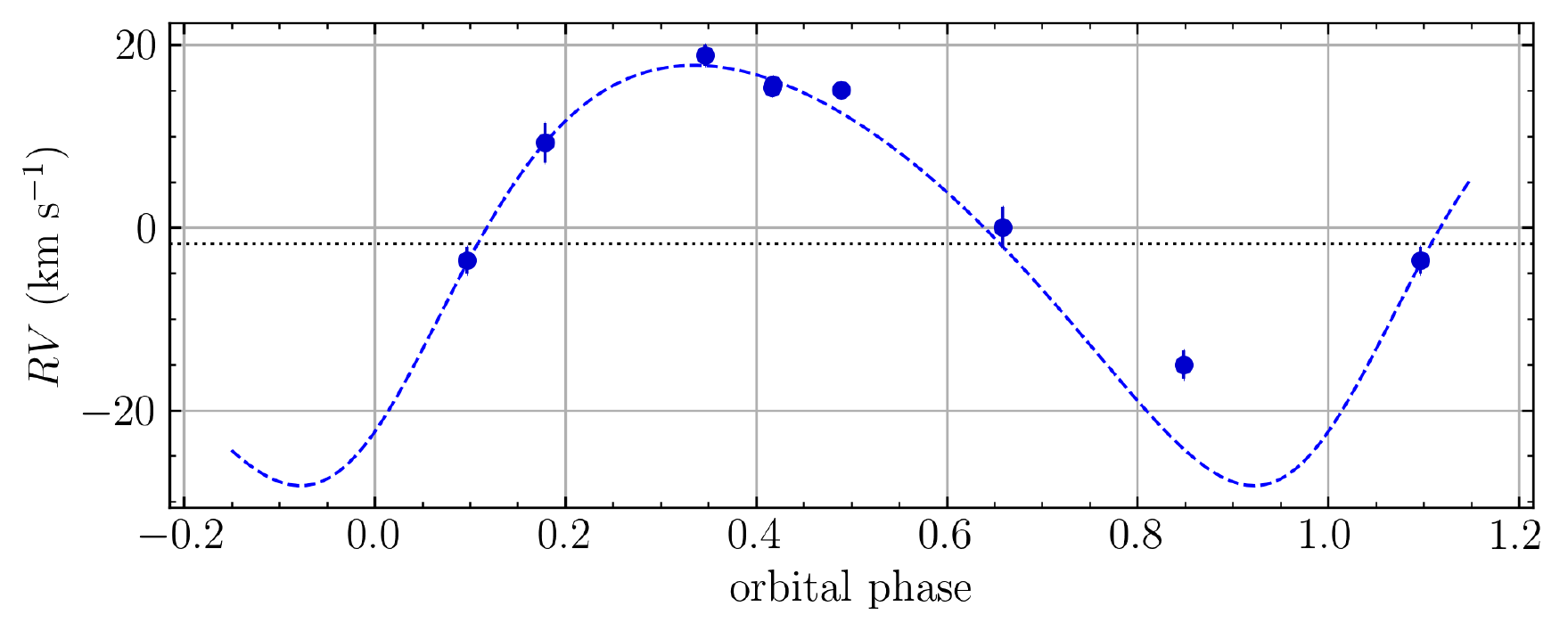} %
    \caption[]%
    {Same figure as Fig.~\ref{app:chiper_2311} but for $h$ Per 1004. }%
    \label{}%
\end{figure}

\begin{figure}%
    \centering
    \includegraphics[width=7cm]{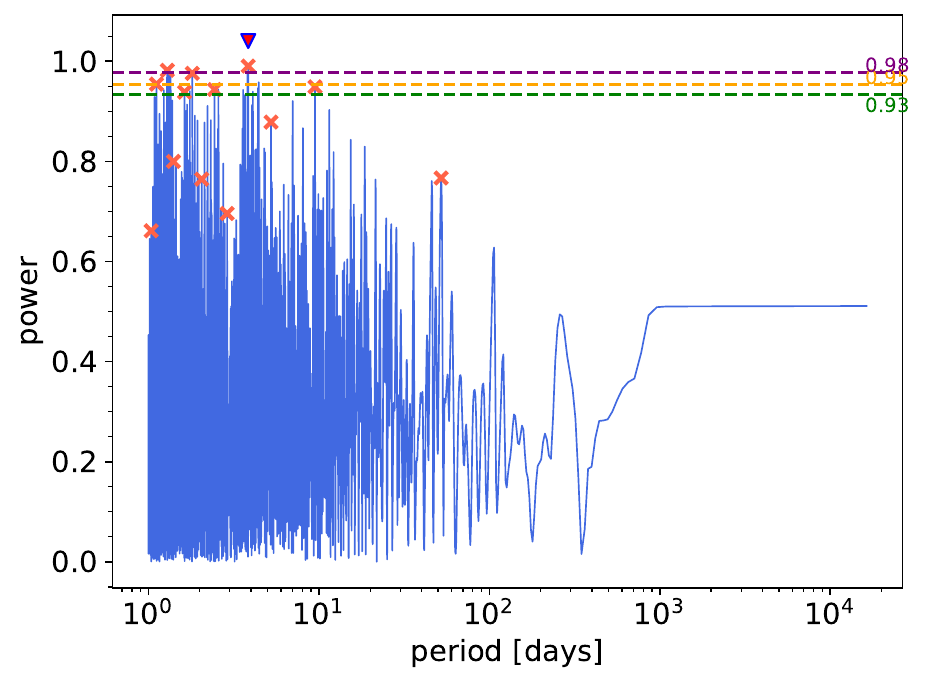} 
    \includegraphics[width=8.7cm]{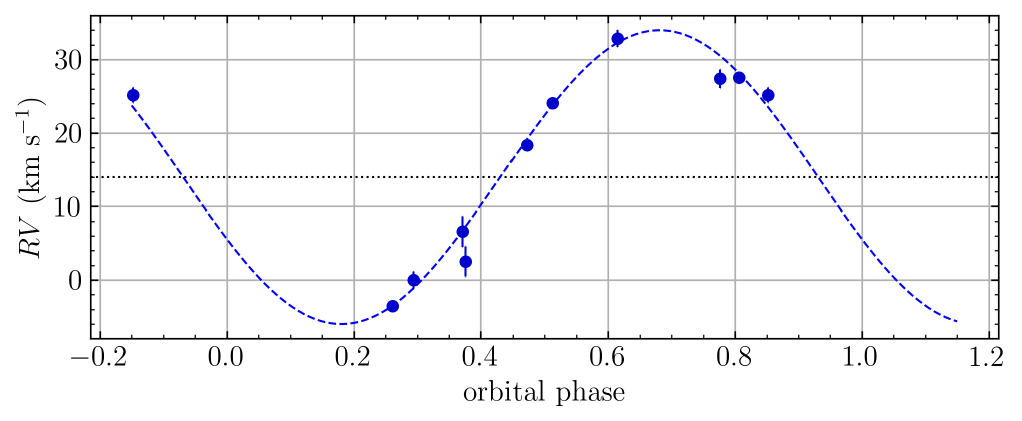} %
    \caption[]%
    {Same figure as Fig.~\ref{app:chiper_2311} but for $h$ Per 1085. }%
    \label{}%
\end{figure}

\begin{figure}%
    \centering
    \includegraphics[width=7cm]{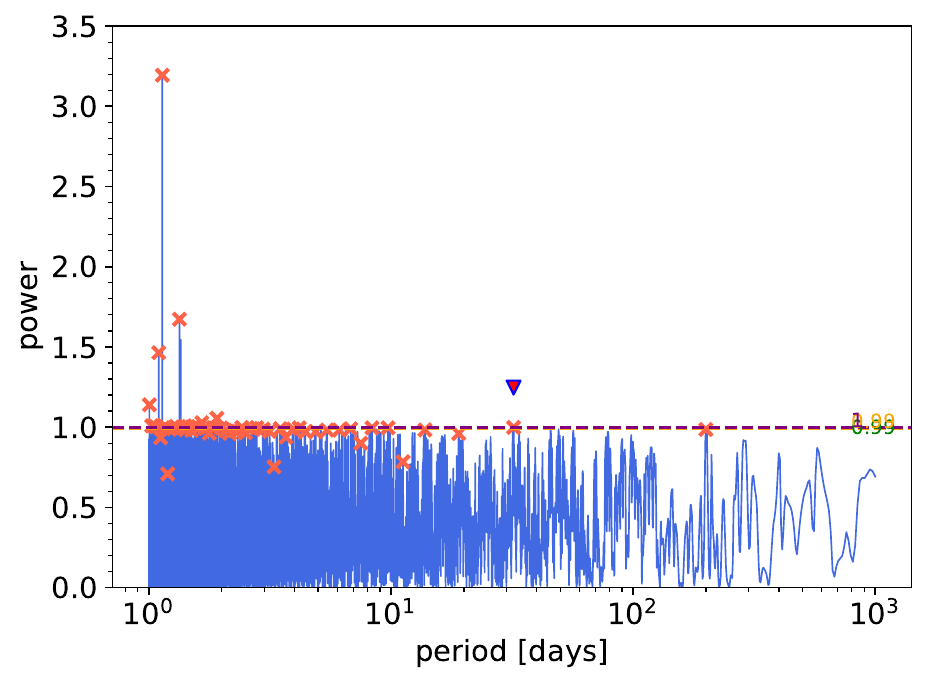} 
    \includegraphics[width=8.7cm]{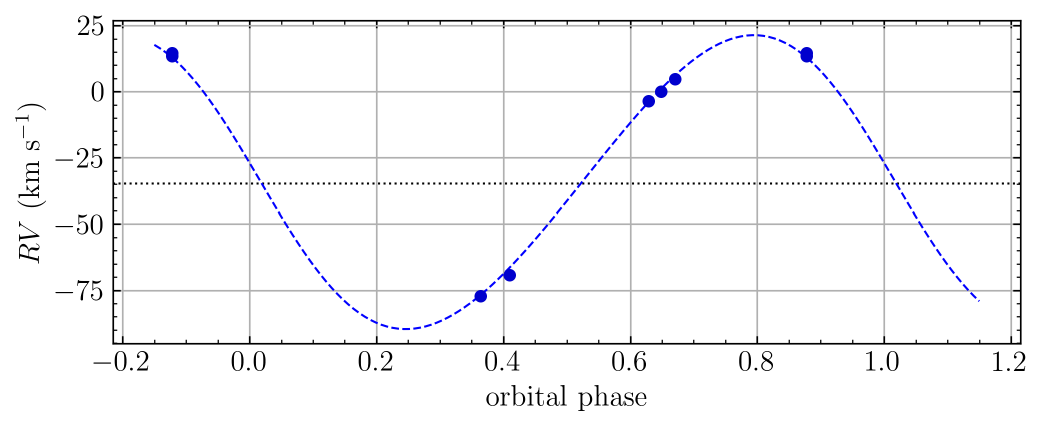} %
    \caption[]%
    {Same figure as Fig.~\ref{app:chiper_2311} but for HD 14052.}%
    \label{}%
\end{figure}

\begin{figure}%
    \centering
    \includegraphics[width=7cm]{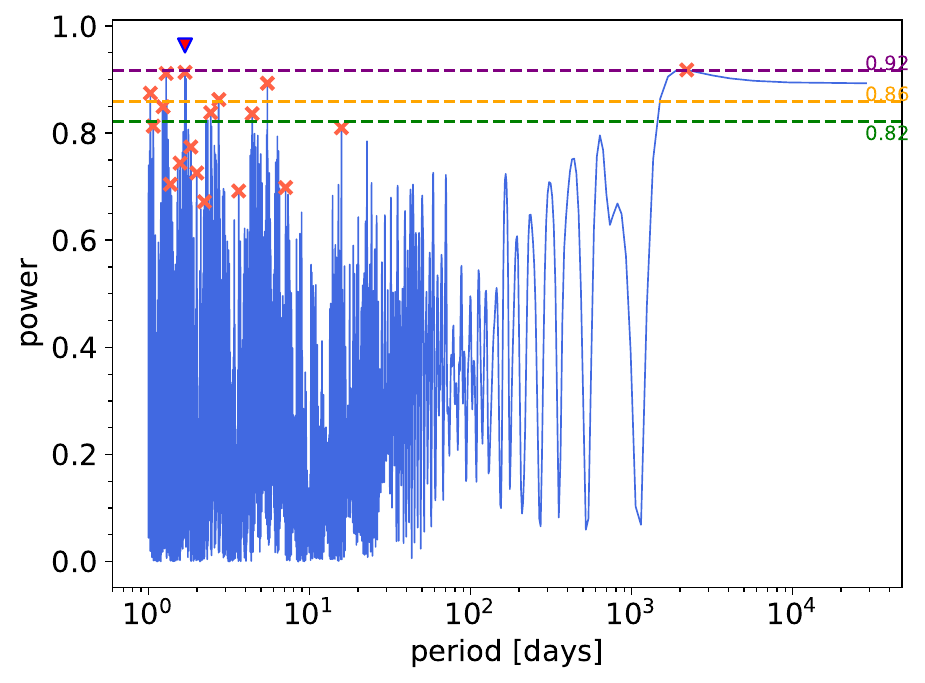} 
    \includegraphics[width=8.7cm]{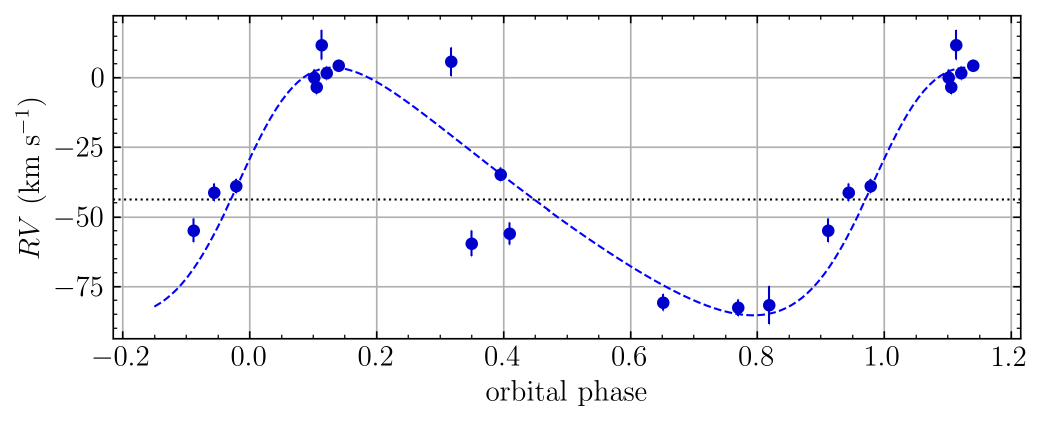} %
    \caption[]%
    {Same figure as Fig.~\ref{app:chiper_2311} but for NGC~457 6. }%
    \label{}%
\end{figure}

\begin{figure}%
    \centering
    \includegraphics[width=7cm]{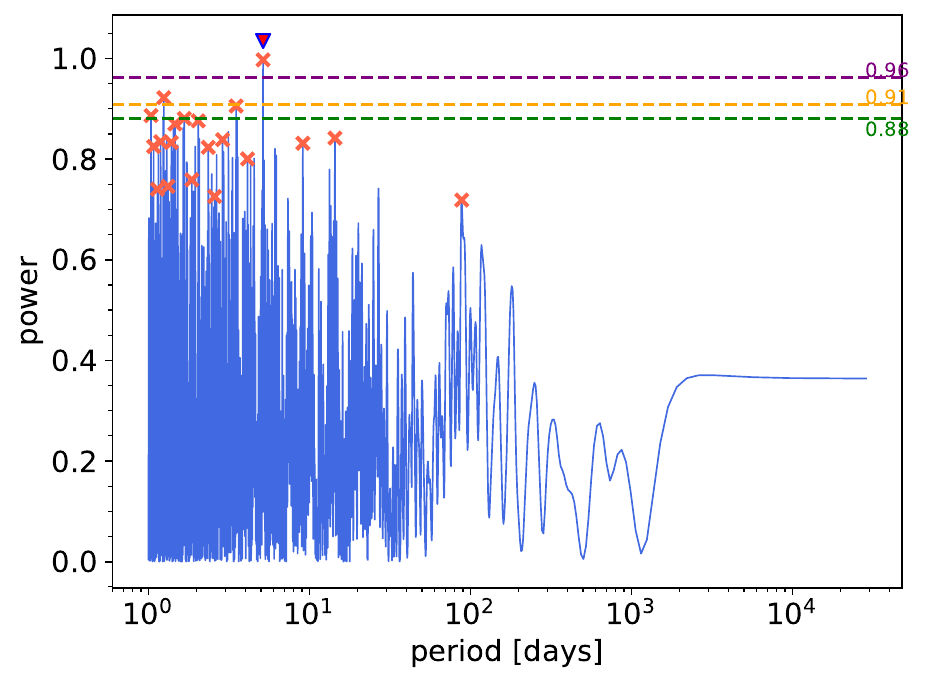} 
    \includegraphics[width=8.7cm]{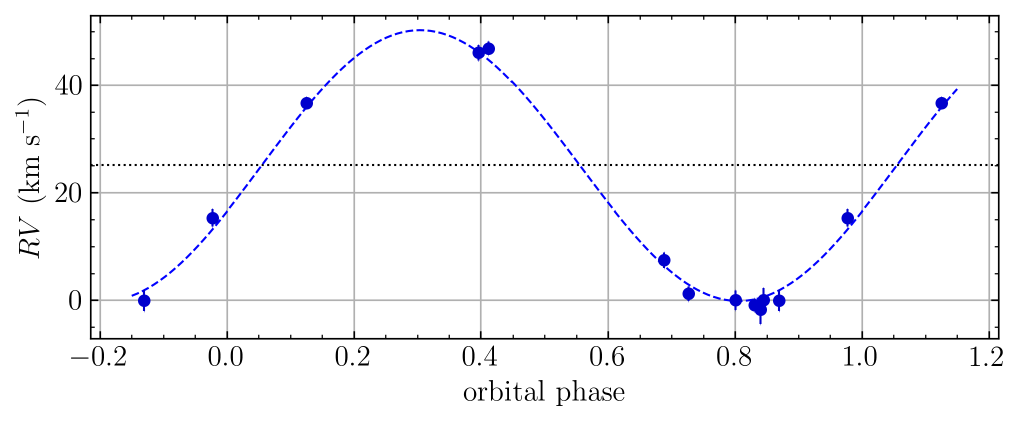} %
    \caption[]%
    {Same figure as Fig.~\ref{app:chiper_2311} but for NGC~457 7. }%
    \label{}%
\end{figure}

\begin{figure}%
    \centering
    \includegraphics[width=7cm]{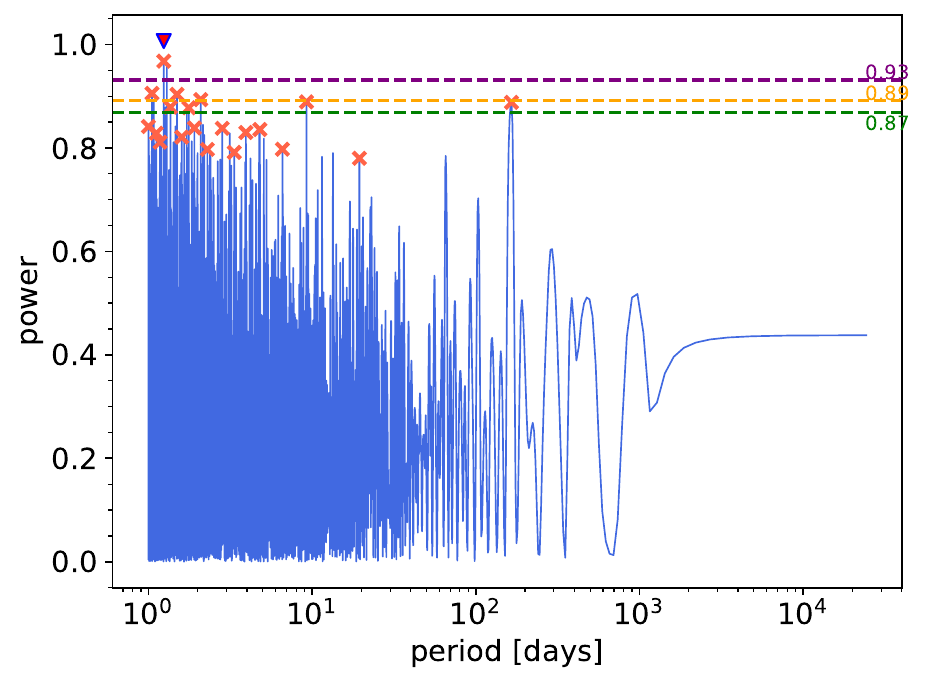} 
    \includegraphics[width=8.7cm]{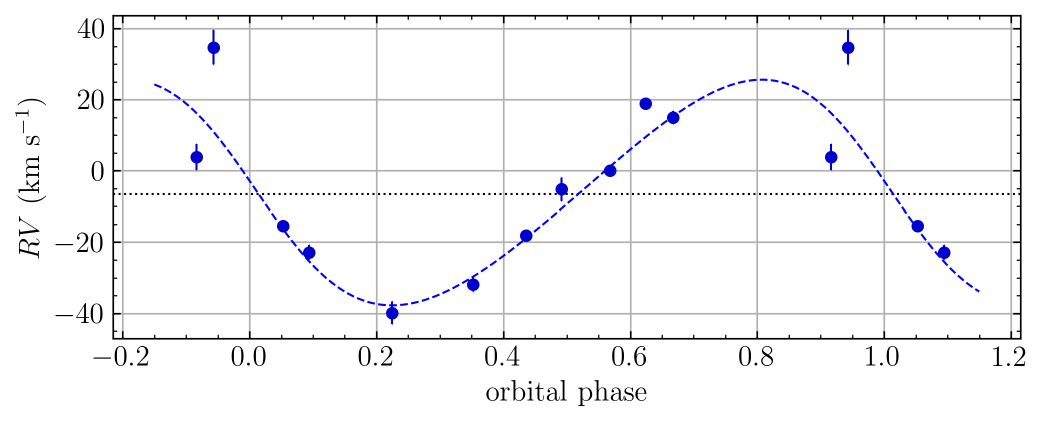} %
    \caption[]%
    {Same figure as Fig.~\ref{app:chiper_2311} but for NGC~457 8. }%
    \label{}%
\end{figure}

\begin{figure}%
    \centering
    \includegraphics[width=7cm]{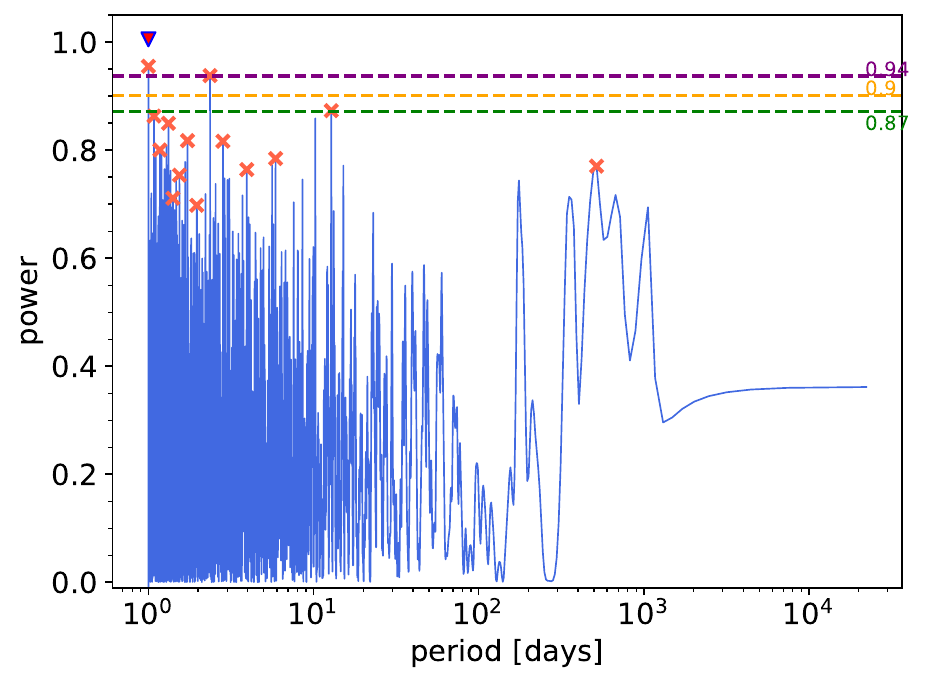} 
    \includegraphics[width=8.7cm]{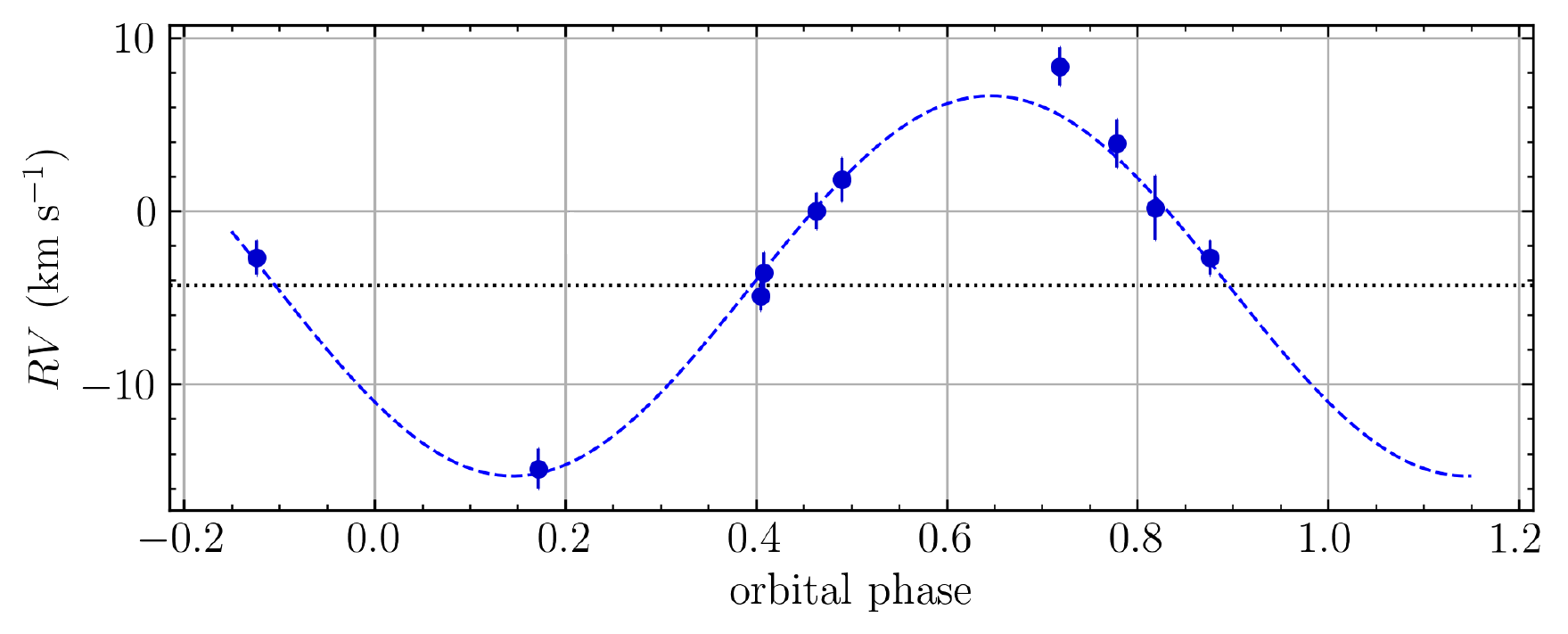} %
    \caption[]%
    {Same figure as Fig.~\ref{app:chiper_2311} but for NGC~457 14.}%
    \label{}%
\end{figure}

\begin{figure}%
    \centering
    \includegraphics[width=7cm]{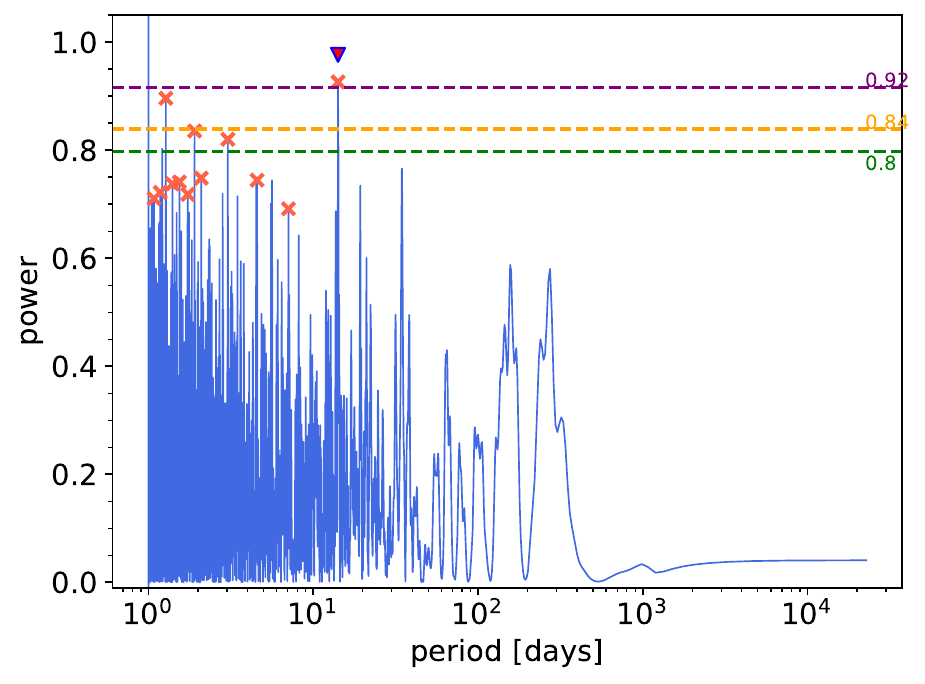} 
    \includegraphics[width=8.7cm]{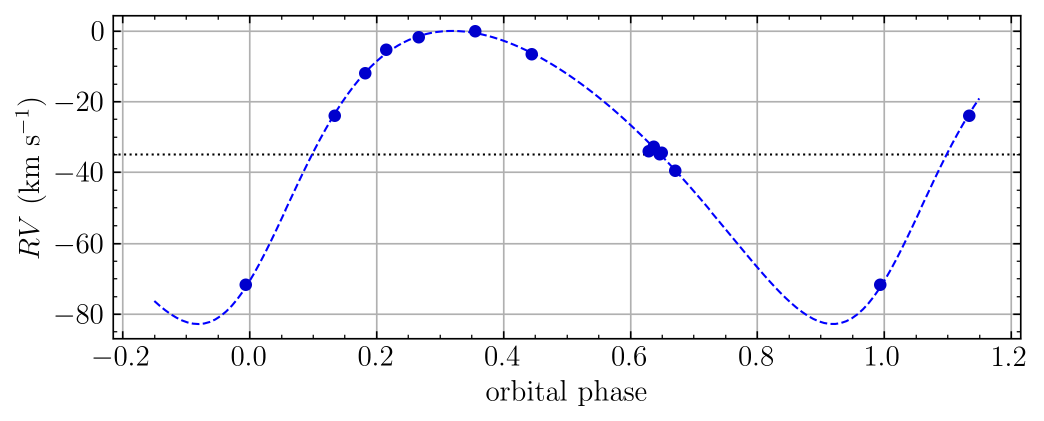} %
    \caption[]%
    {Same figure as Fig.~\ref{app:chiper_2311} but for NGC~457 19. }%
    \label{}%
\end{figure}

\begin{figure}%
    \centering
    \includegraphics[width=7cm]{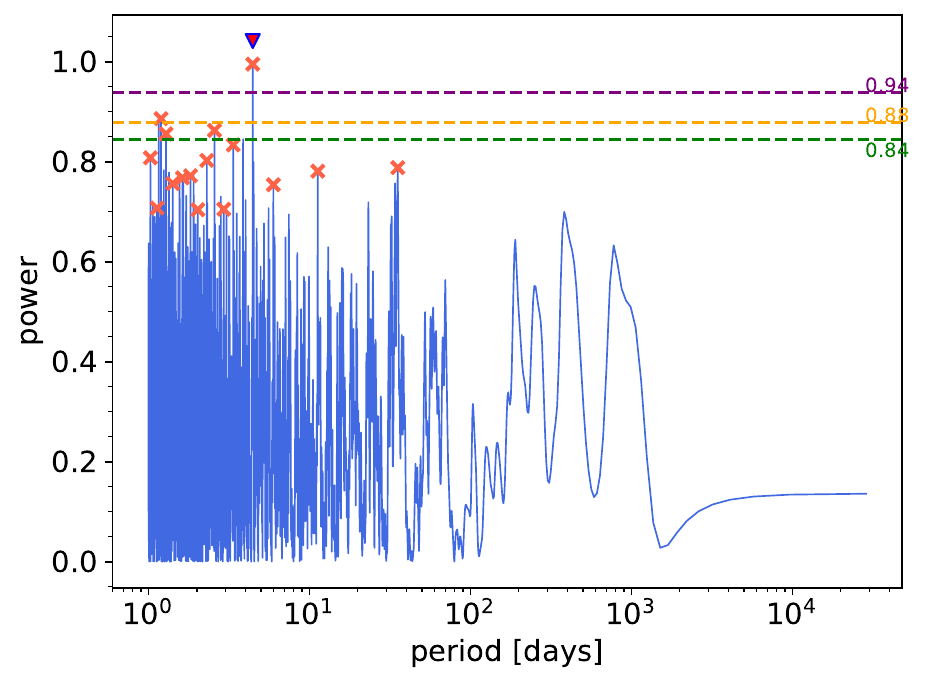} 
    \includegraphics[width=8.7cm]{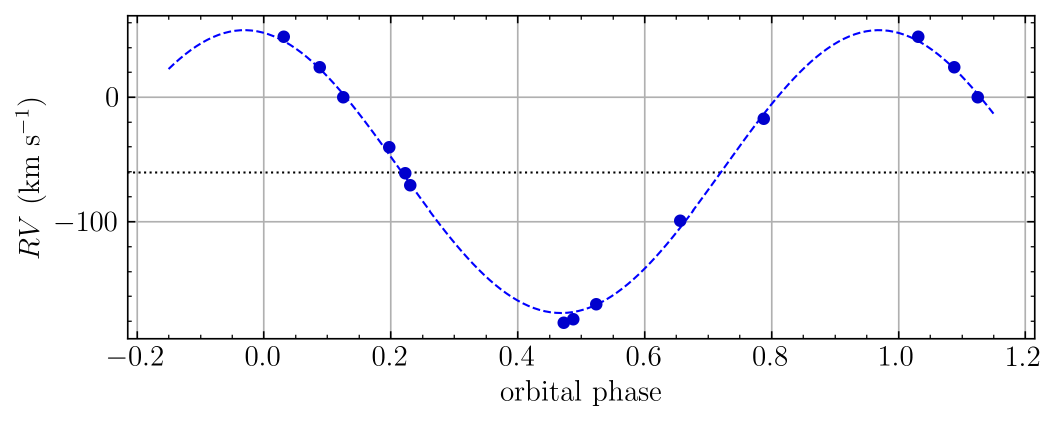} %
    \caption[]%
    {Same figure as Fig.~\ref{app:chiper_2311} but for NGC~457 37. }%
    \label{}%
\end{figure}

\begin{figure}%
    \centering
    \includegraphics[width=7cm]{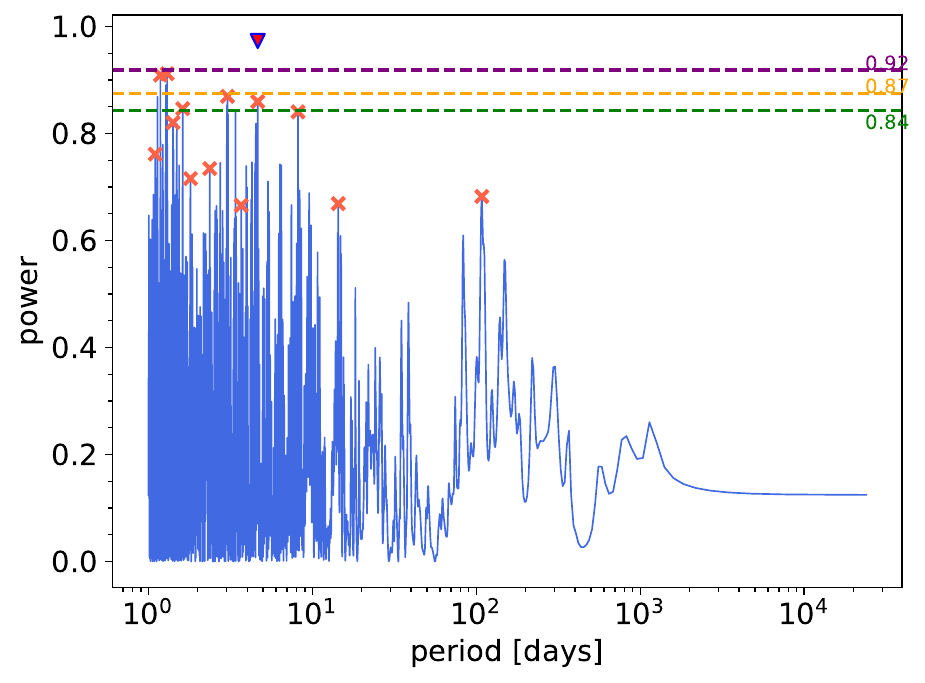} 
    \includegraphics[width=8.7cm]{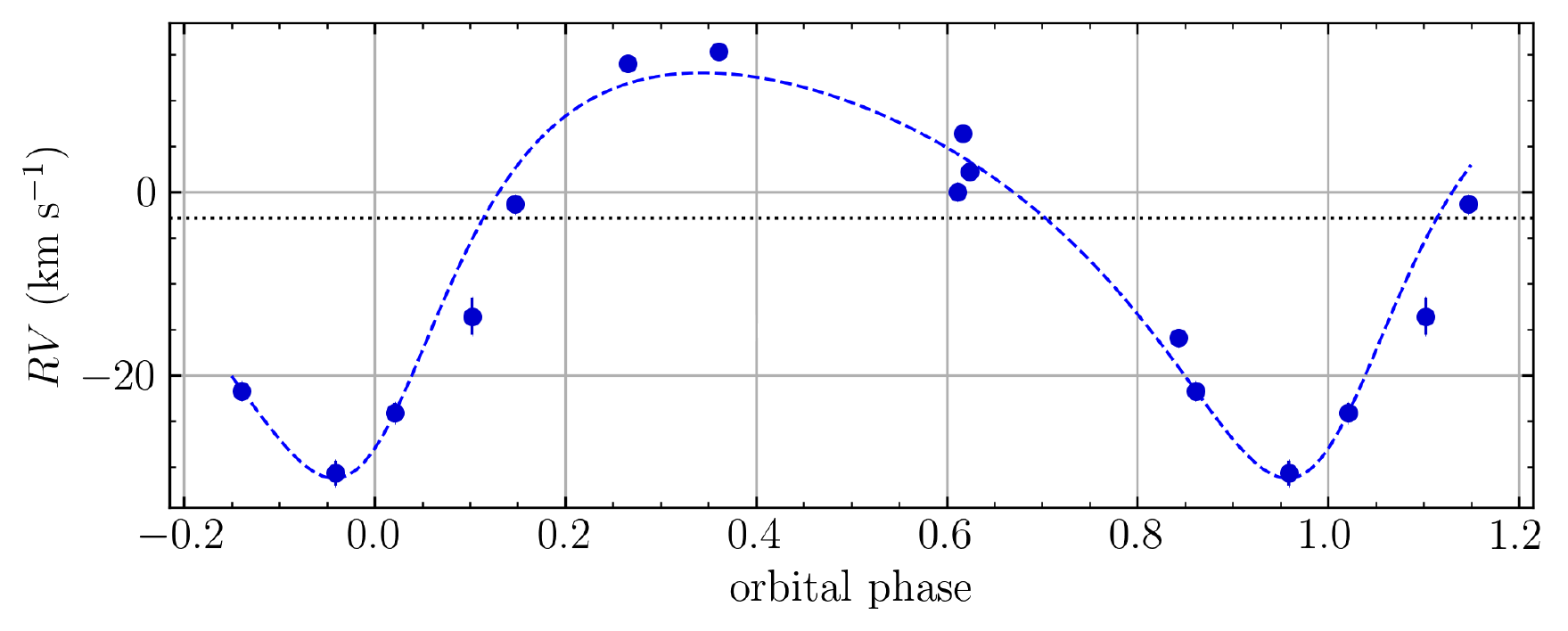} %
    \caption[]%
    {Same figure as Fig.~\ref{app:chiper_2311} but for NGC~457 54. }%
    \label{}%
\end{figure}

\begin{figure}%
    \centering
    \includegraphics[width=7cm]{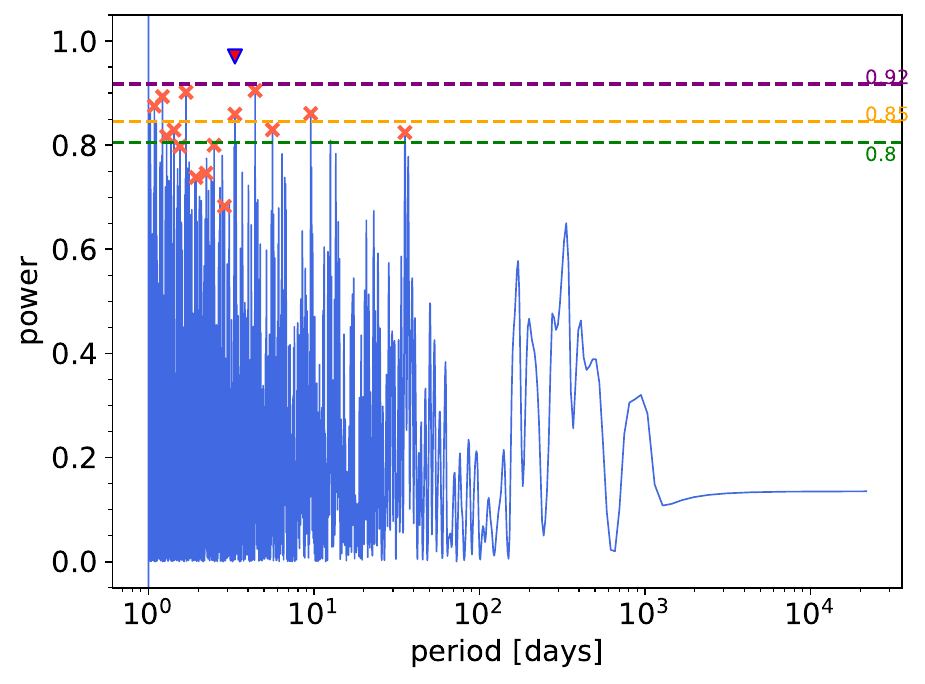} 
    \includegraphics[width=8.7cm]{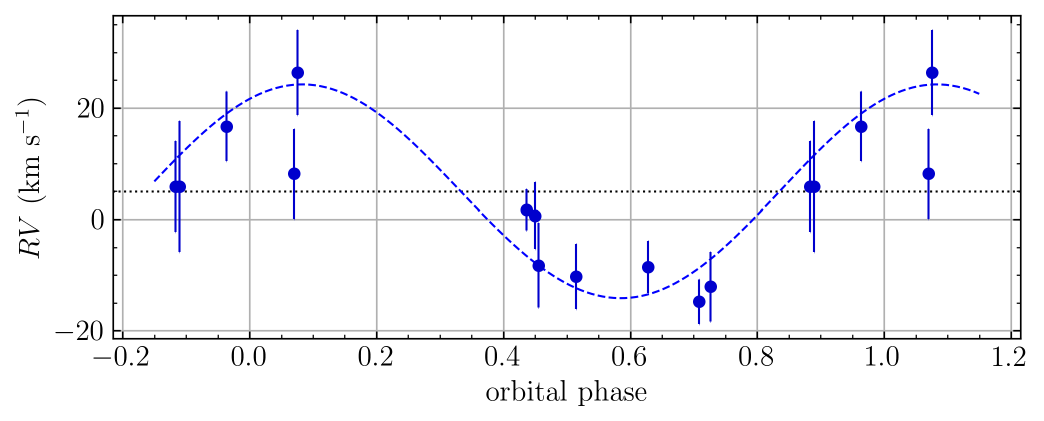} %
    \caption[]%
    {Same figure as Fig.~\ref{app:chiper_2311} but for NGC~457 91. }%
    \label{}%
\end{figure}

\begin{figure}%
    \centering
    \includegraphics[width=7cm]{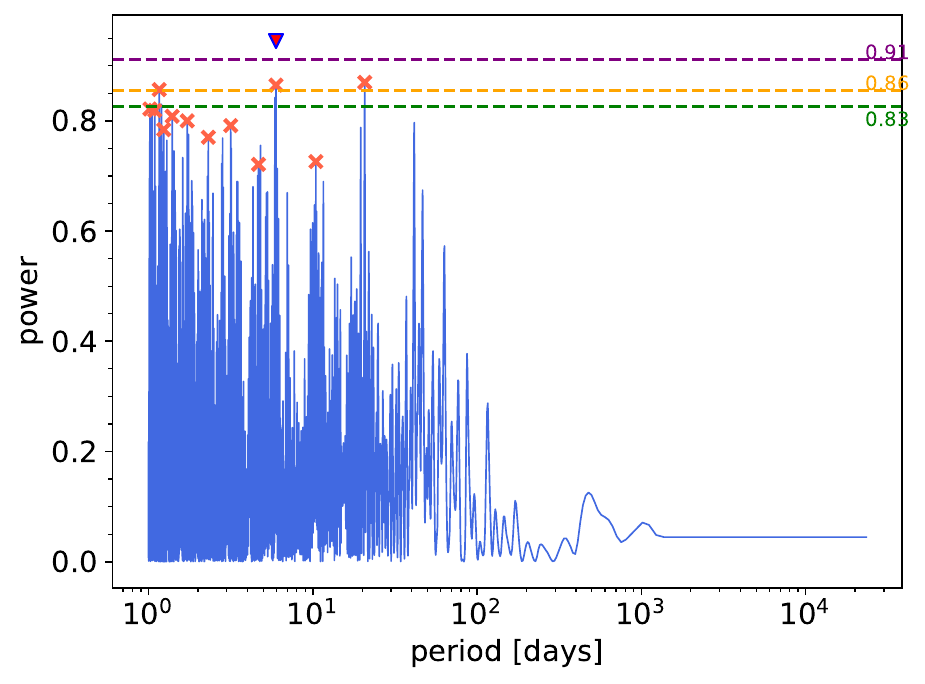} 
    \includegraphics[width=8.7cm]{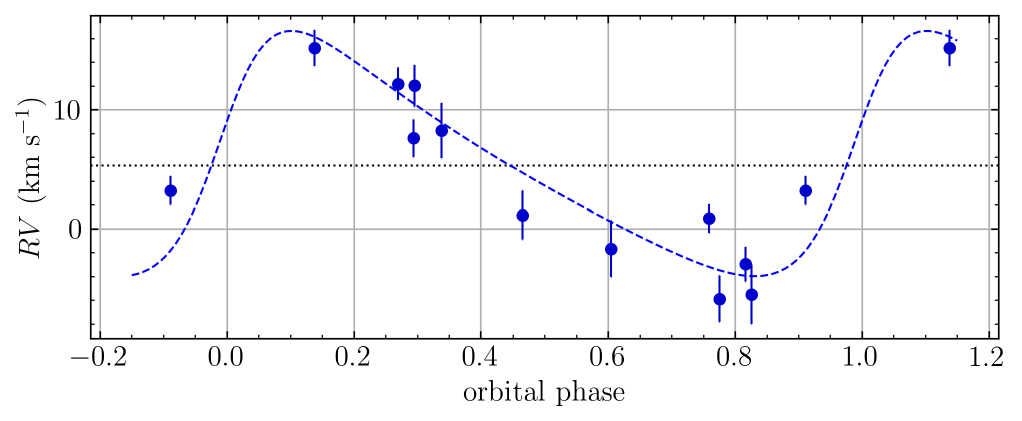} %
    \caption[]%
    { Same figure as Fig.~\ref{app:chiper_2311} but for NGC~457 100. }%
    \label{app:ngc457_100}%
\end{figure}

\begin{figure}%
    \centering
    \includegraphics[width=7cm]{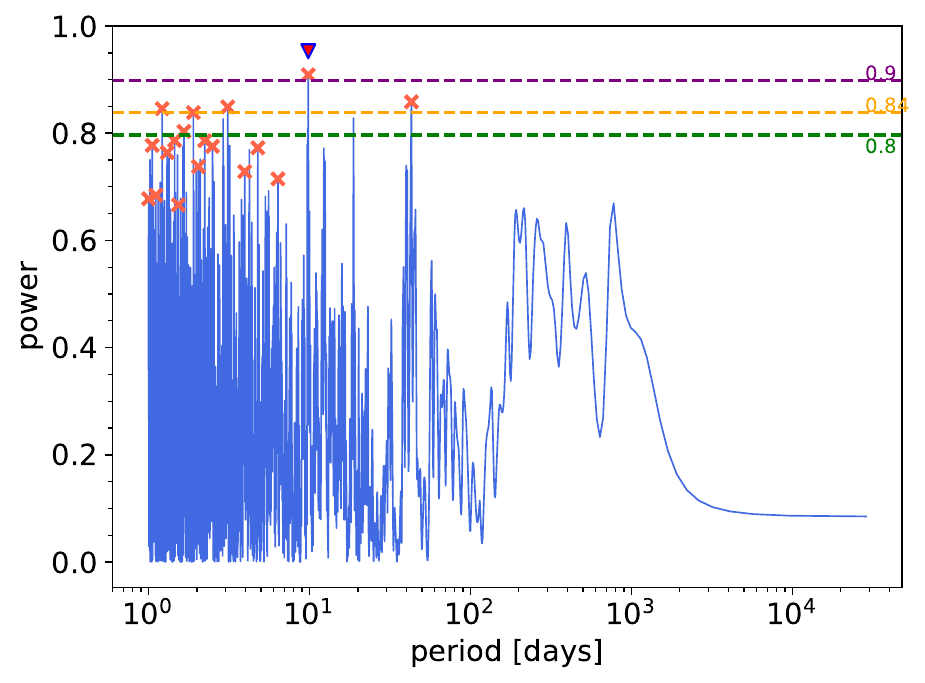} 
    \includegraphics[width=8.7cm]{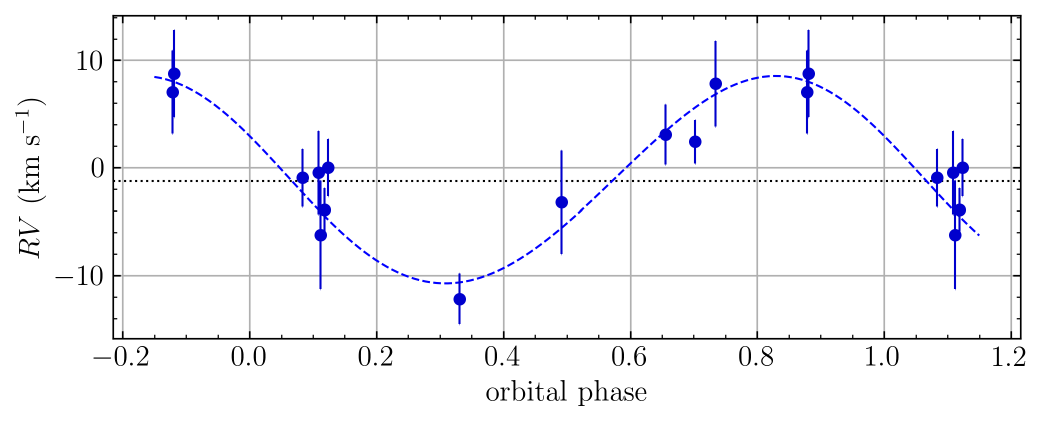} %
    \caption[]%
    {Same figure as Fig.~\ref{app:chiper_2311} but for NGC~457 154. }%
    \label{}%
\end{figure}

\begin{figure}%
    \centering
    \includegraphics[width=7cm]{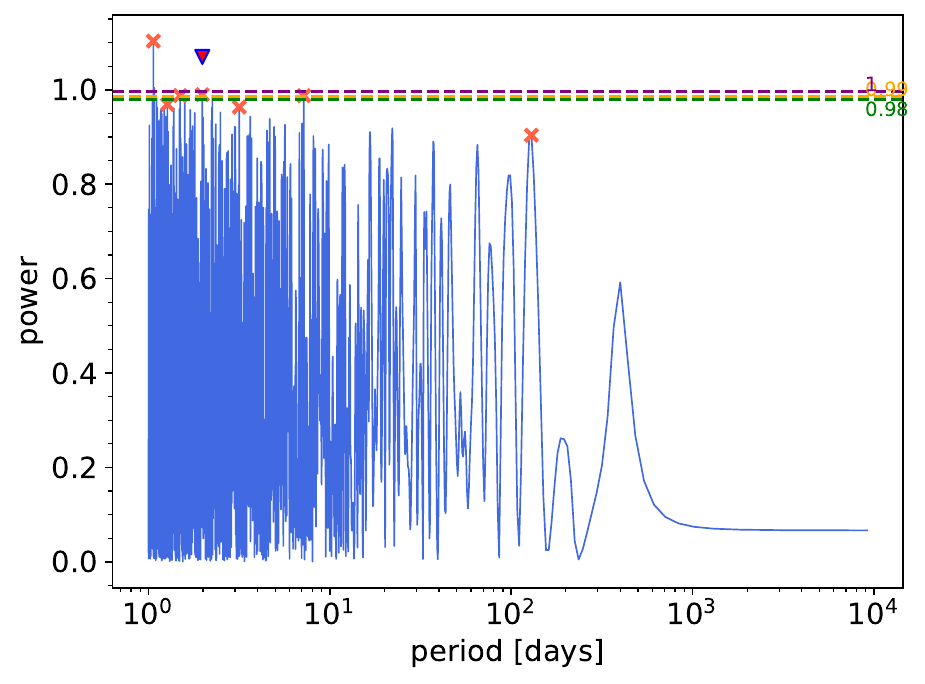} 
    \includegraphics[width=8.7cm]{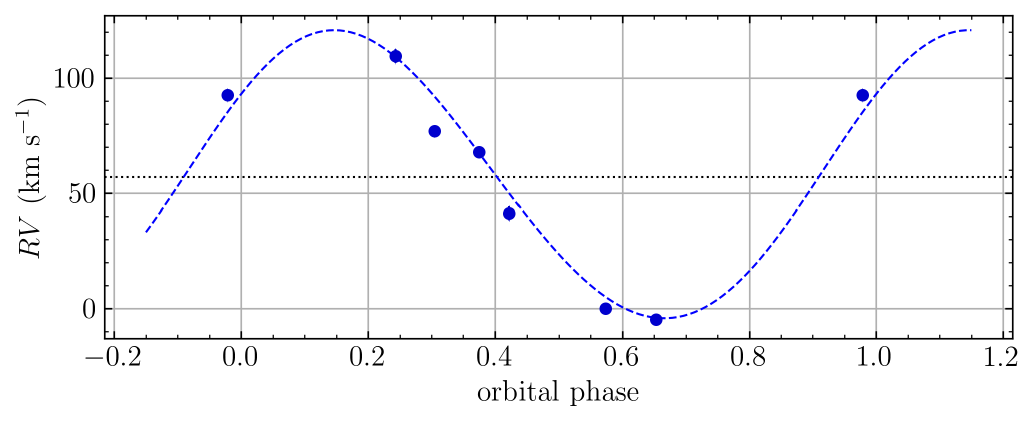} %
    \caption[]%
    {Same figure as Fig.~\ref{app:chiper_2311} but for NGC~581 EM* GGA 52.  }%
    \label{}%
\end{figure}

\begin{figure}%
    \centering
    \includegraphics[width=7cm]{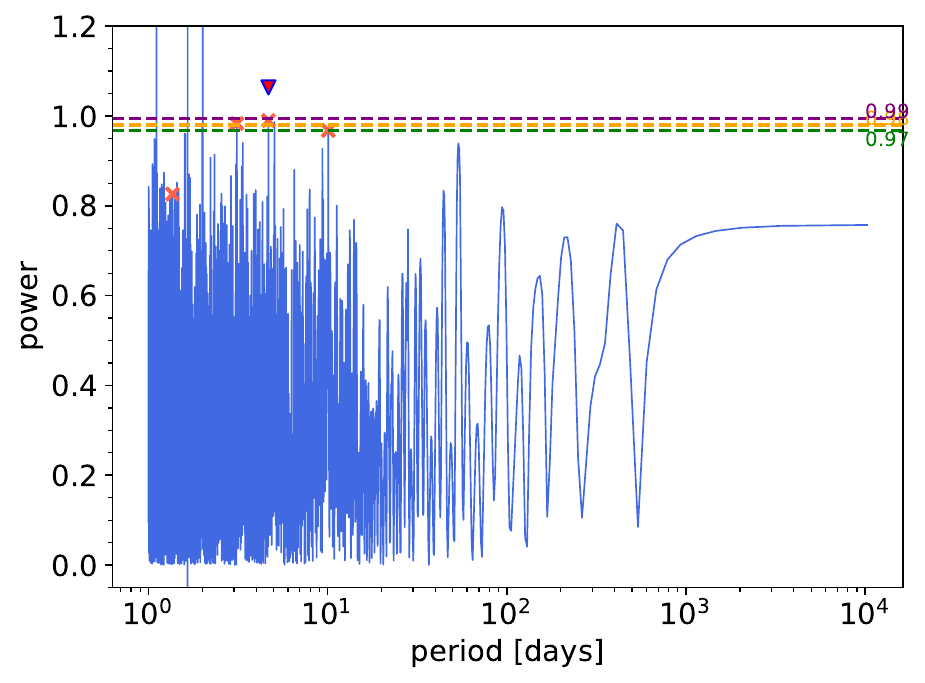} 
    \includegraphics[width=8.7cm]{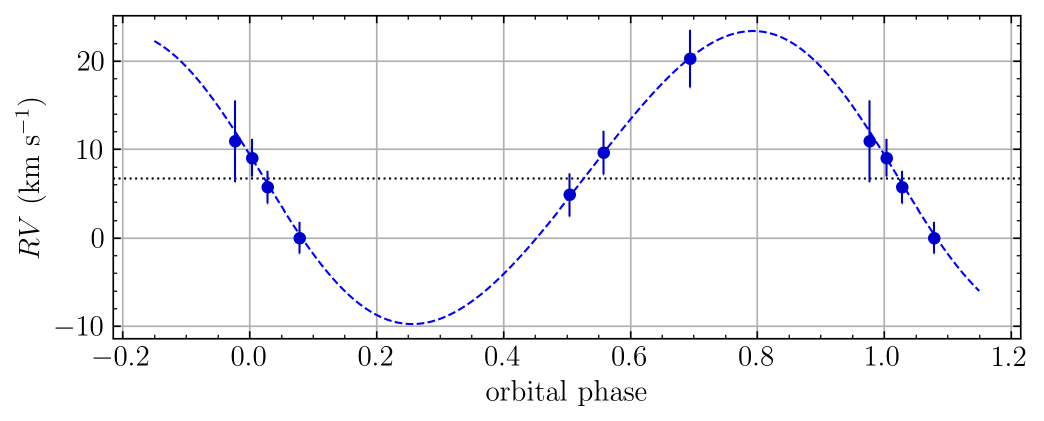} %
    \caption[]%
    {Same figure as Fig.~\ref{app:chiper_2311} but for Cl* NGC~581 ESY 19.  }%
    \label{}%
\end{figure}

\begin{figure}%
    \centering
    \includegraphics[width=7cm]{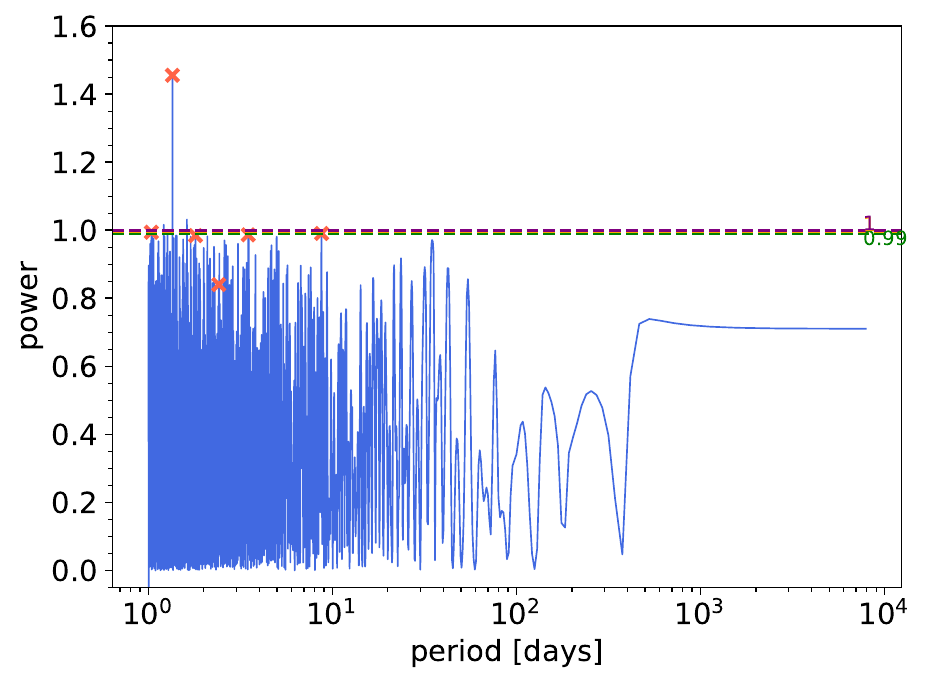} 
    \includegraphics[width=8.7cm]{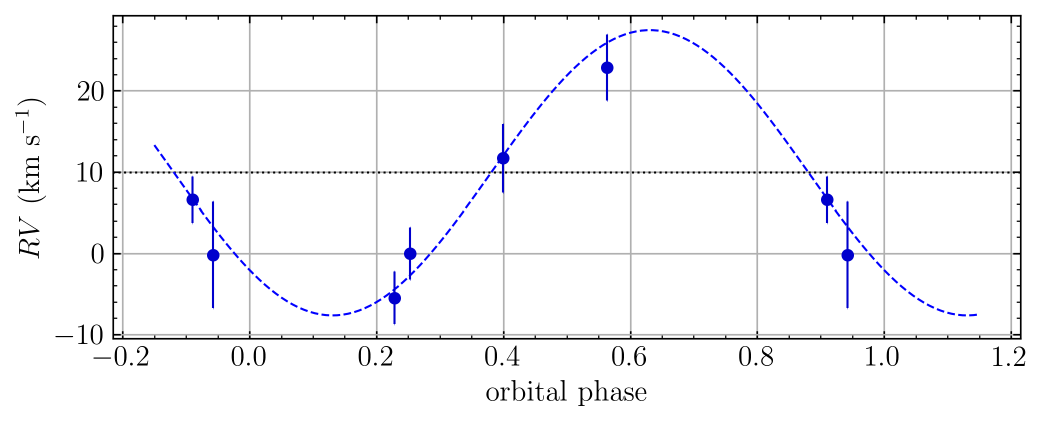} %
    \caption[]%
    {Same figure as Fig.~\ref{app:chiper_2311} but for Cl* NGC~581 ESY 23.  }%
    \label{}%
\end{figure}

\begin{figure}%
    \centering
    \includegraphics[width=7cm]{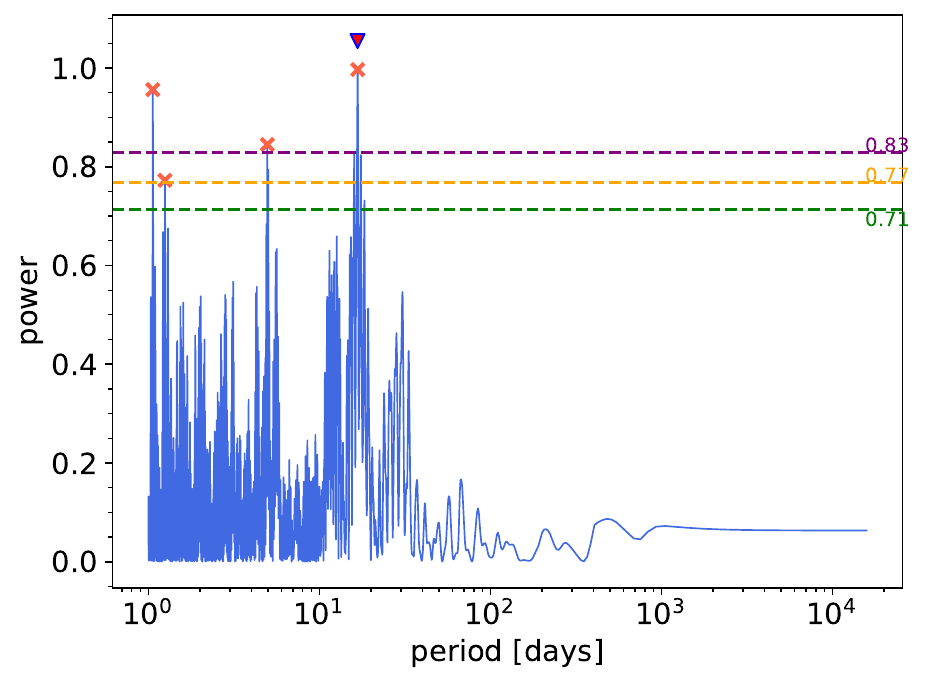} 
    \includegraphics[width=8.7cm]{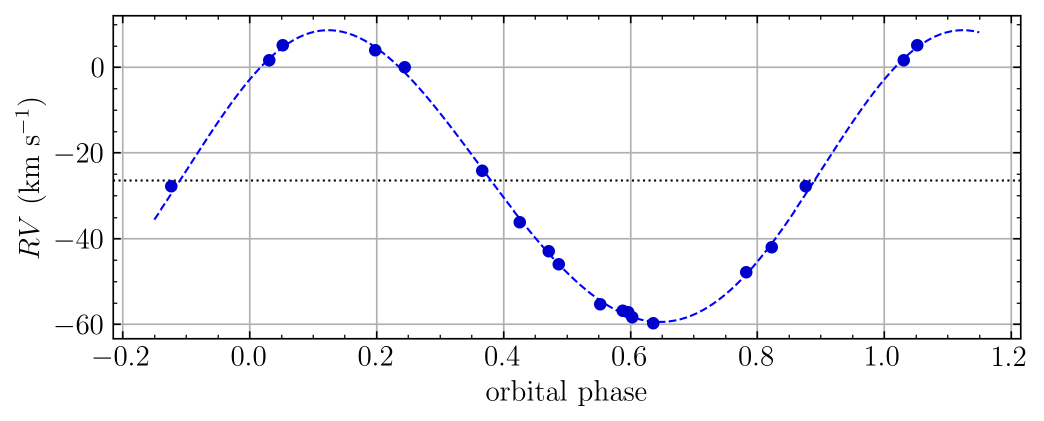} %
    \caption[]%
    {Same figure as Fig.~\ref{app:chiper_2311} but for NGC~1960 009.  }%
    \label{}%
\end{figure}

\begin{figure}%
    \centering
    \includegraphics[width=7cm]{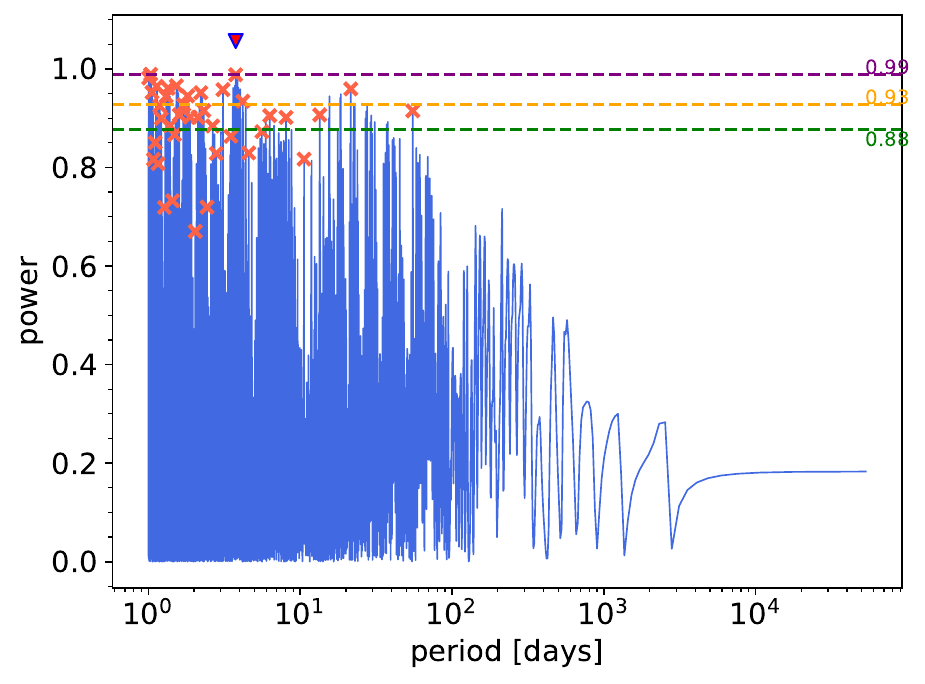} 
    \includegraphics[width=8.7cm]{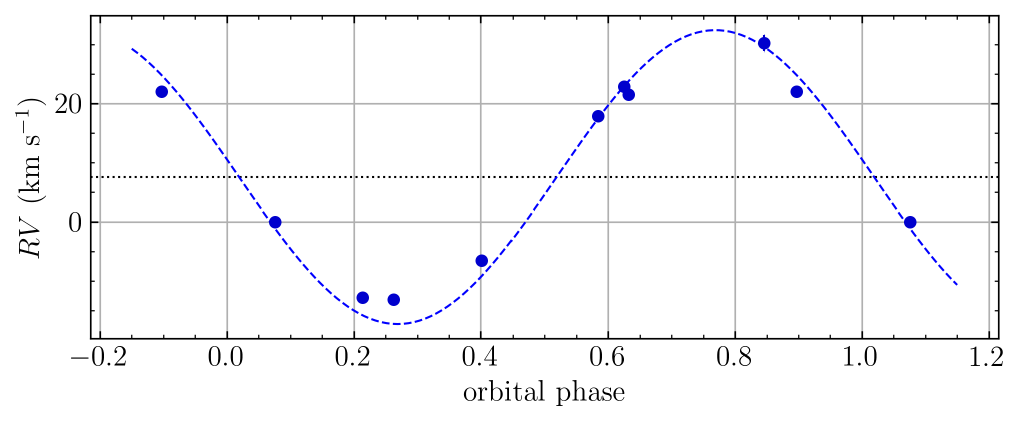} %
    \caption[]%
    {Same figure as Fig.~\ref{app:chiper_2311} but for NGC~1960 016.}%
    \label{}%
\end{figure}

\begin{figure}%
    \centering
    \includegraphics[width=7cm]{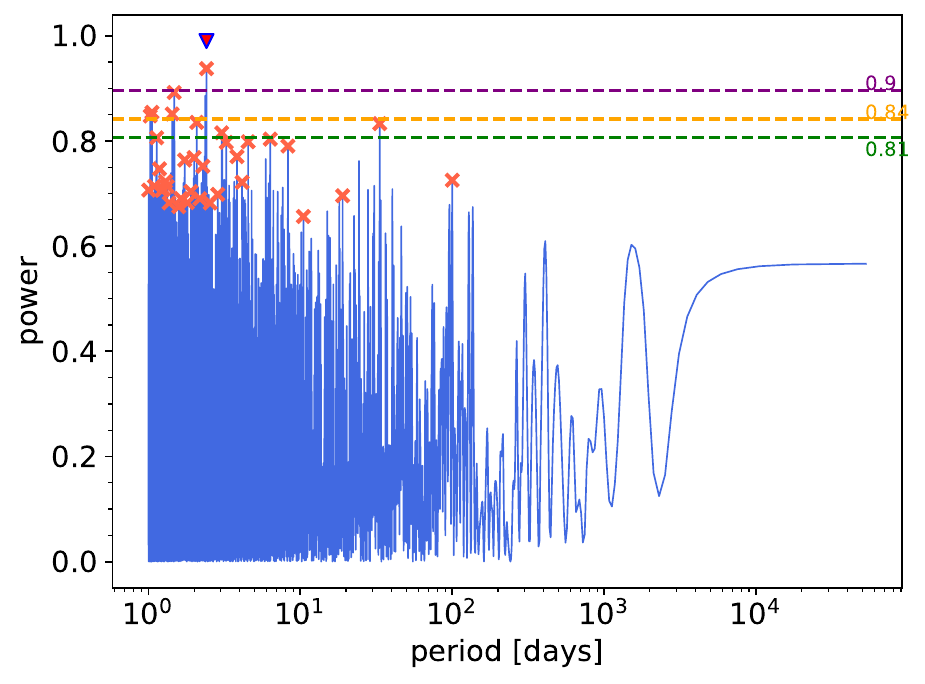} 
    \includegraphics[width=8.7cm]{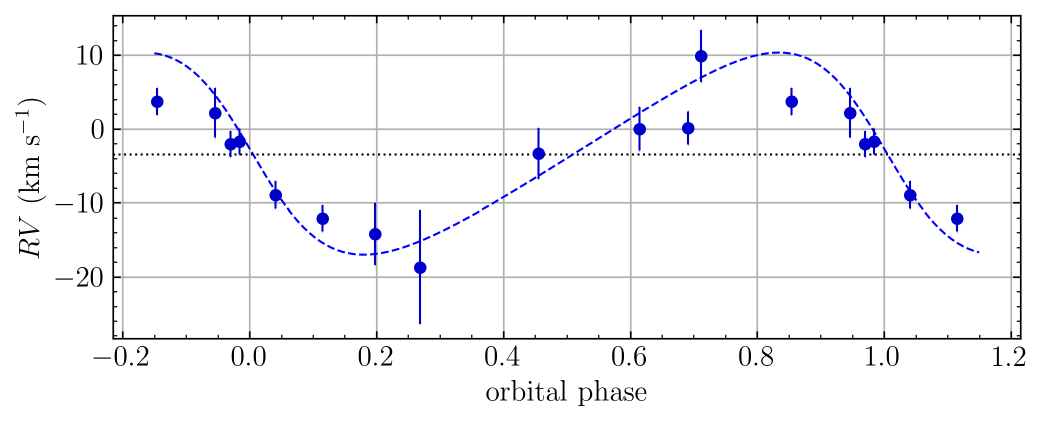} %
    \caption[]%
    {Same figure as Fig.~\ref{app:chiper_2311} but for NGC~1960 027.  }%
    \label{}%
\end{figure}

\begin{figure}
    \centering
    \includegraphics[width=7cm]{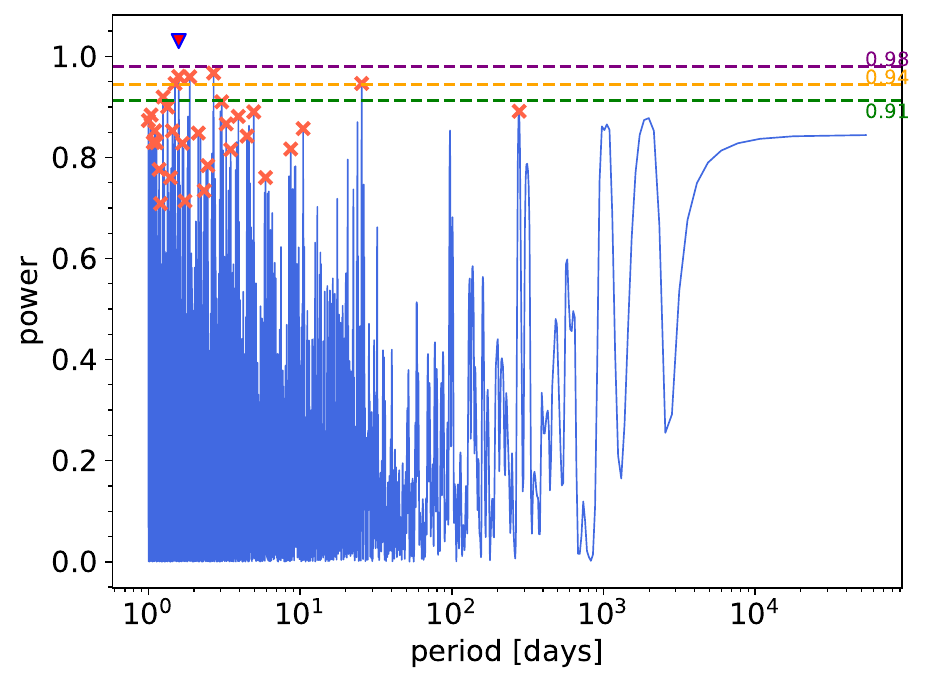} 
    \includegraphics[width=8.7cm]{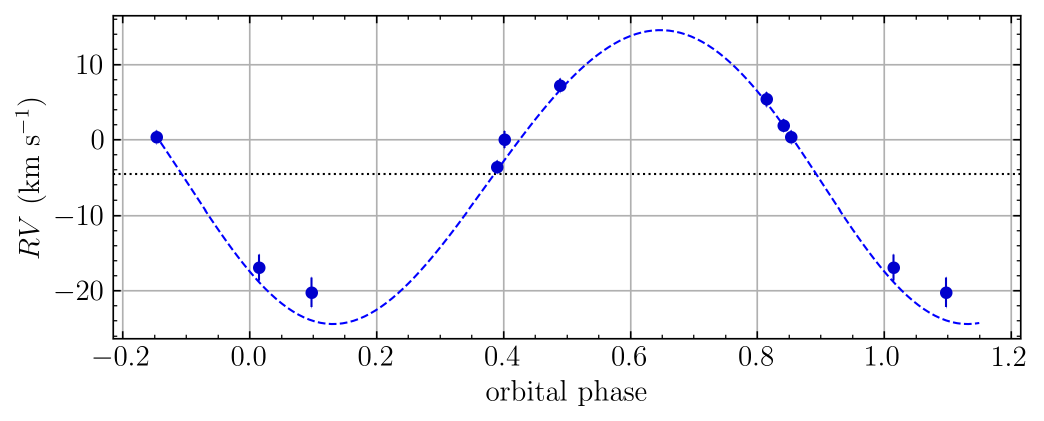} %
    \caption[]%
    {Same figure as Fig.~\ref{app:chiper_2311} but for NGC~1960 109.  }%
    \label{app:ngc1960_109}%
\end{figure}

\begin{figure}%
    \centering
    \includegraphics[width=7cm]{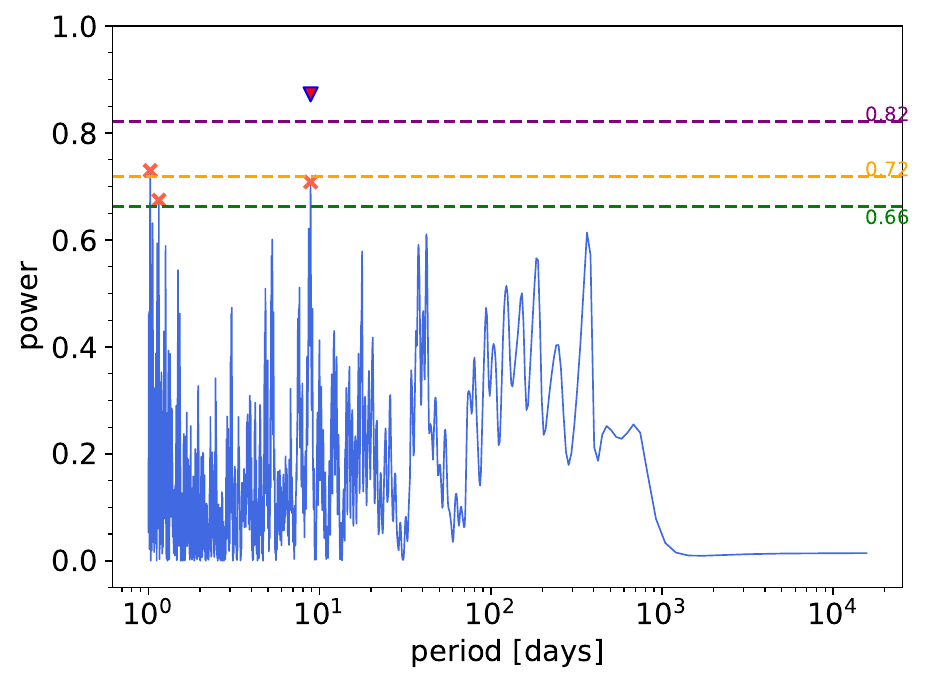} 
    \includegraphics[width=8.7cm]{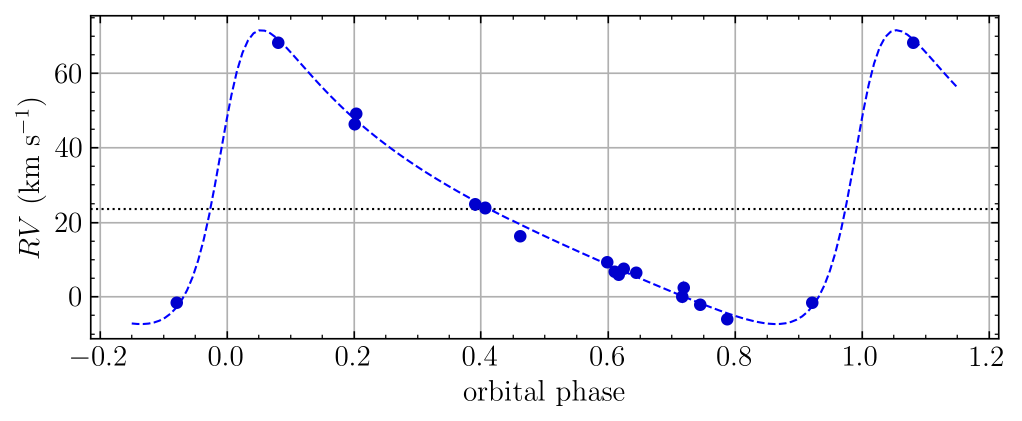} %
    \caption[]%
    {Same figure as Fig.~\ref{app:chiper_2311} but for NGC~1960 134. }%
    \label{}%
\end{figure}

\begin{figure}%
    \centering
    \includegraphics[width=7cm]{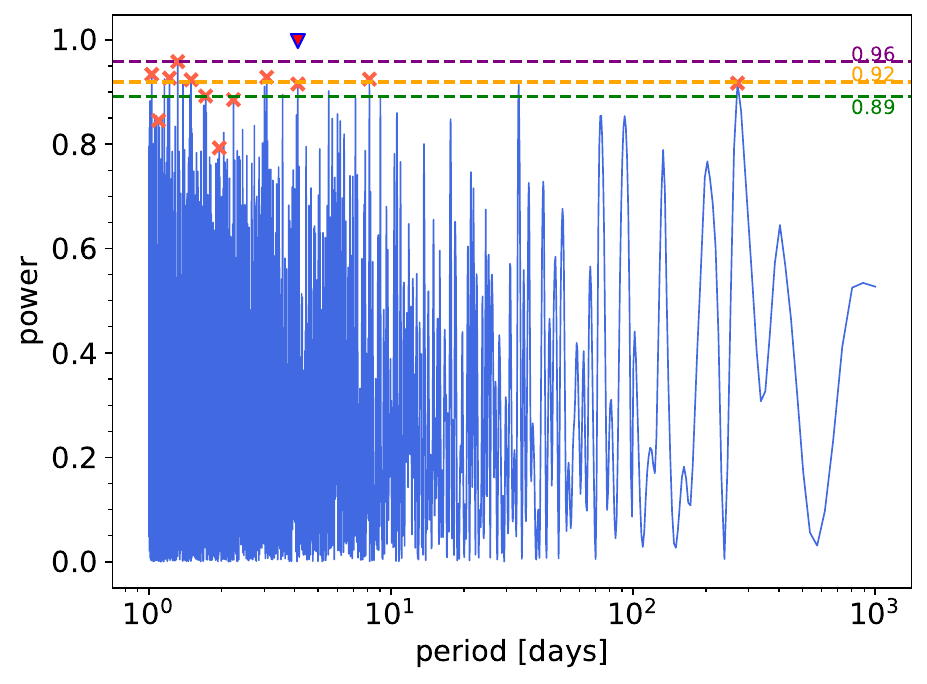} 
    \includegraphics[width=8.7cm]{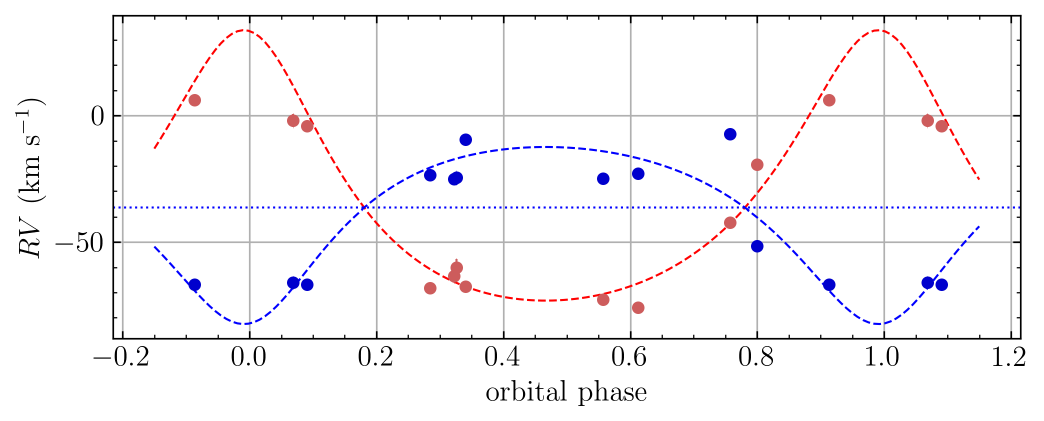} %
    \caption[]%
    {Same figure as Fig.~\ref{app:chiper_2311} but for $\chi$ Per 2392. }%
    \label{}%
\end{figure}

\begin{figure}%
    \centering
    \includegraphics[width=7cm]{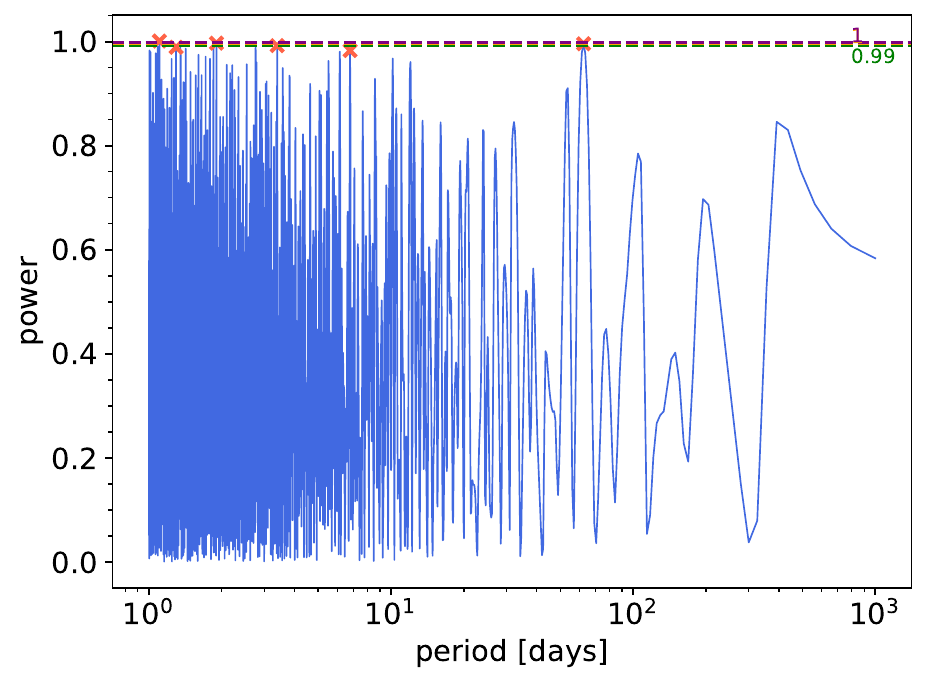} 
    \includegraphics[width=8.7cm]{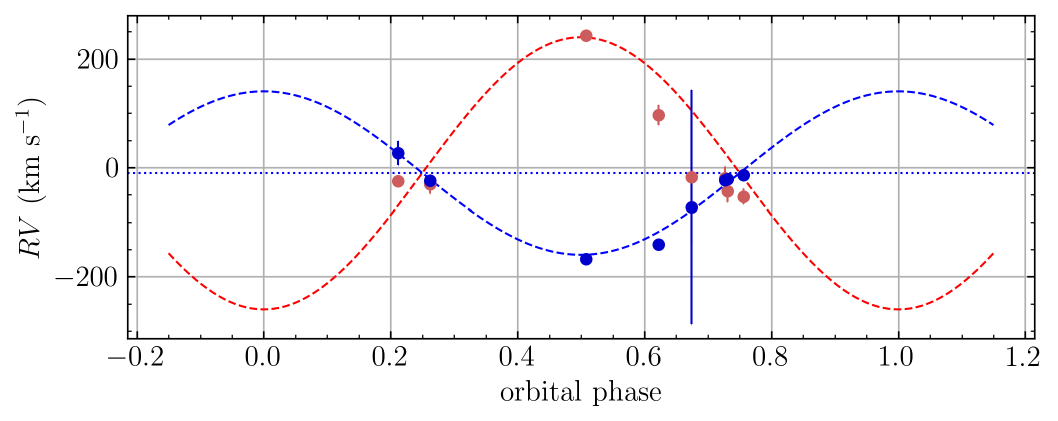} %
    \caption[]%
    {Same figure as Fig.~\ref{app:chiper_2311} but for NGC 457 85. }%
    \label{}%
\end{figure}

\begin{figure}%
    \centering
    \includegraphics[width=7cm]{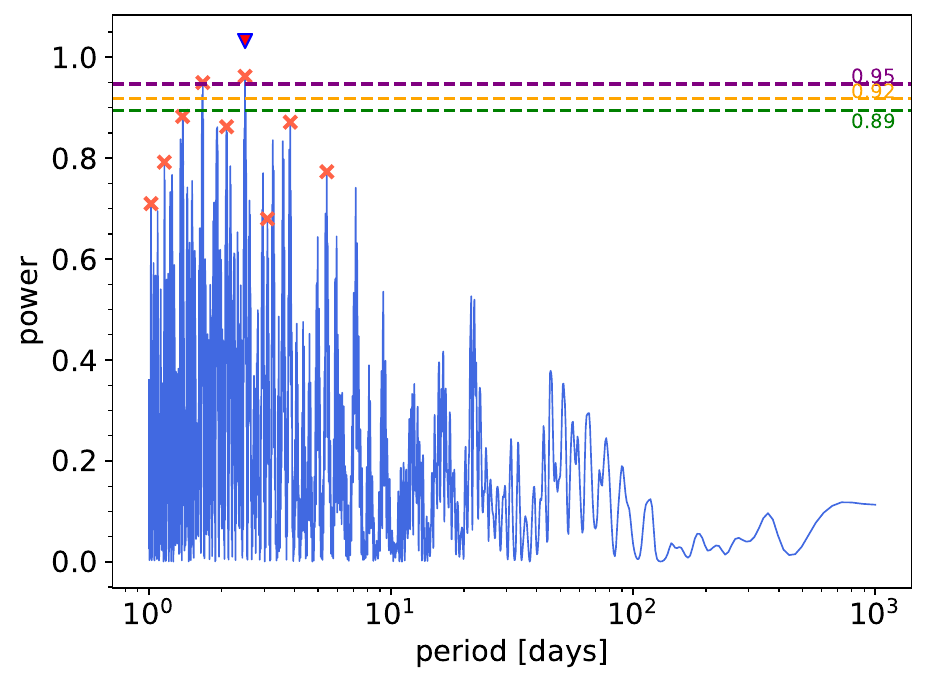} 
    \includegraphics[width=8.7cm]{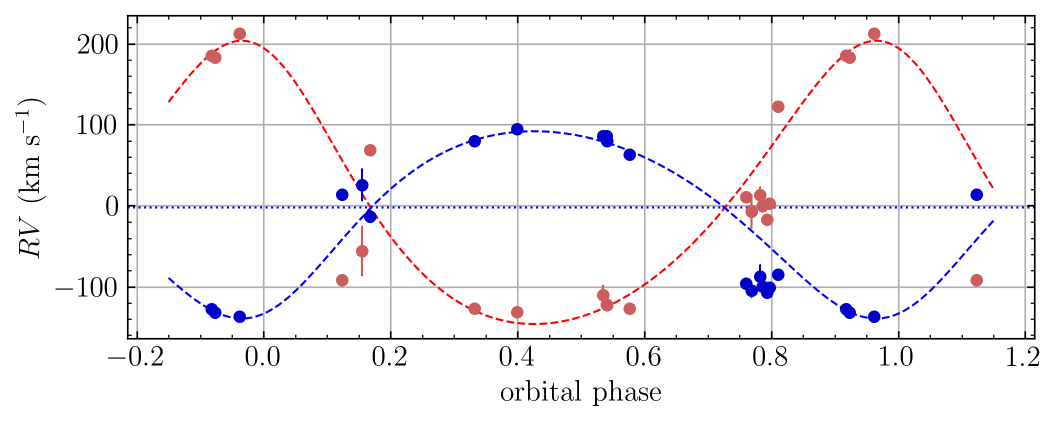} %
    \caption[]%
    {Same figure as Fig.~\ref{app:chiper_2311} but for NGC 1960 008. }%
    \label{}%
\end{figure}



\bsp	
\label{lastpage}
\end{document}